\documentclass[prd,aps,showpacs,floats,letterpaper,floatfix,superscriptaddress]{revtex4}
\usepackage[dvips]{graphicx}

\newcommand{\eqref}[1]{(\ref{#1})}

\newcommand{\hLN}{\hat{\mathbf{L}}_{\mathrm{N}}}
\newcommand{\hL}{\hat{\mathbf{L}}}
\newcommand{\hS}{\hat{\mathbf{S}}}
\newcommand{\vh}{\mathbf{h}}
\newcommand{\hl}{\hat{\mathbf{\lambda}}}
\newcommand{\hn}{\hat{\mathbf{n}}}
\newcommand{\hN}{\hat{\mathbf{N}}}
\newcommand{\hJ}{\hat{\mathbf{J}}}
\newcommand{\vx}{\mathbf{x}}
\newcommand{\vv}{\mathbf{v}}
\newcommand{\va}{\mathbf{a}}
\newcommand{\ve}{\mathbf{e}}
\newcommand{\vT}{\mathbf{T}}
\newcommand{\vJ}{\mathbf{J}}
\newcommand{\vS}{\mathbf{S}}
\newcommand{\vL}{\mathbf{L}}

\newcommand{\beq}{\begin{equation}}
\newcommand{\eeq}{\end{equation}}
\newcommand{\bea}{\begin{eqnarray}}
\newcommand{\eea}{\end{eqnarray}}
\newcommand{\ba}{\begin{array}}
\newcommand{\ea}{\end{array}}

\newlength{\sizeonefig}
\newlength{\sizetwofig}
\begin{document}

\title{Detecting gravitational waves from precessing
binaries of spinning compact objects: Adiabatic limit}

\author{Alessandra Buonanno}

\affiliation{Institut d'Astrophysique de Paris (GReCO, FRE 2435 du
CNRS), 98$^{\rm bis}$ Boulevard Arago, 75014 Paris, France}

\affiliation{Theoretical Astrophysics and Relativity, California
Institute of Technology, Pasadena, CA 91125}

\author{Yanbei Chen}

\affiliation{Theoretical Astrophysics and Relativity, California
Institute of Technology, Pasadena, CA 91125}

\author{Michele Vallisneri}

\affiliation{Jet Propulsion Laboratory,
California Institute of Technology, Pasadena, CA 91109}

\affiliation{Theoretical Astrophysics and Relativity, California
Institute of Technology, Pasadena, CA 91125}

\date{25 November 2002}

\begin{abstract}
Black-hole (BH) binaries with single-BH masses $m=5\mbox{--}20
M_\odot$, moving on quasicircular orbits, are among the most promising
sources for first-generation ground-based gravitational-wave (GW)
detectors. Until now, the development of data-analysis techniques to
detect GWs from these sources has been focused mostly on nonspinning
BHs. The data-analysis problem for the spinning case is complicated by
the necessity to model the precession-induced modulations of the GW
signal, and by the large number of parameters needed to characterize the system, including the initial directions of the spins, and the
position and orientation of the binary with respect to the GW
detector.  In this paper we consider binaries of maximally spinning
BHs, and we work in the adiabatic-inspiral regime to build families of
modulated detection templates that (i) are functions of very few
physical and phenomenological parameters, (ii) model remarkably well
the dynamical and precessional effects on the GW signal, with fitting
factors on average $\gtrsim 0.97$, but (iii) might require
increasing the detection thresholds, offsetting at least partially the
gains in the fitting factors.  Our detection-template families are
quite promising also for the case of neutron-star--black-hole
binaries, with fitting factors on average $\approx 0.93$. For these
binaries we also suggest (but do not test) a further template family,
which would produce essentially exact waveforms written directly in
terms of the physical spin parameters.
\end{abstract}

\maketitle
\vskip 0.2truecm

\section{Introduction}
\label{sec1}

A world-wide network of laser-interferometer gravitational-wave (GW)
detectors, recently built \cite{Inter}, has by now begun
operation. Inspiraling binaries of compact objects, such as black
holes (BHs) and neutron stars (NSs) are among the most promising
astrophysical sources for these detectors. The GWs from the inspirals
are expected to enter the frequency band of good detector sensitivity
during the last few seconds or minutes of evolution of the binaries;
GW scientists plan to track the phase of the signals very accurately,
and to enhance the signal-to-noise ratio by integrating the signals
coherently over their duration in the detector band. This is achieved
by filtering the detector output with a bank of \emph{templates} that
represent our best theoretical predictions for the signals.

Until now, the development of data-analysis techniques has been
focused mostly on binaries containing NSs (whose spins are negligible
for data-detection purposes) and nonspinning BHs~\cite{DA}.
Nonspinning, high-mass BHs pose a delicate problem: the
breakdown of the post--Newtonian (PN) expansion in the last stages of
the inspiral makes it hard to prepare reliable templates for the
detection of binary BHs (BBHs) of relatively high total mass (say, $10\mbox{--}40 M_\odot$) with
LIGO--VIRGO interferometers. Various resummation techniques, such as
Pad\'e approximants \cite{DIS1} and Effective One-Body (EOB)
techniques \cite{BD,EOB3PN} have been developed to extend the validity
of PN formalism~\cite{PN}. Damour, Iyer, and Sathyaprakash
\cite{DIS3} compared the templates generated by different PN
treatments, and found that they can be very different.  In a
companion paper to the present one \cite[henceforth BCV1]{BCV1}, we
investigated this issue for the GW signals emitted by comparable-mass
BBHs with a total mass $M =10\mbox{--}40 M_\odot$. In BCV1 we proposed
a few examples of detection template families (DTFs), built either as
time series or directly in the frequency domain, which try to address
the failure of the PN expansion. The philosophy behind DTFs is to
replace a family of signals that correspond to a specific mathematical
model of the binary with families that can cover a broader range of
plausible signals. Because the direct correspondence with the
mathematical model is lost, DTFs are appropriate for the purpose of
first detecting GW signals, but do not give direct estimates of
physical parameters, such as the masses of the binary constituents.
[Within the EOB framework, see also the recent paper by
Damour, Iyer, Jaranowski, and Sathyaprakash \cite{DIJS}, where
the authors extend 3PN EOB templates with seven \emph{flexibility parameters} and then show that the unextended 3PN templates already \emph{span} the ranges of the flexibility parameters consistent with plausible 4PN effects.]

Very little is known about the statistical distribution of spins
for the BHs in binaries: the spins could very well be
large. Apostolatos, Cutler, Sussman, and Thorne \cite[henceforth
ACST]{ACST94,apostolatos2} have shown that when this is the case, the
evolution of the GW phase and amplitude during the inspiral will be
significantly affected by spin-induced modulations and
irregularities. In a BBH, these effects can become dramatic if the two
spins are large and they are not exactly aligned or antialigned with
the orbital angular momentum. If this happens, there is a considerable
chance that the analysis of interferometer data, carried out without
taking spin effects into accounts, could miss the signals from these
spinning BBHs altogether.
The gravitational waveforms from binaries of spinning compact objects
depend on many parameters: the masses and spins of the objects, the
angles that describe the relative orientations of detector and binary,
and the direction of propagation of GWs to the detector. In practice
it is impossible, due to the extremely high computational cost, to
filter the signals with a template bank parametrized by all of these
parameters.
One strategy is that of providing
\emph{effective} templates that depend on fewer parameters, but that
have still reasonably high overlaps with the expected physical
signals.  An interesting suggestion, built on the results obtained in
Ref.~\cite{ACST94}, came from Apostolatos~\cite{apostolatos2}, who
introduced a modulational sinusoidal term in the frequency-domain
phase of the templates to capture the effects of precession. However,
while Apostolatos' family reduces the number of parameters
considerably, its computational requirements are still very
high. Moreover, using an approximated analytical model of NS--BH waveforms, Grandcl\'ement, Kalogera and Vecchio \cite{GKV} showed that this family fails to capture those waveforms satisfactorily (see however Ref.\ \cite{GK} for a hierarchical scheme that can improve the fit by adding ``spikes?? in the template phasing).

In this paper, complementary to BCV1, we study the data analysis of
GWs from binaries with spinning BHs; for simplicity, we restrict our
analysis to the adiabatic limit, where the two compact objects in the
binary (either two BHs, or a NS and a BH) follow an adiabatic sequence
of \emph{spherical orbits} driven by radiation reaction (RR). The
denomination of spherical orbits comes from the fact that the orbital
plane is not fixed in space, but precesses, so the orbits trace a
complicated path on a (slowly shrinking) spherical surface.
We neglect the problems caused by the failure of PN expansion in these binaries (note that the conservative part of the EOB framework~\cite{BD} has already been extended to the spinning case by Damour~\cite{TD},
providing a tool to move beyond the adiabatic approximation;
we plan to add radiation-reaction effects to this model, and to study the consequences on GW emission and detection elsewhere). 
Here, we carry out a detailed study of PN precessional dynamics and of GW
generation in precessing binaries in the adiabatic limit, and we use the resulting insights
to build a new class of modulated effective templates where
modulational effects are introduced in both the frequency-domain
amplitude and frequency-domain phase of the templates. The
mathematical structure of our templates suggests a way to search
automatically over several of the parameters (in strict analogy to the
automatic search over initial template phase in the data analysis of
nonspinning binaries), reducing computational costs significantly. We
argue that our families should capture very well the expected physical
signals.

We note here a shift in perspective from BCV1. In this paper, we use
the PN equations for the two-body dynamics of spinning compact objects
to build a \emph{fiducial} model (our \emph{target} model) that
represents our best knowledge of the expected physical
signals. Because we cannot use the target model directly for data
analysis (it has too many parameters), we build effective template
families with fewer parameters. These families are then compared with
the target model for a variety of binary parameters, to gauge their
ability to match the physical signals (their \emph{effectualness}~\cite{DIS1}).
On the other hand, in BCV1 we employed several variants of the PN
equations (with diverging behaviors in the late phase of inspiral) to
identify a range of \emph{plausible} physical signals; we then built
our DTFs so that they would match \emph{all} of the PN target models
satisfactorily. This said, we shall still refer to the template
families developed in the present paper as DTFs. We direct the reader
to BCV1 for a simple introduction to matched-filtering techniques and
their use in GW data analysis (developed in the literature
by various authors~\cite{DA,DIS1,DIS3}), and for an explanation of some of the
notation used in this paper.

This paper is organized as follows. In Sec.~\ref{sec2} we define the
target model, and we explain the conventions used to represent the
generation and propagation of GWs.  In Sec.~\ref{sec3} we study the
two-body dynamics of spinning compact objects, looking for the
features that are especially relevant to the data-analysis problem.
Using this insight, in Sec.~\ref{sec4} we formulate our DTFs, and we
also describe two families of standard stationary-phase--approximation
(SPA) templates, to be used as a comparison when evaluating the
performance of the DTFs.  In Sec.~\ref{sec5} we discuss the overlap
and false-alarm statistics of our DTFs.  In Sec.~\ref{sec6} we
evaluate the performance of our DTFs for BBHs and NS--BH binaries, and
we briefly discuss a more advanced (and very promising) template
family for NS--BH systems.  In Sec.~\ref{sec9} we summarize our
conclusions.

Throughout this paper we adopt the noise spectral density for LIGO-I
given by Eq.\ (28) of BCV1. The projected VIRGO noise curve is quite
different (deeper at low frequencies, with a displaced
peak-sensitivity frequency). So our results for high-mass binaries
cannot be applied naively to VIRGO. We plan to repeat this study for
VIRGO in the near future.

\section{Definition of the target model}
\label{sec2}

In this section we define the \emph{target model} used in this paper
as a fiducial representation of the GW signals expected from
precessing binaries of spinning compact objects. We restrict our
analysis to the adiabatic regime where the inspiral of the compact
objects can be represented as a sequence of quasicircular orbits. At
any point along the inspiral, a binary of total mass $M = m_1 + m_2$
and symmetric mass ratio $\eta = m_1 m_2 / M^2$ is completely
described by the orbital angular frequency $\omega$, the orbital phase
$\Psi$, the direction $\hL_N \propto \mathbf{r} \times \mathbf{v}$ of
the orbital angular momentum, and the two spins $\vS_1=\chi_1 m_1^2
\hS_1$ and $\vS_2=\chi_2 m_2^2 \hS_2$, where $\hS_{1,2}$ are unit
vectors and $0<\chi_{1,2}<1$. Throughout this paper we shall use carets
to denote unit vectors, and we shall adopt geometrical units.

In Sec.\ \ref{sec2.1} we write the PN equations that govern the
adiabatic evolution of the binary and the precession of $\hL_N$ and of
$\vS_{1,2}$. All the target waveforms used to test the effectualness~\cite{DIS1}
of our DTFs are obtained by integrating these equations
numerically. The validity of the adiabatic approximation is discussed
in App.\ \ref{appendixA}. In Sec.\ \ref{sec2.2} we discuss our
criterion for stopping the numerical integration of the evolution
equations at the point where the adiabatic approximation ceases to be
valid.  In Sec.\ \ref{sec2.3}, building on Refs.~\cite{FC,ACST94,K},
we describe a formalism for computing
the response of a ground-based detector to the GWs generated by a
spinning binary; the response is not just a function of the trajectory
of the binary, but also of the relative direction and orientation of
binary and detector. The formalism describes also how the precession
of the binary modulates the detector response. Last, in Sec.\
\ref{sec2.4} we give a classification of all the parameters that enter
the expression for the detector response, distinguishing those that
specify the evolution of the binary itself from those that describe
the relative direction and orientation of binary and detector.

\subsection{Equations for an adiabatic sequence of precessing spherical orbits}
\label{sec2.1}

The path of the binary across the sequence of quasicircular orbits is
described by the adiabatic evolution of the orbital angular frequency
$\omega$ up to 3.5PN order~[\onlinecite{2PN,BIWW,2.5PNand3.5PN,BFIJ},\onlinecite{DIS3}] with spin effects included up to 2PN order~[\onlinecite{KWW},\onlinecite{2PN},\onlinecite{K}],
\bea &&
\frac{\dot{\omega}}{\omega^2}=\frac{96}{5}\,\eta\,(M\omega)^{5/3}\Bigg
\{1-\frac{743+924\,\eta}{336}\,(M\omega)^{2/3} -\Bigg
(\frac{1}{12}\sum_{i=1,2}\bigg[\chi_i\left(\hL_N\cdot\hS_i\right)\left(113\frac{m_i^2}{M^2}+75\eta\right)\bigg]
-4\pi\Bigg)(M\omega) \nonumber \\ && + \Bigg
(\frac{34\,103}{18\,144}+\frac{13\,661}{2\,016}\,\eta+\frac{59}{18}\,\eta^2
\Bigg )\,(M\omega)^{4/3}
-\frac{1}{48}\eta\chi_1\chi_2\left[247(\hS_1\cdot\hS_2)-721(\hL_N\cdot\hS_1)(\hL_N\cdot\hS_2)\right]\,(M\omega)^{4/3}
\nonumber \\ && -\frac{1}{672}\,(4\,159 +
15\,876\,\eta)\,\pi\,(M\omega)^{5/3} + \Bigg[
\left(\frac{16\,447\,322\,263}{139\,708\,800}-\frac{1\,712}{105}\gamma_E+\frac{16}{3}\pi^2\right)+
\left(-\frac{273\,811\,877}{1\,088\,640}+\frac{451}{48}\pi^2-\frac{88}{3}\hat\theta
\right)\eta \nonumber \\
&&+\frac{541}{896}\eta^2-\frac{5\,605}{2\,592}\eta^3
-\frac{856}{105}\log\left[16(M\omega)^{2/3}\right] \Bigg] (M\omega)^2
+ \Bigg (
-\frac{4\,415}{4\,032}+\frac{358\,675}{6\,048}\,\eta+\frac{91\,495}{1\,512}\,\eta^2
\Bigg )\,\pi\,(M\omega)^{7/3} \Bigg \}\,,
\label{omegadot}
\eea
where $\gamma_E=0.577\ldots$ is Euler's constant, and where
$\hat{\theta}$ is an arbitrary parameter that enters the GW flux at
3PN order~\cite{BFIJ} and that could not be fixed in the
regularization scheme used by the authors of Ref.~\cite{BFIJ}. Note
that in Eq.~(\ref{omegadot}) we set the static parameter
$\omega_s=0$~\cite{DJSd}.
(Note for v4 of this paper on gr-qc: Eq.\ (1) is now revised as per Ref.\ \cite{errata};
the parameter $\hat{\theta}$ has been determined to be 1039/4620 \cite{thetapar}.)
The precession equations for the two spins
are (see, for instance, Eqs.~(4.17b,c) of Ref.~\cite{K} or
Eqs.~(11b,c) of Ref.~\cite{ACST94})
\bea
\label{S1dot}
\dot{\vS}_1&=&
\frac{(M\omega)^2}{2M}
\left\{ \eta\,(M\omega)^{-1/3}\,\left(4+3\frac{m_2}{m_1}\right)\hL_N
+ \frac{1}{M^2}\,\left[\vS_2-3(\vS_2\cdot\hL_N)\hL_N\right]\right\}\times\vS_1
\,,\\
\label{S2dot}
\dot{\vS}_2&=&
\frac{(M\omega)^2}{2M}
\left\{ \eta\, (M\omega)^{-1/3}\left(4+3\frac{m_1}{m_2}\right)\hL_N
+ \frac{1}{M^2}\,\left[\vS_1-3(\vS_1\cdot\hL_N)\hL_N\right]\right\}\times\vS_2
\,,
\eea
where we have replaced $r$ and $|\vL_N|$ by their leading-order
Newtonian expressions in $\omega$,
\beq
\label{rLofomega}
r=\left(\frac{M}{\omega^2}\right)^{1/3}\,, \quad \quad
|\vL_N|=\mu\, r^2\omega=\eta\,M^{5/3}\omega^{-1/3}\,.
\eeq
This approximation is appropriate because the next spin-precession
term is ${\cal O}(\omega^{1/3})$ higher than the leading order, while
next terms in the expressions of $r$ and $|\vL_N|$ are ${\cal
O}(\omega^{2/3})$ higher.

The precession of the orbital plane (defined by the normal vector
$\hL_N$) can be computed as follows. From Eqs.~(4.7) and (4.11) of
Ref.~\cite{K} we see that the total angular momentum $\vJ$ and its
rate of change $\dot{\vJ}_{\rm RR}$ (due to RR) depend on $\omega$,
$\hLN$ and $\vS_{1,2}$ (schematically) as ($\vS=\vS_1+\vS_2$)
\beq
\label{JtotPN}
\vJ=\vL+\vS=
\underbrace{\eta\,M^2\, (M \omega)^{-1/3}\,
\hLN\,\left[1+{\cal O}(\omega^{2/3})\right] -
\eta\,(M\omega)^{2/3}\,\vS_{\rm eff}}_{\vL}
 +\vS\,,
\eeq
\beq
\label{JdotRR}
\dot{\vJ}_{\rm RR}=
-\frac{32}{5}\eta^2 M (M\omega)^{7/3}\, \hL_N \left[1+{\cal
    O}(\omega^{2/3})\right]  +
{\cal O}(\omega^{10/3})\,\hS_1 +
{\cal O}(\omega^{10/3})\,\hS_2\,,
\eeq
where the combination
\beq
\label{Seff}
\vS_{\rm eff} \equiv \left ( 1 + \frac{3}{4}\,\frac{m_2}{m_1} \right)
\,\vS_1 + \left ( 1 + \frac{3}{4}\,\frac{m_1}{m_2} \right )\,\vS_2
\eeq
is known as \emph{effective spin}~\cite{TD}. Note that both terms
in the $\mathbf{L}$ brace of Eq.~(\ref{JtotPN}) originate from
orbital angular momentum (the second term comes from the spin-orbit coupling).
Taking the time derivative of (\ref{JtotPN}), we obtain
\beq
\label{JdotPN}
\dot{\vJ} = \eta\, M^2\,(M\omega)^{-1/3}\,\dot{\hL}_{\rm N}
\left[1+{\cal O}(\omega^{2/3})\right] - {\cal O}(\omega^{2/3})\, \dot{\vS}_{\rm eff} +
\dot{\vS} + \left[{\cal O}(\omega^{7/3})\,\hLN - {\cal O}(\omega^{10/3})\,\vS_{\rm eff}\right]\,,
\eeq
where to get the last term on the right-hand side we have used $\dot{\omega} = {\cal O}(\omega^{11/3})$. Comparing Eqs.~(\ref{JdotPN}) and (\ref{JdotRR}), projecting out only the direction perpendicular to $\hLN$, and keeping only the terms up to the leading and next-to-leading orders, we get
\bea
\label{Lhdot}
\dot{\hL}_N= -\frac{(M\omega)^{1/3}}{\eta M^2}\,\dot{\vS}
=\frac{\omega^2}{2M}
\Bigg\{&& \!\!\!\!\!\! \left[\left(4+3\frac{m_2}{m_1}\right)\,\vS_1+\left(4+3\frac{m_1}{m_2}\right)\,\vS_2\right]\times\hL_N
\nonumber \\
&& \!\!\!\!\!\! -\frac{3\,\omega^{1/3}}{\eta\,M^{5/3}}\left[(\vS_2\cdot\hL_N)\,\vS_1+
(\vS_1\cdot\hL_N)\,\vS_2\right]\times\hL_N \Bigg\}\,.
\eea
Thus, we now have
the set of four equations (\ref{omegadot})--(\ref{S2dot}) and
(\ref{Lhdot}) for the four variables $\omega$, $\vS_1$, $\vS_2$, and
$\hL_N$. We follow Ref.~\cite{K}, Eq.~(4.15), in defining the
\emph{accumulated orbital phase} $\Psi$ as
\beq
\label{orbitalpsi}
\Psi \equiv\int_{t_i}^{t}\omega \,dt=\int_{\omega_i}^{\omega }\,\frac{\omega}{\dot{\omega}}\,d\omega\,.
\eeq
This phase describes the position of the two compact objects along the
instantaneous circular orbits of the adiabatic sequence; the phase
of the GW waveforms, as detected by a ground-based detectors, differs
from this by precessional effects, as explained below in Sec.\ \ref{sec2.3}.

\subsection{Endpoint of evolution}
\label{sec2.2}

The orbital energy of the two-body system at 2PN and 3PN orders,
expressed as a function of $\omega$, and assuming the static parameter
$\omega_s=0$~\cite{DJS,DJSd}, reads~[\onlinecite{2PN},\onlinecite{KWW},\onlinecite{BFIJ}]
\bea
\label{s1}
E_{\rm 2PN}(\omega) &=& -\frac{\mu}{2}\,(M\omega)^{2/3}\,\left \{
1 - \frac{(9+\eta)}{12}\,(M\omega)^{2/3} + \frac{8}{3} \hL_N \cdot \frac{\vS_{\rm eff}}{M^2}\,
(M\omega)+
\frac{1}{24}(-81 + 57 \eta - \eta^2)\,(M\omega)^{4/3} \right . \nonumber \\
&& \left. + \frac{1}{\eta}\,\left [\frac{\vS_1}{M^2}\cdot \frac{\vS_2}{M^2} - 3 \left(\hL_N \cdot \frac{\vS_1}{M^2}\right) \left(\hL_N \cdot \frac{\vS_2}{M^2}\right)
 \right ]\,(M\omega)^{4/3} \right \}\,,\\
\label{s2}
E_{\rm 3PN}(\omega) &=& E_{\rm 2PN}(\omega) -\frac{\mu}{2}\,(M\omega)^{2/3}\,\left \{
\left [-\frac{675}{64} + \left (\frac{34445}{576}-\frac{205}{96}\pi^2 \right )\eta
-\frac{155}{96}\eta^2-\frac{35}{5184}\eta^3 \right ]\,(M\omega)^{2}\right \}\,.
\eea
In the context of our adiabatic approximation, it is natural to stop
the integration of Eqs.~(\ref{omegadot})--(\ref{S2dot}) and
(\ref{Lhdot}) at the point (the Minimum Energy Circular Orbit, or
MECO) where the energy $E_{n\rm PN}$ reaches a minimum,
\beq
\label{MECO}
{\rm MECO:} \quad \frac{dE_{n\rm PN}}{d\omega}=0\,;
\eeq
after this point the adiabatic approximation breaks down \cite{note37}.
(The MECO is discussed by
Blanchet~\cite{LB} for nonspinning binaries under the name ICO, for
Innermost Circular Orbit.)  However, if we find that $\dot{\omega}=0$
(which implies certainly that the adiabatic approximation has become
invalid) before the MECO is reached, we stop the evolution there. In
BCV1 we noticed that for nonspinning binaries this behavior occurs for
the 2.5PN evolutions, but not at 2PN, 3PN and 3.5PN orders.

Throughout this paper, we shall call the
instantaneous frequency of GWs at the endpoint of evolution the \emph{ending frequency}, which, up to a correction that arises from precessional effects, is
twice the instantaneous orbital frequency defined in this section.
It so happens (see BCV1) that a knowledge of
the ending frequency is important to cut off the candidate detection
templates at the point where we know too little about the physical
signals to model them further.  In Sec.~\ref{sec3.1} we study the
dependence of the ending frequency on the spins of the binary.

\subsection{Gravitational waveforms}
\label{sec2.3}

\begin{figure}[t]
\begin{center}
\includegraphics{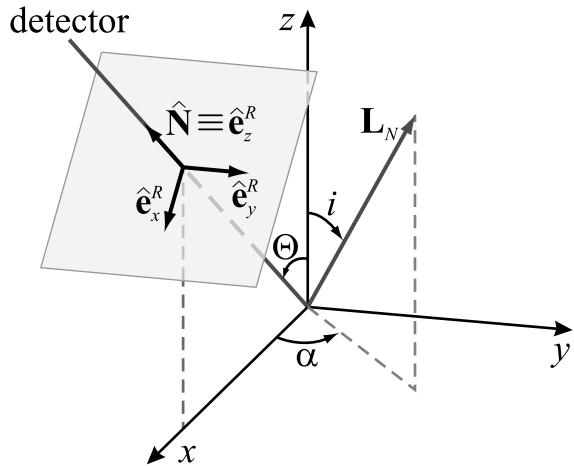}
\caption{\label{Fig1} Source and radiation frames in the
FC convention~\protect\cite{FC}.}
\end{center}
\end{figure}

As we have seen, the trajectory of the inspiraling binary is obtained
by integrating Eqs.\ \eqref{omegadot}--\eqref{S2dot} and \eqref{Lhdot}
for the time evolution of $\omega(t)$, $\vS_1(t)$, $\vS_2(t)$ and
$\hL_N(t)$.  To determine the corresponding gravitational waveforms,
we need to choose a specific coordinate system.  We follow the
convention proposed by Finn and Chernoff \cite[henceforth FC]{FC}, and
also adopted by Kidder~\cite{K}. FC employ a fixed (\emph{source})
coordinate system with unit vectors $\{\ve_x^S$, $\ve_y^S$,
$\ve_z^S\}$ (see Fig.~\ref{Fig1}).  For a circular orbit, the
leading-order mass-quadrupole waveform is (throughout this paper, we
use geometrical units)
\beq
\label{hij}
h^{ij}=\frac{2\mu}{D}\,\left(\frac{M}{r}\right)\,Q_{c}^{ij}\,,
\eeq
where $D$ is the distance between the source and the Earth, and where
$Q_c^{ij}$ is proportional to the second time derivative of the
mass-quadrupole moment of the binary,
\beq
\label{Qc}
Q_c^{ij}=2\left[\lambda^i\,\lambda^j-n^i\,n^j\right]\,,
\eeq
with $n^i$ and $\lambda^i$ the unit vectors along the separation
vector of the binary $\mathbf{r}$ and along the corresponding relative
velocity $\mathbf{v}$. These unit vectors are related to the adiabatic
evolution of the dynamical variables by
\bea
\label{nhat_lambdahat}
\hn=\ve_1^{S}\,\cos\Phi_S+\ve_2^{S}\,\sin\Phi_S\,, \quad
\hl=-\ve_1^{S}\,\sin\Phi_S+\ve_2^{S}\,\cos\Phi_S\,;
\eea
the vectors $\ve_{1,2}^S$ form an orthonormal basis for the instantaneous orbital plane, and in the FC convention they are given by
\beq
\label{e12s}
\ve_1^{S}=\frac{\ve_z^S\times\hL_N}{\sin i}\,,\quad
\ve_2^{S}=\frac{\ve^S_z-\hL_N\,\cos i}{\sin i}\,.
\eeq
The vector $\ve_1^S$ points in the direction of the ascending node of
the orbit on the $(x,y)$ plane. The quantity $\Phi_S$ is the orbital
phase with respect to the ascending node; its evolution is given by
\beq
\dot{\Phi}_S=\omega-\dot{\alpha}\,\cos i\,,
\label{Phidot}
\eeq
where $i$ and $\alpha$ are the spherical coordinates of $\hLN$ in
the source frame, as
shown in Fig.~\ref{Fig1}.
\begin{figure}[t]
\begin{center}
\includegraphics{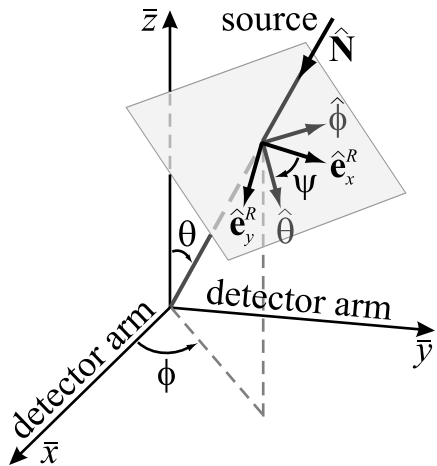}
\caption{\label{Fig2} Detector and radiation
frames in the FC convention~\protect\cite{FC}.}
\end{center}
\end{figure}
Using Eqs.~(\ref{hij}) and (\ref{nhat_lambdahat}), we can write
Eq.~(\ref{Qc}) as
\beq
\label{Qcij_e}
Q_c^{ij}=-2([\ve_+^S]^{ij}\,\cos2\Phi_S+[\ve_\times^S]^{ij}\,\sin2\Phi_S)\,,
\eeq
where the polarization tensors $\ve_+^S$ and $\ve_\times^S$ are given by
\beq
\label{epx}
\ve_+^S\equiv \ve_1^S \otimes \ve_1^S - \ve_2^S \otimes \ve_2^S\,,\quad
\ve_\times^S \equiv \ve_1^S \otimes \ve_2^S + \ve_2^S \otimes \ve_1^S\,.
\eeq
For a detector lying in the direction
$\hN=\ve_z^S\,\cos\Theta+\ve_x^S\,\sin\Theta$, it is expedient to
express GW propagation in the \emph{radiation} coordinate system with
unit vectors \{$\ve_x^R$,$\ve_y^R$,$\ve_z^R$\} (see our
Fig.~\ref{Fig1} together with, for instance, Eq.~(4.22) of
Ref.~\cite{K}), given by
\bea
\label{erx}
\ve^{R}_x &=& \ve_x^S\,\cos\Theta-\ve_z^S\,\sin\Theta \,,\\
\label{ery}
\ve^{R}_y &=& \ve_y^S \,,\\
\label{erz}
\ve^{R}_z &=& \ve_x^S\,\sin\Theta+\ve_z^S\,\cos\Theta =\hN \,.
\eea
In writing Eqs.~(\ref{erx})--(\ref{erz}) we used the fact that for a
generic binary-detector configuration, the entire system consisting of
the binary and the detector can be always rotated along the $z$ axis,
in such a way that the detector will lie in the $(x,z)$ plane.  Later in this paper (in Sec.~\ref{sec4}) we shall find it convenient to
conserve the explicit dependence of our formulas on the azimuthal
angle $\varphi$ that specifies the direction of the detector.

In the transverse-traceless (TT) gauge, the metric perturbations are
\beq
\vh^{TT}=h_+\,{\bf T}_{+} + h_{\times}\,{\bf T}_{\times}\,,
\eeq
where
\beq
\label{Tpx}
{\bf T}_+ \equiv \ve^{R}_x \otimes \ve^{R}_x-\ve^{R}_y \otimes
\ve^{R}_y\,,\quad {\bf T}_{\times} \equiv \ve^{R}_x \otimes
\ve^{R}_y+\ve^{R}_y \otimes \ve^{R}_x \,, \eeq
and
\beq
\label{hpc}
h_+=\frac{1}{2}\, h^{ij}\,[\vT_+]_{i j}\,,\quad
h_\times = \frac{1}{2}\, h^{ij}\,[\vT_\times]_{ij}\,.
\eeq
The response of a ground-based, interferometric detector (such as LIGO or VIRGO) to the GWs is~\cite{FC}
\bea h_{\rm resp} &=& F_+\,h_+ + F_\times\,h_\times \nonumber \\
&=& -\frac{2\mu}{D}\frac{M}{r}\,\left[e_+^{S\,ij}\,\cos2\Phi_S +
e_\times^{S\,ij}\,\sin2\Phi_S\right]([\vT_+]_{ij}\,F_+ +
[\vT_\times]_{ij}\,F_\times)\,,
\label{h}
\eea
where $F_+$ and $F_\times$ are the \emph{antenna patterns}, given by
\beq
\label{Fpxgeneral}
F_{+,\,\times}=\frac{1}{2}[\bar{\mathbf{e}}_{x}\otimes\bar{\mathbf{e}}_{x}€
-\bar{\mathbf{e}}_{y}€\otimes \bar{\mathbf{e}}_{y}€
]^{ij}[\mathbf{T}_{+,\,\times}]_{ij}\,,
\eeq
with $\bar{\mathbf{e}}_{x,\,y}$ the unit vectors along the orthogonal
interferometer arms. For the geometric configuration shown in
Fig.~\ref{Fig2}, with detector orientation parametrized by the angles
$\theta$, $\phi$ and $\psi$, we have
\bea
\label{Fplus}
F_+ &=& \frac{1}{2}(1+\cos^2 \theta)\,\cos 2\phi\,\cos 2 \psi -
\cos \theta\,\sin 2\phi\,\sin2\psi \,,\\
F_\times &=& \frac{1}{2}(1+\cos^2 \theta)\,\cos 2\phi\,\sin 2 \psi +
\cos \theta\,\sin 2\phi\,\cos2\psi\,.
\label{Fcross}
\eea
Inserting Eqs.~\eqref{e12s}, \eqref{epx},
\eqref{erx}--\eqref{erz}, and \eqref{Tpx} into Eq.~\eqref{h},
we get the final result~\cite{K},
\beq
\label{cqsqwaveforms}
h_{\rm resp} = C_Q\,\cos 2 \Phi_S + S_Q\,\sin 2\Phi_S\,,
\eeq
where
\bea
\label{cqeq}
C_Q &=& -\frac{4 \mu}{D}\,(M\,\omega)^{2/3}\,\left [ C_+\,F_+ +
C_\times\,F_\times \right ]\,,\\
S_Q &=& -\frac{4 \mu}{D}\,(M\,\omega)^{2/3}\,\left [ S_+\,F_+ +
S_\times\,F_\times \right ]\,,
\eea
and
\bea
C_+&=&\frac{1}{2}\cos^2\Theta(\sin^2\alpha-\cos^2i\cos^2\alpha)
+\frac{1}{2}(\cos^2i\sin^2\alpha-\cos^2\alpha) \nonumber \\
&&-\frac{1}{2}\sin^2\Theta\sin^2i
-\frac{1}{4}\sin2\Theta\sin2i\cos\alpha\,,
\\
S_+&=& \frac{1}{2}(1+\cos^2\Theta)\cos
i\sin2\alpha+\frac{1}{2}\sin2\Theta\sin i\sin\alpha\,,\\
C_{\times}&=&
-\frac{1}{2}\cos\Theta(1+\cos^2i)\sin2\alpha-\frac{1}{2}\sin\Theta\sin2i\sin\alpha\,,\\
S_{\times}&=&-\cos\Theta\cos i\cos2\alpha-\sin\Theta\sin i\cos\alpha\,.
\label{steq}
\eea

\begin{figure}[t]
\begin{center}
\includegraphics{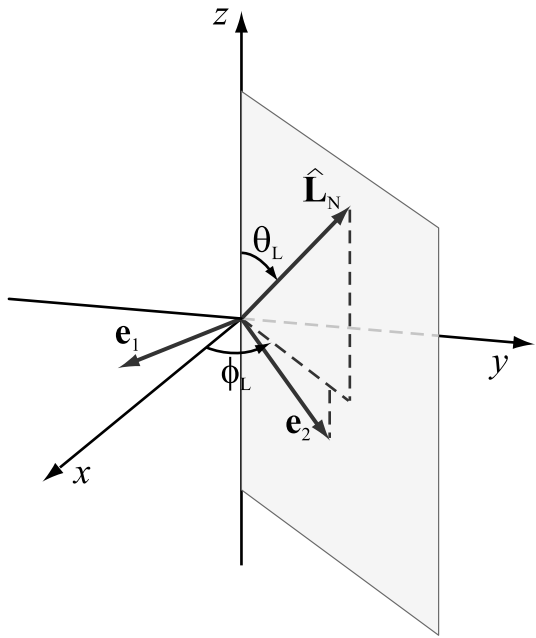}
\end{center}
\caption{Specification of the initial Newtonian orbital
angular momentum in the source frame $\{\ve_x,\ve_y,\ve_z\}$.\label{fig:LNh}}
\end{figure}

\begin{figure}[t]
\begin{center}
\includegraphics{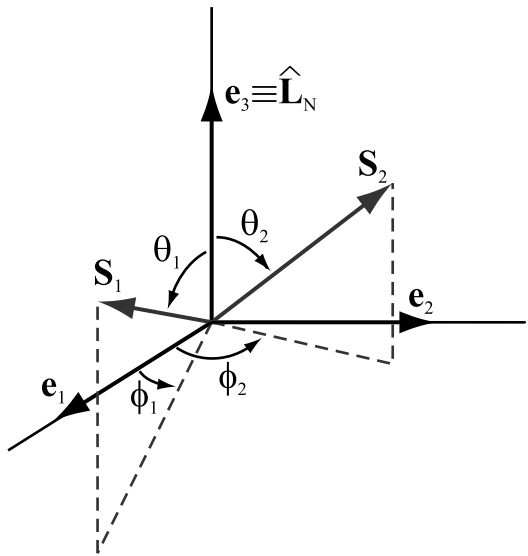}
\caption{Specification of the initial directions of the spins with
respect to the FC orthonormal basis $\{\ve_1,\ve_2,\ve_3\}$ [Eq.~(\protect{\ref{eq:e1e2}})].
\label{fig:S12def}}
\end{center}
\end{figure}

\subsection{Binary and detector parameters}
\label{sec2.4}

We shall refer to the total mass $M$, to the mass ratio $\eta =
m_1\,m_2/M^2$, and to the magnitudes of the two BH (or NS) spins $S_1$
and $S_2$ as the {\it basic parameters} of the binary.  Once these are
set, we complete the specification of a binary configuration by giving
the initial orbital phase and the components of the orbital and spin
angular momenta \emph{in the source frame}, for a given initial
frequency.  In our convention, the initial orbital angular momentum is
determined by the angles $(\theta_{\rm L_{\rm N}} \equiv i, \phi_{\rm L_{\rm
N}} \equiv \alpha)$, as shown in Fig.~\ref{fig:LNh}.
[In the rest of this paper, we shall use the notation $(\theta_{\rm L_{\rm N}},\phi_{\rm L_{\rm
N}})$ to emphasize that these angles define the direction of the orbital angular momentum.]
The directions of the spins are
specified by the angles $(\theta_{S_1},\phi_{S_1})$ and
$(\theta_{S_2},\phi_{S_2})$, defined with respect to an orthonormal
basis aligned with $\hLN$,
\beq
\label{eq:e1e2}
\ve_1 \equiv \frac{\hLN\times\ve_z^S}{|\hLN\times\ve_z^S|}\,,\quad
\ve_2 \equiv \hLN \times \ve_1\,,\quad \ve_3\equiv\hLN\,,
\eeq
shown in Fig.~\ref{fig:S12def}. We then have
\begin{table*}
\begin{tabular}{c|c|c|c|c}
\hline
\multicolumn{3}{c|}{Binary} &
\multicolumn{1}{c}{GW propagation} &
\multicolumn{1}{|c}{Detector orientation} \\
$M$, $\eta$, $S_1$, $S_2$ & $\theta_{S_1}$, $\theta_{S_2}$,
$\phi_{S_1}-\phi_{S_2}$ & $\theta_{\rm L_{\rm N}}$, $\phi_{\rm L_{\rm
    N}}$, $\phi_{S_1}+\phi_{S_2}$ & $\Theta$, $\varphi$ & $\theta$,
$\phi$, $\psi$ \\
\multicolumn{1}{c|}{Basic} &
\multicolumn{1}{c|}{Local} &
\multicolumn{3}{c}{Directional} \\
\hline
\end{tabular}
\caption{Classification of binary, GW-propagation, and detector
parameters.\label{tab:parameters}}
\end{table*}
\bea
\hS_1 &=&
 \ve_1\, \sin\theta_{S_{1}}\, \cos\phi_{S_{1}}
+\ve_2\, \sin\theta_{S_{1}}\, \sin\phi_{S_{1}}
+\ve_3\, \cos\theta_{S_{1}} \,, \\
\hS_2 &=&
\ve_1\, \sin\theta_{S_{2}}\,\cos\phi_{S_{2}}
+\ve_2\, \sin\theta_{S_{2}}\,\sin\phi_{S_{2}}
+\ve_3\, \cos\theta_{S_{2}} \,.
\eea
Among the six angles $(\theta_{\rm L_{\rm N}}, \phi_{\rm L_{\rm N}})$,
$(\theta_{S_1},\phi_{S_1})$, and $(\theta_{S_2},\phi_{S_2})$, only
three are intrinsically relevant to the evolution of the binary:
$\theta_{S_1}$, $\theta_{S_2}$ and $\phi_{S_1}-\phi_{S_2}$. We shall
refer to them as {\it local} parameters. The other three independent
parameters, which are relevant to the computation of the
waveform, describe the rigid rotation of the binary as a whole in
space, and we shall refer to them as {\it directional} parameters. In
fact, there are five more directional parameters: $\Theta$ and
$\varphi$ specify the direction to the detector in the source frame,
and $\theta$, $\phi$, and $\psi$ specify the orientation of the
detector with respect to the radiation frame. All these parameters
have already been introduced in the previous section. Our
classification of the 15 binary and detector parameters is summarized in
Tab.~\ref{tab:parameters}.

\section{Analysis of precessional dynamics}
\label{sec3}

In a seminal paper \cite{ACST94}, ACST investigated in detail the
evolution of binaries of spinning compact objects, focusing on orbital
precession and on its influence on the gravitational waveforms.  In
this section, we build on their analysis to discuss several aspects of
quasicircular precessional dynamics that are especially important to
the formulation of a reliable DTFs for these systems. Note also that
Wex \cite{Wex} has derived analytic solutions for quasielliptical solutions 
to the 2PN conservative dynamics, including spin-orbit effects.

We complement ACST's analytical arguments with the empirical evidence
obtained by studying the orbits generated by the numerical integration
of Eqs.~\eqref{omegadot}--\eqref{S2dot} and \eqref{Lhdot}.  We select
the following typical binaries: BBHs with masses $(m_1 + m_2)$ given
by $(20 + 10)M_\odot$, $(15 + 15)M_\odot$, $(20 + 5)M_\odot$, $(10 +
10)M_\odot$, $(7 + 5)M_\odot$; and NS--BH binaries with masses $m_1 =
10 M_\odot$ (BH) and $m_2 = 1.4 M_\odot$ (NS). The BHs are always
chosen to be maximally rotating ($S = m^2$), while the NSs are assumed
to be nonspinning. There are neither astrophysical data nor theoretical
results which suggest that maximal spins are preferred. However, in
this paper we decide to investigate the most pessimistic (in terms of
precessional effects) scenario.
The initial GW frequency is chosen at $30\,{\rm
Hz}$ for binaries with total mass larger than $20\,M_\odot$, and
$40\,{\rm Hz}$ for all the other cases. For each set of masses, we
consider 1000 (or, when the numerical study is very computationally
expensive, only 200) orbital evolutions obtained with random initial
orientations of the orbital and spin angular momenta. (These initial
configurations are taken from the pseudorandom sequence specified in
Sec.~\ref{montecarlo} and used in Sec.~\ref{sec8} to evaluate the
effectualness~\cite{DIS1} of our DTF in matching the target signals.)

In Sec.\ \ref{sec3.0} we introduce the ACST results, and in particular
the distinction between simple and transitional precession.  In Sec.\
\ref{sec3.1} we study the dependence of the GW ending frequency
(defined in Sec.\ \ref{sec2.2}) on the initial values of spins and on
their evolution, and we link this dependence to the conservation of
certain functions of the spins through evolution. As mentioned above,
a knowledge of the ending frequencies of our target model is important
to decide what extension each of the detection templates should have
in the frequency domain.  In Sec.\ \ref{energyout} we examine the
value of the binding energy and of the total angular momentum at the
end of evolution, and we estimate the amount of GWs that must be
emitted during plunge, merger and ringdown to reduce the spin of the
final BH to the maximal value.  In Sec.\ \ref{sec3.2} we discuss,
largely on the basis on numerical evidence, the effects of spin on the
accumulated orbital phase $\Psi$ [defined by Eq.\ \eqref{orbitalpsi}];
we argue that these effects are mainly \emph{nonmodulational}, and
that, for data-analysis purposes, they can be treated in the same way
as the standard PN corrections to the orbits of nonspinning
binaries. It follows that the precession of the orbital angular
momentum is the primary source of modulations in the signal (as
already emphasized by ACST for particular classes of binaries).  In
Sec.\ \ref{sec3.3} we show, again on the basis of numerical evidence,
that transitional precession has little relevance to the data-analysis
problem under consideration.  In Sec.\ \ref{sec3.4} we discuss the
power-law approximations introduced by ACST to describe the precession
of the orbital plane as a function of frequency in particular
binaries, and we show that they are appropriate in general for the
larger class of binaries under consideration in this paper. These
approximations are a basic building block of the effective template
families developed by Apostolatos \cite{apostolatos2} and, indeed, of
our generalized and improved families.

\subsection{The ACST analysis}
\label{sec3.0}

In their paper \cite{ACST94} on precessing binaries of compact
objects as GW sources, ACST chose to work at the leading order in both
the orbital phasing and the precessional effects to highlight the main
features of dynamical evolution.  For orbital evolution, they retained
only the first term on the right-hand side of Eq.~(\ref{omegadot}): as
a consequence, the precession of the orbital plane is the only source
of GW modulation considered in the analysis.  [The resulting
accumulated orbital phase $\Psi$, given by Eq.~(\ref{orbitalpsi}), is
known as {\it Newtonian Chirp}.] For the precession of the orbital
angular momentum and of the spins, ACST retained only the first terms
(the \emph{spin-orbit} terms) in Eqs.~(\ref{S1dot}), (\ref{S2dot}) and
(\ref{Lhdot}).  On the basis of these approximations, and in the
context of binaries with either $m_1 \approx m_2$ (and spin-spin terms
neglected) or $S_2\approx 0$,
ACST classified the possible evolutions of spinning binaries into two
categories: \emph{simple precession} and \emph{transitional
precession}.

The vast majority of evolutions is characterized by simple precession,
where the direction of the total angular momentum remains roughly
constant, and where both the orbital angular momentum and the total
spin $\vS = \vS_1 + \vS_2$ precess around that direction. ACST
provided a simple analytic solution for the evolutions in this class. They
also showed that the orbital precession angle, expressed as a function
of the orbital frequency, follows approximately a power law (see
Eq.~(45) of Ref.~\cite{ACST94}).

Transitional precession happens when, at some point during the
evolution, the orbital angular momentum and the total spin become
antialigned and have roughly the same magnitude, so the total angular
momentum is almost zero,
\beq
\vJ = \vL + \vS \approx 0\,.
\label{transP}
\eeq
When this condition is satisfied, the total angular momentum is liable
to sudden and repeated changes of direction. The evolutions in this
class cannot be easily treated analytically, but they occur only for a
small portion of the possible initial conditions.

In this paper, we shall refer to the special cases investigated by
ACST (with either $m_1\approx m_2$, or $S_2\approx 0$) as \emph{ACST
configurations}. NS--BH binaries and BBHs with $m_1 \gg m_2$ are
astrophysically relevant cases among ACST configurations, because for
both we can set $S_2\approx 0$. The ACST formalism can also describe well BBHs with equal masses, but where spin-spin effects are
negligible.

\subsection{Conservation laws and GW ending frequencies}
\label{sec3.1}

For the ACST configurations, both the total spin and its
projection on the orbital angular momentum are
constants of the motion:
\bea \left[\hL_{\rm N}(t) \cdot \vS(t) \right]_{\mbox{\scriptsize
ACST}}&=&\mathrm{const} \,; \\
\left[\vS^2(t)\right]_{\mbox{\scriptsize ACST}}&=&\mathrm{const}\,.
\eea
For generic non--ACST configurations (as discussed, for instance, by
Damour~\cite{TD}), the \emph{effective spin} $\vS_{\rm eff}$
[Eq.~\eqref{Seff}] can, to some extent, replace the total spin in
these conservation laws.

From Eqs.~(\ref{S1dot}), (\ref{S2dot}), and (\ref{Lhdot}), we see also
that if we ignore the spin-spin effects in the precession equations,
then the projection
\beq
\label{kappaeff}
\kappa_{\rm eff} \equiv \frac{\hLN \cdot \vS_{\rm eff}}{M^2}
\eeq
of the effective spin onto the Newtonian orbital angular momentum is a
constant of motion,
\beq
\left[\kappa_{\rm eff}(t)\right]_{\rm SO} = \mathrm{const}\,
\eeq
(where the subscript ``SO'' stands for the inclusion of spin-orbit effects only); on the other hand, neither $\vS^2(t)$ nor
$\vS^2_{\rm eff}(t)$ are conserved.

The conservation of $\kappa_{\rm eff}$ has important consequences for
the endpoints of evolution, defined in Sec.~\ref{sec2.2} by
Eq.~\eqref{MECO} for the MECO.  In the nonspinning case, as discussed
in BCV1, if the dynamics was known at all PN orders, then the MECO
would agree with the Innermost Stable Spherical Orbit (ISCO), defined
as the orbit beyond which circular orbits become dynamically unstable.
When only spin-orbit (henceforth, SO) effects are included, the
conservation of $\kappa_{\rm eff}$ preserves this correspondence
between MECO and ISCO, because the leading-order SO term in the energy
is proportional to $\kappa_{\rm eff}$: in fact, the frequency of the
MECO has a precise functional dependence on $\kappa_{\rm eff}$ [see
Eqs. \eqref{s1}--\eqref{MECO}].
\begin{figure*}[t]
\begin{center}
\includegraphics[width=0.95\textwidth]{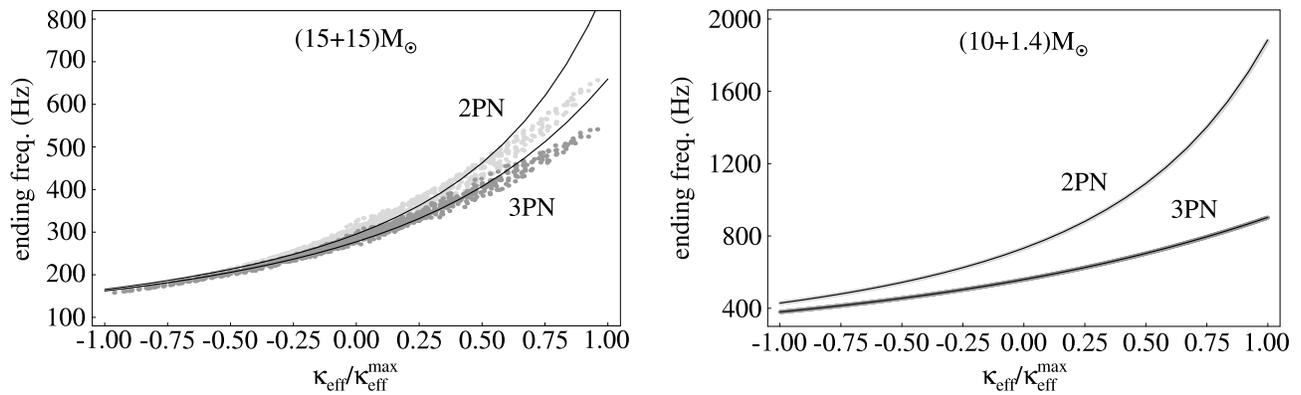}
\caption{\label{finalf} Binary ending frequencies (gray dots) as
functions of the initial value of $\kappa_{\rm eff}/\kappa_{\rm eff}^{\rm max}$, for 1000 initial
spin configurations of $M=(15 + 15)M_\odot$ BBHs (in the left panel),
and $M=(10 + 1.4)M_\odot$ NS--BH binaries (in the right panel), at 2PN
and 3.5PN orders. The solid lines plot the SO-only predictions for the
ending frequencies.}
\end{center}
\end{figure*}

When spin-spin (henceforth, SS) couplings are also included,
$\kappa_{\rm eff}$ is no longer conserved, and the MECOs (and
therefore the ending frequencies) of binaries with the same initial
$\kappa_{\rm eff}$ become smeared around their SO-only values, which
are functions only of $\kappa_{\rm eff}$. In addition to this
smearing, the SS contribution to the energy introduces also a bias. In
the end, however, the SS correction is not very important for the
ending frequencies, as we can see in the following examples.  In the
left panel of Fig.~\ref{finalf}, we plot the ending frequency at 2PN
and 3.5PN orders \cite{note39} versus the initial value of
$\kappa_{\rm eff}$ for BBHs with $M=(15 + 15)M_\odot$ (in gray dots),
as compared to the SO-only predictions (in solid lines). The smearing
of the ending frequencies is relatively mild, and so is the systematic
deviation from the SO-only predictions. We have checked that this
behavior characterizes all the mass configurations enumerated just
before Sec.~\ref{sec3.1}, at both 2PN and 3.5PN orders. In the right
panel of Fig.~\ref{finalf}, we plot the ending frequencies for NS--BH
binaries [with $M=(10+1.4)M_\odot$]. The ending frequencies follow
exactly the expected functional dependence on $\kappa_{\rm eff}$.

The mildness of these deviations can be understood (in part) by
looking at the variation of $\kappa_{\rm eff}$ during the
evolution. For example, for the $(15+15)M_\odot$ BBHs shown in
Fig.~\ref{finalf}, the maximum deviation of $\kappa_{\rm eff}$ from
being a constant (measured as $\mathrm{maxdev}(\kappa_{\rm eff}) =
[\max(\kappa_{\rm eff})-\min(\kappa_{\rm eff})]/2$) is $0.036$, to be
compared with the maximum kinematically allowed deviation, $0.875$;
for $(20+5)M_\odot$ BBHs at 2PN order, $\mathrm{maxdev}(\kappa_{\rm
eff}) = 0.033$, to be compared with the maximum kinematically allowed
deviation $0.92$.

As we can infer from Fig.~\ref{finalf}, the ending frequencies depend
also on the PN order, and the difference between 2PN and 3.5PN orders
is more striking for NS--BH binaries than for BBHs. This trend is
present also in the nonspinning case (see BCV1): for nonspinning
($\chi_1=\chi_2=0$) equal-mass BBHs, we have $\omega_{\rm MECO}^{\rm
2PN} = 0.137 M^{-1}$ and $\omega_{\rm MECO}^{\rm 3PN} =0.129
M^{-1}$. To give a few numbers, for a $(10+10) M_\odot$ BBH, we have
$f_{\rm GW,2PN}^{\rm MECO} = 443\,{\rm Hz}$, and $f_{\rm GW,3PN}^{\rm
MECO} = 416\,{\rm Hz}$; for a $(15+15)M_\odot$ BBH, $f_{\rm GW,
2PN}^{\rm MECO} = 295\,{\rm Hz}$, and $f_{\rm GW, 3PN}^{\rm MECO} =
277\,{\rm Hz}$; on the other hand, for a $(10 + 1.4)\,M_\odot$ NS--BH
binary, we have $f_{\rm GW,2PN}^{\rm MECO} = 734\,{\rm Hz}$, and
$f_{\rm GW,3PN}^{\rm MECO} = 559\,{\rm Hz}$. For the second and third
binaries, these values can be read off from the solid lines of
Fig.~\ref{finalf}, by setting $\kappa_{\rm eff}=0$ (no spins).

\begin{figure}[t]
\begin{center}
\includegraphics[width=\sizeonefig]{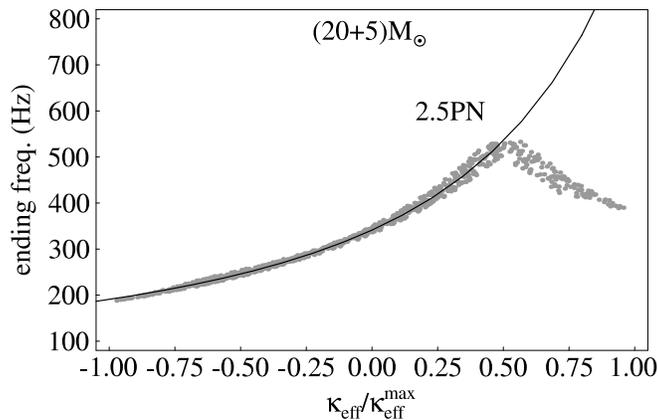}
\caption{\label{finalf2.5}Binary ending frequencies (gray dots) as
functions of the initial value of $\kappa_{\rm eff}/\kappa_{\rm eff}^{\rm max}$,
for 1000 initial spin configurations of $M=(20 + 5)M_\odot$ BBHs, at 2.5PN order. The
solid lines plot the SO-only predictions for the ending frequencies.}
\end{center}
\end{figure}

Finally, in Fig.~\ref{finalf2.5} we show the ending frequencies for
$(20+5)\,M_\odot$ BBHs, when Eq.~\eqref{omegadot} (which rules the
evolution of the orbital phase) is evaluated at 2.5PN order. In this
case, if $\kappa_{\rm eff} \geq 0.5$, then $\dot{\omega}$ goes to zero
before the MECO can be reached. The resulting ending frequencies
deviate considerably from SO-only predictions. As already discussed in
BCV1, $\dot{\omega}$ goes to zero because at 2.5PN order the
gravitational flux goes to zero for high orbital velocities; since
this very nonphysical behavior happens systematically, we then choose
to exclude the 2.5PN order from our analysis.

\subsection{Energy radiated during inspiral and (estimated)
total angular-momentum emitted
 after inspiral}
\label{energyout}
\begin{figure*}
\begin{center}
\includegraphics{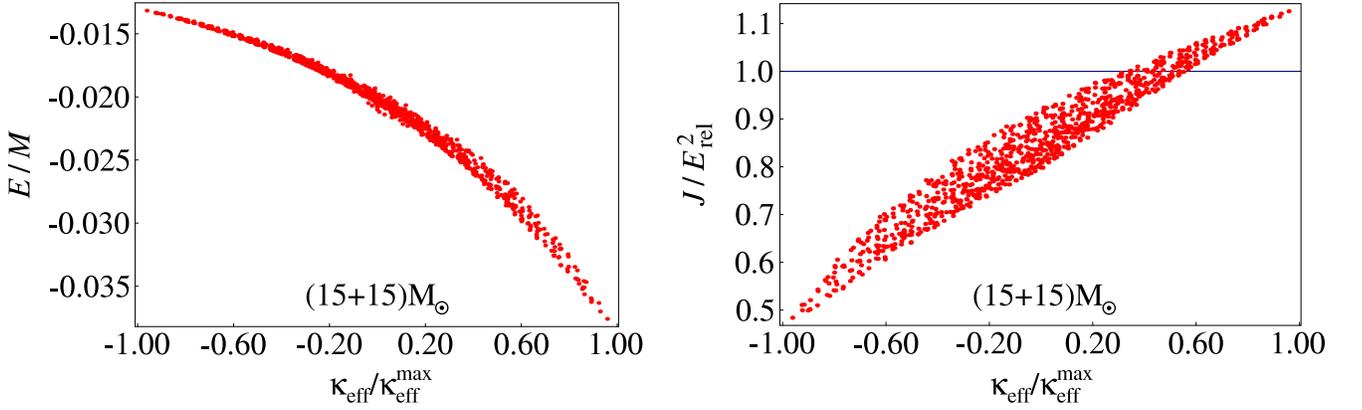}
\end{center}
\caption{For 1000 $(15+15)M_\odot$ BBHs with different initial spin
configurations, in the left panel we plot the ratio between the
(nonrelativistic) 2PN energy [Eq.~(\protect{\ref{s1}})] at the ending frequency and
the mass-energy initially available $M$, versus the initial
$\kappa_{\rm eff}/\kappa_{\rm eff}^{\rm max}$; in the right panel we plot the ratio between the
total angular momentum $J$ at 2PN order and the 
square of the (relativistic) 2PN energy
[Eq.~(\protect{\ref{s1}})] at the ending frequency, versus the initial
$\kappa_{\rm eff}/\kappa_{\rm eff}^{\rm max}$.
\label{fig:EJ}}
\end{figure*}

It is interesting to evaluate how much energy can be radiated in
GWs before the final plunge, especially for binaries whose inspiral end in the LIGO--VIRGO frequency band.
In the left panel of Fig.~\ref{fig:EJ}, for $M=(15 + 15)
M_\odot$ BBHs, we plot the ratio between the 2PN (nonrelativistic) energy, given by Eq.~(\ref{s1}) and evaluated at the endpoint
of evolution (as defined in Sec.\ \ref{sec2.2}), and the
total mass--energy initially available, $M$. Depending on the
initial relative orientation between the spins and the orbital angular
momentum (as expressed by the initial $\kappa_{\rm eff}/\kappa_{\rm eff}^{\rm max}$), the energy
that can be released in GWs during the inspiral ranges between $\sim 1.5$ and $3.5\%$ of $M$.  More energy can be emitted when the initial spins are aligned with the orbital angular momentum. We find similar results for all the other BBHs investigated, and similar results were also obtained by Damour in the EOB framework (see Fig. 1 of Ref.~\cite{TD}).

It is also interesting to estimate how much total angular momentum can be radiated during the coalescence phases that follow the inspiral (plunge and merger), especially when those phases fall in the LIGO--VIRGO band. In general, we have
\beq
\label{Jrad}
 \vJ_{\rm rad} = \vJ - \vS_{\rm BH}\,,
\eeq
where $\vJ_{\rm rad}$ is the angular momentum radiated during
plunge--merger, $\vJ$ is the total angular momentum of
the binary at the end of the inspiral, and $\vS_{\rm BH}$ is the
spin of the final black hole. A 
lower limit on the angular momentum radiated
in these phases can be obtained using the fact 
that the magnitude of the final spin can be at most $M^2_{\rm BH}$ (where $M_{\rm
BH}$ is the mass of the final black hole).
To derive this lower limit we follow Flanagan and
Hughes~\cite{FH}, and we write, using Eq.~(\ref{Jrad}),
\beq
|\vJ_{\rm rad}| \ge |\vJ|-|\vS_{\rm BH}| \ge |\vJ| - M_{\rm BH}^2
\ge |\vJ|- E_{\rm rel}^2\,,
\label{spininequality}
\eeq
where $E_{\rm rel} = M + E$ is the relativistic energy of the
binary at the end of inspiral; in deriving Eq.\
\eqref{spininequality} we used the relation $|\vS_{\rm BH}| \le
M_{\rm BH}^2 \le E_{\rm rel}^2$. It is straightforward to see from
Eq.~(\ref{spininequality}) that this
lower limit is nontrivial (that is, greater than zero) only when
$|\vJ|>E_{\rm rel}^2$.

In the right panel of Fig.~\ref{fig:EJ}, for $M=(15 + 15)M_\odot$
BBHs, we plot $|\vJ|/E_{\rm rel}^2$, where the angular momentum is
evaluated at 2PN order~[\onlinecite{KWW},\onlinecite{K}],
\bea
\vJ/M^2 &=& \eta \,(M\omega)^{-1/3}\,\hL_N\,\left \{ 1 +
\frac{(9+\eta)}{6}\,(M\omega)^{2/3} - \frac{7}{3}\,
\hL_N\cdot \frac{\vS_{\rm eff}}{M^2}\,(M\omega) +
\left[ \frac{1}{24}(81-57\eta + \eta^2) \right.
\right. \nonumber \\
&& \left. \left. -
\frac{1}{\eta}\,\left(\frac{\vS_1}{M^2}\cdot \frac{\vS_2}{M^2} - 3 \left(\hL_N\cdot \frac{\vS_1}{M^2}\right) \left(\hL_N\cdot \frac{\vS_2}{M^2}\right)\right) \right ]\,
(M\omega)^{4/3} \right \} - \eta\,(M\omega)^{2/3}\,\frac{\vS_{\rm eff}}{M^2} + \frac{\vS}{M^2}\,.
\eea
We see that $J/E_{\rm rel}^2$ is generally less than one,
except when $\kappa_{\rm eff} \ge 0.4$ (which happens for $13\%$ of all the initial spin configurations); the
maximum value of $|\vJ|/E_{\rm rel}^2$ is $1.13$. (For a similar plot
obtained in the EOB framework see Fig. 2 of Ref.~\cite{TD}.) Such large
values of $\kappa_{\rm eff}$ imply large ending frequencies [for
the $(15+15)M_\odot$ BBHs shown, larger than 400 Hz], which 
do not lie in the LIGO--VIRGO band of good interferometer sensitivity, 
unless the BBHs have higher masses; then all the frequencies are scaled down. In any case, for $\kappa_{\rm eff} = 1$ (spins and orbital momenta
initially aligned), in the high-mass binaries investigated, Eq.\ \eqref{spininequality} suggests the lower limit
\beq
|\vJ_{\rm rad}| \ge 0.13 \, E_{\rm rel}^2 \sim 0.1 \, M^2,
\eeq
to be compared with the value $0.4 M^2$ obtained by Flanagan and
Hughes \cite{FH} using BH spins aligned with the orbital angular
momentum (estimated to be $\sim 0.9M^2$).

A (trivial) upper limit for $\vJ_{\rm rad}$ is obtained by setting $\vS_{\rm BH}=0$:
\beq
|\vJ^{\rm rad}| \le |\vJ|\,.
\eeq
For different values of $\kappa_{\rm eff}$, the upper limit for our $(15+15)M_\odot$ binary is $\sim 0.5$ -- $1.1 M^2$.  However,
in order for the inspiral to end within the LIGO--VIRGO band of good interferometer sensitivity (which requires a MECO frequency lower than 400\,Hz), we need $\kappa_{\rm eff} < 0.4$, which corresponds to an upper limits $\sim 0.5$ -- $0.7\,M^2$.

To put this section into context, we point out that 
most reliable PN estimates for the energy and the angular momentum
radiated after the MECO can be achieved only with models that include
information about the plunge phase, such as the model that can be
built on Damour's spinning-EOB equations \cite{TD}.

\subsection{Spin-orbit and spin-spin effects on the accumulated orbital phase}
\label{sec3.2}
\begin{table*}
\begin{center}
\newcommand{\llba}{$\langle \Delta \Psi^{\rm res} \rangle_{200}$}
\newcommand{\llbb}{$\Delta \Psi^{\rm res}_{90\%\,(200)}$}
\newcommand{\llbc}{$\max_{200} \Delta \Psi^{\rm res}$}
\newcommand{\llca}{$\langle \varepsilon_\mathrm{orbital} \rangle_{200}$}
\newcommand{\llcb}{$\varepsilon_\mathrm{orbital\,90\%\,(200)}$}
\newcommand{\llcc}{$\max_{200} \varepsilon_\mathrm{orbital}$}
\begin{tabular}{l|ccccc|c}
\hline
& \multicolumn{6}{c}{Maximum modulational correction $\Delta \Psi^\mathrm{res}$} \\
& $(20+10)M_\odot$ & $(15+15)M_\odot$ & $(20+5)M_\odot$ & $(10+10)M_\odot$ &
\multicolumn{1}{c}{$(7+5)M_\odot$} & $(10+1.4)M_\odot$ [NS--BH] \\
\hline \hline
\llba &  0.0247 &  0.0214 &  0.0450 &  0.0402 &  0.0828 &  0.1228\\ 
\llbb &  0.0460 &  0.0411 &  0.0676 &  0.0787 &  0.1504 &  0.1884\\ 
\llbc &  0.0680 &  0.0523 &  0.1227 &  0.1186 &  0.2196 &  0.1895\\ 

\hline
\end{tabular}
\end{center}
\caption{Maximum modulational effects in the accumulated orbital phase
$\Psi$. We give the average over the 200 samples, the 90\% quantile of
the distribution, and the maximum value for the diagnostic $\Delta \Psi^{\rm res}$, defined
in Eq.~(\protect{\ref{psires}}).
\label{tab:resmod}}
\end{table*}
While for nonspinning binaries the accumulated orbital phase [defined
by Eq.\ \eqref{orbitalpsi}] coincides with (half) the GW phase at the
detector, for spinning binaries the two phases differ by precessional
effects; in the FC convention, these are found \emph{in part} in the
relation
\begin{equation}
\dot{\Phi}_S = \dot{\Psi} - \dot{\alpha} \cos i,
\end{equation}
where $\Phi_S$ is the orbital phase with respect to the ascending node
of the orbit, which appears in Eq.\ \eqref{cqsqwaveforms} for the
detector response to GW; and in part in the explicit time dependence
of the coefficients $C_Q$ and $S_Q$ on $\alpha$ and $i$ [see Eqs.\
\eqref{cqeq}--\eqref{steq}]. In this section, we are going to argue
that the evolution of the accumulated orbital phase is very similar in
spinning and nonspinning binaries; and that, as a consequence, the
effect of spins on detector response \emph{through the accumulated
orbital phase} can be reproduced using nonspinning-binary templates,
such as those studied in BCV1 [see also Eqs.\
\eqref{psispa}--\eqref{SS}]. Of course, precessional effects \emph{do}
enter the detector response through the other dependences mentioned
above, and these cannot be neglected when building templates to detect
physical signals.

Both the spin-orbit and spin-spin couplings can affect the accumulated
orbital phase $\Psi$ through the 1.5PN and 2PN terms in
Eq.~(\ref{omegadot}). However, as we shall discuss in this section,
this effect is largely {\it nonmodulational}.
For each binary configuration, we introduce three different functions
of time: (a) the accumulated orbital phase $\Psi^{\rm full}$, obtained by solving the
\emph{full} set of Eqs.~\eqref{omegadot}--\eqref{S2dot} and
\eqref{Lhdot}, including the SO and SS couplings; (b) the accumulated
orbital phase $\Psi^{\rm fix}$, obtained by using the initial orbital
angular momentum and spins \emph{at all times} in the SO and SS
couplings; and (c) the accumulated orbital phase $\Psi^{\rm nospin}$
for a nonspinning binary, obtained by dropping the SO and
SS couplings altogether.

In general, $\Psi^{\rm fix}$ and $\Psi^{\rm nospin}$ are quite
different for the same set of binary masses.  However, the difference
$\Psi^{\rm fix} - \Psi^{\rm nospin}$ is not a strongly oscillating
function (that is, it does not show any modulation), and it can be
reduced considerably by modifying the 1.5PN and 2PN coefficients in
the phasing equation for the nonspinning binary.  It is then
reasonable to assume that such a nonmodulational effect could be
captured by the nonspinning DTFs constructed in BCV1. Moreover,
the difference between $\Psi^{\rm full}$ and $\Psi^{\rm fix}$ is due
to the nonconservation of the SO and SS terms that appear in
Eq.~\eqref{omegadot} for $\dot\omega$. These terms have relatively
high PN orders, so we expect that they will be small.

Thus, we expect that $\Psi^{\rm full}$ can be well described by
a nonmodulational phasing of the kind
\beq
\Psi^{\rm nonmod}(f) = {\cal C}_0+ {\cal C}_1\, f +
\frac{{\cal C}_2}{f^{5/3}} + \frac{{\cal C}_3}{f^{2/3}}\,,
\label{fit}
\eeq
which looks rather like the frequency-domain phasings employed in the
DTFs of BCV1. [Here $\mathcal{C}_2$ and $\mathcal{C}_3$ can be seen
as actual (intrinsic) template parameters, whereas $\mathcal{C}_0$ and
$\mathcal{C}_1$ represent, respectively, the initial phase and the
time of arrival of the GW signal, both of which are extrinsic
parameters in the sense discussed in BCV1.] To verify this
hypothesis, we first evaluate $\Psi^{\rm full}$ in the
frequency range $50\,{\rm Hz} \mbox{--} 250\,{\rm Hz}$ (which is
appropriate for first-generation ground-based GW detectors), using
Eqs.\ \eqref{omegadot}--\eqref{S2dot} and \eqref{Lhdot} at 2PN order,
for all the BBH and NS--BH configurations considered earlier [(5 + 1)
masses $\times$ 200 angles]. We then (least-square) fit $\Psi^{\rm full}$
with functions of the form (\ref{fit}). A measure of the goodness
of the fit, given by
\beq
\Delta\Psi^{\rm res} = \max_{50\,{\rm Hz}<f<250\,{\rm Hz}}
\left|\Psi^{\rm full}(f)-\Psi^{\rm nonmod}(f)\right|\,,
\label{psires}
\eeq
is shown in Tab.~\ref{tab:resmod}. The maximum deviations are all
smaller than $\sim 0.1\,{\rm rad}$, except for the lighter $(7+5)M_\odot$ BBH and $(10+1.4)M_\odot$ NS--BH systems (where however the \emph{average} deviations are still $\sim 0.1\,{\rm rad}$).
This suggests that templates with
phasing expressions similar to \eqref{fit} (such as those proposed in
BCV1) could already approximate rather well the full target model
studied in this paper.

\subsection{Simple and transitional precession of total angular momentum}
\label{sec3.3}

\begin{table*}
\begin{center}
\begin{tabular}{c|ccccc|c}
\hline
& \multicolumn{6}{c}{Percentage of binary configurations where $\exists t : \hJ(t) \cdot \hJ(0) < 1 - \epsilon_J$}
\\
& $(20+10)M_\odot$ & $(15+15)M_\odot$ & $(20+5)M_\odot$ & $(10+10)M_\odot$ & \multicolumn{1}{c}{$(7+5)M_\odot$} & $(10+1.4)M_\odot$ [NS--BH] \\
\hline \hline
$\epsilon_{\rm J}=0.05$ & $17.5\%$ & $6.0\%$ & $33.5\%$ & $7.0\%$ & $3.5\%$ & $0.0\%$ \\
$\epsilon_{\rm J}=0.10$ & $2.5\%$ & $0.0\%$ & $11.0\%$ & $0.0\%$ & $0.0\%$ & $0.0\%$ \\
\hline
\end{tabular}
\end{center}
\caption{Deviation of the total angular momentum $\hJ$ from its initial
direction. This table shows the percentage of the binary
configurations where $\hJ(t) \cdot \hJ(0)$ goes below $1 - \epsilon_{\rm J}$, for
the $\epsilon_{\rm J}$ given in the first column.\label{tab:Jtot}}
\end{table*}

For most of the binary configurations investigated, we find, in
analogy with the ACST analysis, that the direction of total angular
momentum does not change much during evolution. In other words,
transitional precession does not occur.  Table~\ref{tab:Jtot} shows
the fraction of configurations that yield
\beq
\displaystyle \min_t \hJ(t) \cdot \hJ_0 < 1 - \epsilon_{\rm J}\,,
\eeq
when $\epsilon_{\rm J}=0.05$ and $0.10$.  Let us now try to understand
the numbers of Tab.~\ref{tab:Jtot} in more detail.

We first focus on the columns two to six, which deal with binaries of
maximally spinning BHs. For BBHs with single masses
$m=5\mbox{--}20\,M_\odot$, the total spin is not usually large enough
to satisfy the transitional-precession condition \eqref{transP}, as we
can prove easily by using all the evolution equations at the leading
PN order: during the evolution, the magnitude of the orbital angular
momentum decreases with the GW frequency $f$, as in
\beq
|\vL| \approx |\vL_{\rm N}|=\eta\,(\pi M f)^{-1/3} M^2 \,,
\eeq
while the total spin is bounded by
\beq
|\vS|<|\vS_1|+|\vS_2|=m_1^2 + m_2^2 =(1-2\eta)M^2\,.
\eeq
In order for transitional precession to occur, we need at the
very least $|\vL_{\rm N}|=|\vS|$ [see Eq.~(\ref{transP})], which
requires
\beq
\eta (\pi M f)^{-1/3} < (1-2\eta)\,,
\eeq
or
\beq
f > f_{\rm trans}^{\rm min} \equiv \frac{\eta^3}{\pi M (1-2\eta)^3}\,.
\label{eq:ftrans}
\eeq
For transitional precession to occur before we reach the Schwarzschild
ISCO frequency $f_{\rm Schw}=1/\sqrt{6^3}\pi M$, we then need
\beq
\label{transcondition}
\frac{f_{\rm trans}^{\rm min}}{f_{\rm
Schw}}=\left(\frac{\sqrt{6}\eta}{1-2\eta}\right)^{3} \gtrsim 1
\quad \Rightarrow \quad \eta \gtrsim 0.22
\,.
\eeq
Although the ending frequencies obtained within our target model are
usually higher than $f_{\rm Schw}$, the very configurations that can
have transitional precession (those with nearly antialigned total spin
and orbital angular momenta) have always \emph{lower} ending
frequencies, making 0.22 too large an estimate for the critical value
of $\eta$.

\begin{figure*}
\begin{center}
\includegraphics{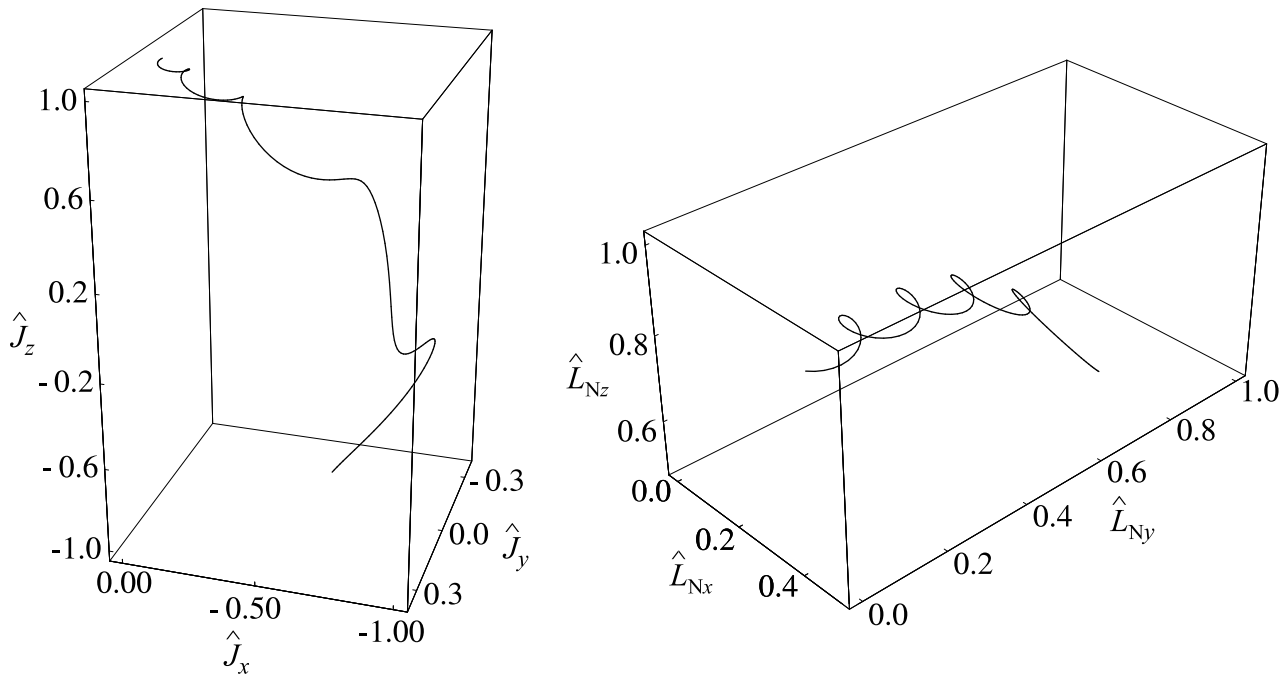}
\end{center}
\caption{Transitional precession. Evolution of the direction of total
angular momentum (left panel) and of Newtonian orbital angular momentum (right
panel) in the transitionally precessing $(20+5)M_\odot$ BBH with
initial angles $\theta_{S_1}=175.4^\circ$, $\theta_{S_2}=105.4^\circ$,
and $\phi_{S_1}-\phi_{S_2}=92.0^\circ$ (at $f_{\rm GW}=30\,{\rm Hz}$).
\label{fig:extran}}
\end{figure*}

As a consequence, among all the configurations we have
considered, only $(20+5)M_\odot$ and $(20+10)M_\odot$
BBHs can then have \emph{observable} transitional-precession
phases. These latter binaries are characterized by significantly
larger changes in $\vJ$ [see Tab.~\ref{tab:Jtot}].  However,
$(20+10)M_\odot$ BBHs still require $f>f_{\rm tran}^{\rm
min}=138\,{\rm Hz}$, which is very close to the relevant ending
frequency; so the change in $\vJ$ is smaller, and we never observed
episodes of transitional precession in the 200 initial configurations
analyzed.  On the contrary, we observed a few for $(20+5)M_\odot$
BBHs; one example follows from the initial configuration given by
$\theta_{S_1}=175.4^\circ$, $\theta_{S_2}=105.4^\circ$, and
$\phi_{S_1}-\phi_{S_2}=92.0^\circ$ (at $f_{\rm GW}=30\,{\rm Hz}$). In
this configuration the initial spin of the more massive body is almost
exactly antialigned with the orbital angular momentum.  The
trajectories of $\hJ$ and $\hLN$ during this evolution are shown,
respectively, in the left and right panels of Fig.~\ref{fig:extran}.

By contrast, none of the NS--BH configurations examined exhibits
transitional precessions. This is because the BH is taken as maximally
spinning, so $\vS$ is always much larger than $\vL$ in the frequency
band under consideration.

\subsection{Apostolatos' power law for orbital precession}
\label{sec3.4}
\begin{figure*}
\begin{center}
\includegraphics{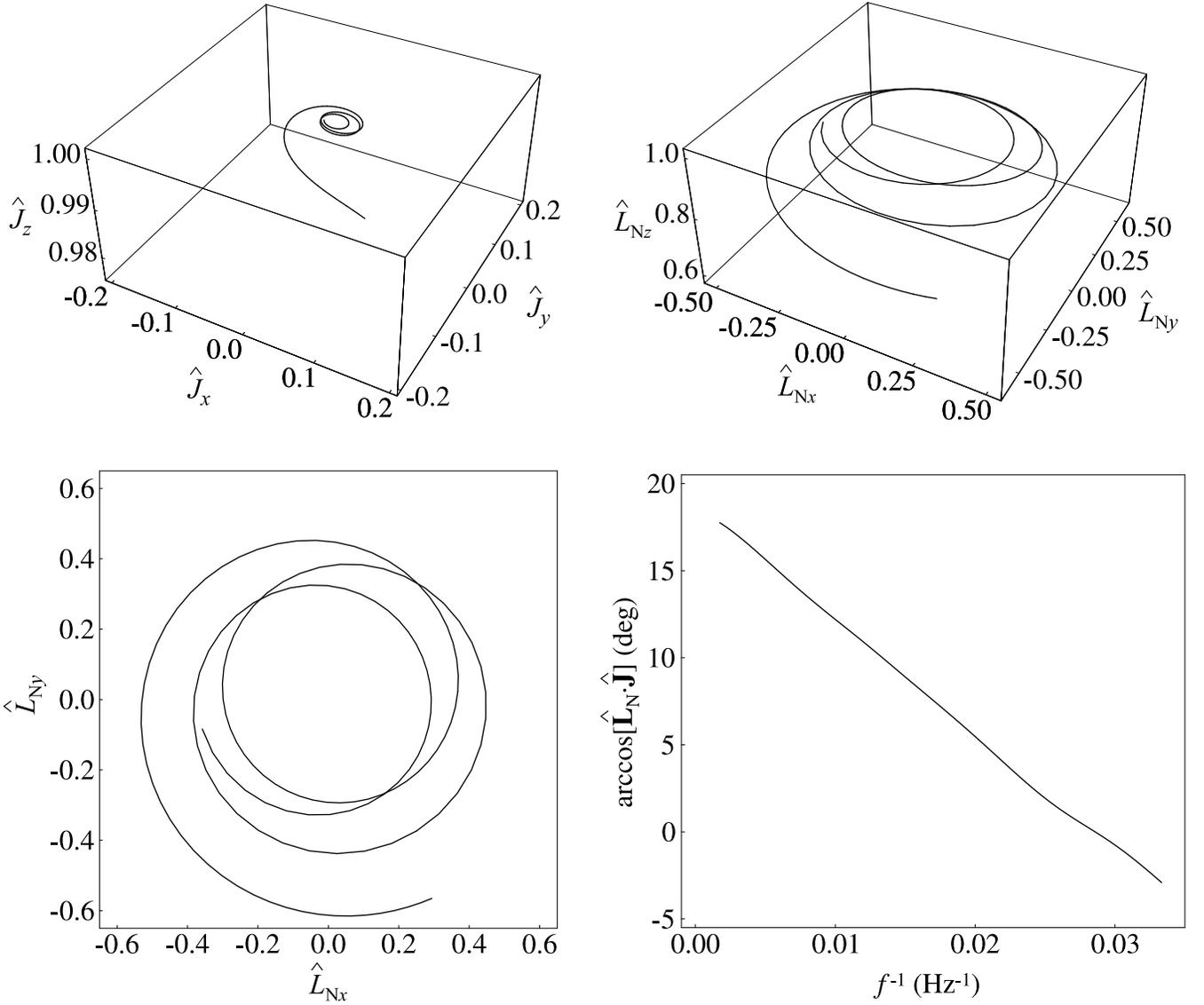}
\end{center}
\caption{Simple precession. The upper graphs show the evolution of the
direction of total angular momentum $\hJ$ (left), and of Newtonian
orbital angular momentum $\hLN$ (right), in the case of the simply precessing
$(20+5)M_\odot$ BBH with initial angles $\theta_{S_1}=44.6^\circ$,
$\theta_{S_2}=101.0^\circ$, and $\phi_{S_1}-\phi_{S_2}=-39.7^\circ$
(at $30\,{\rm Hz}$). The lower graphs show the projection of $\hLN$
onto the plane perpendicular to the initial $\hJ$ (left), and the
angle between $\hLN$ and $\hJ$, plotted as a function of inverse GW
frequency (right). The BBH was rotated in space so that the initial
direction of $\hJ$ would be parallel to the $z$
axis.\label{fig:exsimp}}
\end{figure*}
As discussed in the previous section, the vast majority of binary
configurations undergoes simple precession, where $\hJ$ remains
constant, while $\hLN$ and $\vS_{1,2}$ precess around it.
For ACST configurations
($m_1\approx m_2$ and negligible SS interactions, or $S_2\approx 0$),
both $\hL_N$ and $\hS$ precess around $\vJ$ with the precession
frequency \cite[Eq.~(42)]{ACST94}
\beq
\Omega_{\rm p} \equiv \frac{d \alpha_{\rm p}}{d t}=
\left (2 + \frac{3}{2}\,\frac{m_2}{m_1} \right ) J \omega^2 \,.
\label{omegap}
\eeq

ACST identified two regimes where
the evolution of $\alpha_{\rm p}$ can be approximated very well by a power law in $\omega$ (or $f$). For $L_N \gg S$, the total angular-momentum $J \approx L_N \sim \omega^{-1/3}$; using $\dot{\omega} \sim \omega^{11/3}$, it is straightforward to derive from Eq.~(\ref{omegap}) that $\alpha_{\rm p}$ is approximated well by a linear function of $f^{-1}$,
\beq
\label{Ap_alpha_1}
\alpha_{\mathrm{p}(-1)}^\mathrm{fit}(f) \approx \frac{{\cal
B}_1}{f}+{{\cal B}_2}\,,
\eeq
where ${\cal B}_1$ and ${\cal B}_2$ are constant coefficients.
Since $L_N/S \sim \eta\,\omega^{-1/3}$, the condition
$L_N \gg S$ corresponds to comparable-mass binaries ($\eta \sim 1/4$)
or to large separations. For $L_N \ll S$, we have $J \approx S$; in this case we derive from Eq.~(\ref{omegap}) that $\alpha_{\rm p}$ is approximated well by
a linear function of $f^{-2/3}$,
\beq
\label{Ap_alpha_2}
\alpha^\mathrm{fit}_{\mathrm{p}(-2/3)}(f)\approx \frac{{\cal
B}'_1}{f^{2/3}}+{{\cal B}'_2}\,,
\eeq
where ${\cal B}'_1$ and ${\cal B}'_2$ are constant coefficients.
The condition $L_N \ll S$ corresponds to $m_1 \ll m_2$ or to
small separations (late inspiral).

It turns out that Eqs.~(\ref{Ap_alpha_1}) and (\ref{Ap_alpha_2})
apply also to a large fraction of the BBHs and NS--BH binaries studied in this paper. This can be tested semiquantitatively by the
following procedure. For each configuration, we take the precession
angle $\alpha_{\rm p}(f)$ and we fit it with a function
$\alpha_{\mathrm{p}(-1,-2/3)}^\mathrm{fit}(f)$ of the form
(\ref{Ap_alpha_1}) or (\ref{Ap_alpha_2}), for frequencies in
the range $50$--$250\,{\rm Hz}$. We then evaluate
the maximum difference
\beq
\label{alphamax}
\Delta\alpha_{\mathrm{max}(-1,-2/3)}\equiv \max_{50\,{\rm Hz} < f< 250\,{\rm Hz}}
\left|\alpha_{\rm p}(f)-\alpha_{\mathrm{p}(-1,-2/3)}^{\rm fit}(f)\right|\,.
\eeq
In Tab.~\ref{tab:Dalpha}, we show the values of
$\Delta\alpha_{\mathrm{max}(-1)}^{90\%}$ (that is, the 90\% percentile
of $\Delta\alpha_{\mathrm{max}(-1)}$) and
$\Delta\alpha_{\mathrm{max}(-2/3)}^{90\%}$, for $(15+15)M_\odot$,
$(20+10)M_\odot$, $(10+10)M_\odot$, and $(7+5)M_\odot$ BBHs, and
for $(10+1.4)M_\odot$ NS--BH binaries.
The numbers show that Eqs.~(\ref{Ap_alpha_1}) and (\ref{Ap_alpha_2}) yield (roughly) comparable approximation.
This result is confirmed also by the more detailed analyses discussed later in this paper.

Figure~\ref{fig:exsimp} plots the 2PN evolutions of $\hJ$ (upper left panel) and $\hLN$ (upper right panel) for a $(20+5)M_\odot$ BBH with initial conditions $\theta_{S_1}=44.6^\circ$, $\theta_{S_2}=101.0^\circ$, and $\phi_{S_1}-\phi_{S_2}=-39.7^\circ$ (at $30\,{\rm Hz}$). The figure plots also the projection of $\hLN$ onto the plane perpendicular to the initial $\hJ$ (lower left panel), and the precession angle $\alpha_{\rm p}$ between $\hLN$ and $\hJ$, plotted as a function of inverse GW frequency $f^{-1}$ (lower right panel), and showing a very nearly linear dependence.
\begin{table*}
\begin{center}
\begin{tabular}{l|ccccc|c}
\hline & \multicolumn{6}{c}{90\% percentiles of error in precession
angle, $\Delta \alpha_\mathrm{max}$} \\ & $(15+15)M_\odot$ &
$(20+10)M_\odot$ & $(20+5)M_\odot$ & $(10+10)M_\odot$ &
\multicolumn{1}{c}{$(7+5)M_\odot$} & $(10+1.4)M_\odot$ [NS--BH]\\
\hline \hline $\Delta\alpha_{\mathrm{max}(-1)}^{90\%}$ & $0.30$ &
$0.24$ & $0.23$ & $0.34$ & $0.64$ & $0.61$ \\
$\Delta\alpha_{\mathrm{max}(-2/3)}^{90\%}$ & $0.52$ & $0.48$ & $0.50$
& $0.68$ & $1.14$ & $0.72$ \\ \hline
\end{tabular}
\end{center}
\caption{Approximation of binary precession histories using best-fit
parameters $\mathcal{B}_1$ and $\mathcal{B}_2$ in
Eqs.~\eqref{Ap_alpha_1} and \eqref{Ap_alpha_2}. This table shows the
90\% percentiles of $\Delta\alpha_{\mathrm{max}(-1)}$
[Eq.~\eqref{alphamax}] and $\Delta\alpha_{\mathrm{max}(-2/3)}$ in
the BBH and NS--BH populations studied throughout this section.
\label{tab:Dalpha}}
\end{table*}

Building on the results obtained by ACST,
Apostolatos~\cite{apostolatos2} conjectured (quite reasonably) that
orbital precession will modulate the gravitational waveforms with
functional dependencies given by Eqs.~(\ref{Ap_alpha_1}) and
(\ref{Ap_alpha_2}). On the basis of this conjecture and of the
observation that, in matched-filtering techniques, matching the phase
of signals is more important than matching their amplitudes,
Apostolatos proposed a family of detection templates
\cite{apostolatos2} obtained by modifying the phasing of nonspinning
PN templates as in
\beq
\label{ansatzone}
\mbox{Apostolatos' ansatz:} \quad \psi_{\rm spinning}
\rightarrow \psi_{\rm non\,spinning} + {\cal C}\cos(\delta+{\cal B}f^{-2/3})\,,
\eeq
while keeping a Newtonian amplitude $f^{-7/6}$.  Recently,
Grandcl\'ement, Kalogera and Vecchio~\cite{GKV} applied Apostolatos'
suggestion to an approximated analytical model of NS--BH binaries and low-mass BBHs: whereas the addition
of phase modulations according to Eq.\ \eqref{ansatzone} \emph{did}
increase the effectualness~\cite{DIS1} of nonspinning PN templates, the resulting
DTF family was still not good enough to recommend its application when
trying to capture the real modulated waveforms. Moreover, this DTF
requires three additional \emph{intrinsic} parameters (${\cal C},
\delta$, and ${\cal B}$) on top of the two BH (or NS) masses. The
resulting GW searches would then be plagued by an extremely high
computational cost.

In the rest of this paper, we shall propose a better template family,
inspired by old and new insight on the precessional effects that
appear in the gravitational waveforms. As we shall see, Apostolatos'
ansatz can be improved to build DTFs that have both high
effectualness~\cite{DIS1} and low computational requirements.

\section{Definition of modulated DTFs for precessing binaries}
\label{sec4}

We are now going to bring together all the observations reported in
Sec.\ \ref{sec3} to build DTFs that perform well in capturing the
detector response to the GWs emitted by precessing binaries of NSs and
spinning BHs (at least as long as the actual physical signals are
modeled faithfully enough by the adiabatic target model described in
Sec.\ \ref{sec2}).

In Sec.\ \ref{newconv} we develop a new (as far as we know) convention
for the generation and propagation of GW from spinning binaries; this
convention has the desirable property of factorizing the waveform into
a \emph{carrier signal} whose phase is essentially the accumulated
orbital phase of the binary, and a \emph{modulated amplitude} term
which is sensitive to the precession of the orbital plane.  In Sec.\
\ref{newdtf} we then use the results of Sec.\ \ref{sec3.2} to build an
approximation of the carrier signal, and the results of Secs.\
\ref{sec3.1}, \ref{sec3.3}, and \ref{sec3.4} to build an approximation
to the modulated amplitude; using these terms together, we define
three families of detection templates. In Sec.\ \ref{refspa} we describe
two standard families of nonspinning-binary templates; in Sec.\
\ref{sec6} we shall compare their performance with the performance of
our DTFs, to evaluate the performance improvements brought about by
our treatment of precession.

\subsection{A new convention for GW generation in spinning binaries}
\label{newconv}

At least two conventions are used to express the gravitational
waveforms generated by binaries of spinning compact objects, as
computed in the quadrupolar approximation \cite{note40}: the ACST convention~\cite{ACST94}, which uses a rotating
reference frame, and the FC convention~\cite{FC}, which uses a
nonrotating reference frame. We discussed the FC convention in
Sec.~\ref{sec2.3}, and we used it throughout this paper to generate
gravitational waveforms from the numerical integration of the
equations of motion of the target model.
Before going to the specific conventions, we shall first sketch a
generic procedure to write the gravitational waveform.

In general, the unit vector along the separation vector of the
binary, $\hn(t)$, and the unit vector along the corresponding relative
velocity, $\hl(t)$, can be written as
\beq
\label{n_lambda}
\hn(t)=\mathbf{e}_1(t)\,\cos\Phi(t)+\mathbf{e}_2(t)\,\sin\Phi(t)\,,\quad
\hl(t)=-\mathbf{e}_1(t)\,\sin\Phi(t)+\mathbf{e}_2(t)\,\cos\Phi(t)\,,
\eeq
where $\mathbf{e}_1(t)$,
$\mathbf{e}_2(t)$, and $\mathbf{e}_3(t) \equiv \hL_N(t)$ are
orthonormal vectors, and $\mathbf{e}_{1,2}(t)$ forms a basis for the
instantaneous orbital plane [see Fig.~\ref{fig:S12def}]; the
quantity $\Phi(t)$ is then the orbital phase with respect to $\mathbf{e}_{1,2}(t)$. The definition of $\ve_{1,2}(t)$ and of $\Phi(t)$ is not unique: an arbitrary function of time can be added to
$\Phi(t)$, and then compensated by a time-dependent rotation of
$\ve_{1,2}€(t)$ around $\hL_{N}(t)$, leaving $\hn(t)$
and $\hl(t)$ unchanged. In nonspinning binaries the orbital plane (and therefore $\hL_N$) does not precess, so the natural choice is to keep $\ve_{1,2}$ constant. In spinning binaries $\hL_{N}(t)$ precesses, and different, but nonetheless meaningful, conventions can be given for $\ve_{1,2}(t)$ and $\Phi(t)$. Note that $\Phi(t)$ is \emph{not}, in general, the same as the accumulated orbital phase $\Psi(t) = \int \omega(t) \, dt$.
Given a convention for $\ve_{1,2}€(t)$ and $\Phi(t)$,
the tensor $Q_c^{ij}$ that appears in Eq.~(\ref{hij}) can be written as
\beq
Q^{ij}_c=-2\left([\mathbf{e}_+]^{ij} \cos2\Phi + [
\mathbf{e}_\times]^{ij} \sin2\Phi\right)
\eeq
where
\beq
\mathbf{e}_+=\mathbf{e}_1 \otimes \mathbf{e}_1 - \mathbf{e}_2 \otimes \mathbf{e}_2\,,\quad
\mathbf{e}_\times=\mathbf{e}_1 \otimes \mathbf{e}_2 + \mathbf{e}_2 \otimes \mathbf{e}_1\,.
\eeq
With the detector lying along the direction $\hN$, one goes on to
define a radiation frame, formed by orthonormal vectors
$\mathbf{e}_{x}^{R}€$, $\mathbf{e}_{y}^{R}€$ and
$\mathbf{e}_{z}^{R}€=\hN$. The GW response is then given by
\beq
\label{generich}
h_{\rm resp}=-\frac{2\mu}{D}\frac{M}{r}\,
\underbrace{
\left( [\ve_+]^{ij}\,\cos2\Phi+[\ve_\times]^{ij}\,\sin2\Phi \right)
}_\mathrm{factor\;Q:\;quadrupole\;moment}
\underbrace{
\left( [\mathbf{T}_+]_{ij}\,F_+ + [\mathbf{T}_\times]_{ij}\,F_\times \right)
}_\mathrm{factor\;P:\;detector\;projection}
\,,
\eeq
where the tensors $[\mathbf{T}_{+,\times}]_{ij}$ are given by  (\ref{Tpx}), namely
\beq
\label{Tpxnew}
{\bf T}_+ \equiv \ve^{R}_x \otimes \ve^{R}_x-\ve^{R}_y \otimes
\ve^{R}_y\,,\quad {\bf T}_{\times} \equiv \ve^{R}_x \otimes
\ve^{R}_y+\ve^{R}_y \otimes \ve^{R}_x \,,
\eeq
and where $F_+$ and $F_\times$ are given by Eq.\ \eqref{Fpxgeneral},
namely
\beq
F_{+,\,\times}=\frac{1}{2}[\bar{\mathbf{e}}_{x}\otimes\bar{\mathbf{e}}_{x}€
-\bar{\mathbf{e}}_{y}€\otimes \bar{\mathbf{e}}_{y}€
]^{ij}[\mathbf{T}_{+,\,\times}]_{ij}\,,
\eeq
with $\bar{\mathbf{e}}_{x,\,y}$ the unit vectors along the orthogonal
arms of the interferometer.  Again, $\ve_{x}^{R}$ and $\ve_{y}^{R}$
are not uniquely defined, because they can be rotated at will around
$\hN$, of course changing the values of $F_+$ and $F_\times$.

ACST refer $\Phi(t)$ to the
direction $\hN$ of GW propagation, by imposing that
$\mathbf{e}^{\rm ACST}_1(t) \propto \hN\times\hL_N(t)$; they also set
$\mathbf{e}_{x}^{R}(t) \propto \pm \hN \times \hLN(t)$.
Although the ACST convention has allowed some insight into the waveforms, it is rather inconvenient for the purpose of data analysis,
because almost all the quantities that come into Eq.~(\ref{generich}) [$\ve_{1,\,2}$, $\mathbf{T}_{+,\,\times}$, and $F_{+,\,\times}$] depend both on the time evolution of the binary and on the direction to the detector. Using the terminology introduced in Sec.~\ref{sec2.3} and Tab.\ \ref{tab:parameters}, under the ACST convention the local and directional parameters are entangled in a time-dependent manner.

FC introduce the fixed source axes $\{\mathbf{e}^S_x,\mathbf{e}^S_y,\mathbf{e}^S_z\}$ [see
Sec.~\ref{sec2.3}], and they impose that $\mathbf{e}^S_1(t) \propto
\mathbf{e}^S_z \times \hL_N(t)$ [see Eq.~\eqref{e12s}].  The radiation
frame does not change with time [see
Eqs.~\eqref{erx}--\eqref{erz}]. As a consequence, the factors Q and P in Eq.~(\ref{generich}) become disentangled:
factor Q expresses the components of the quadrupole moment, which depend only on the evolution of the binary inside the source frame; factor P expresses the projection of the quadrupole moment onto the radiation frame and onto the antisymmetric mode of the detector, which depend only on the relative orientation between the source frame and the detector. However, for our purposes there are still two shortcomings in the FC convention:
\begin{enumerate}
\item The FC convention defines $\ve_{1,2}(t)$ and $\Phi(t)$ in terms of
the fixed source frame $\ve^{S}_{x,y,z}$, which is quite artificial, because only the relative orientation between binary and detector affects the detector response $h_{\rm resp}$.
\item In Sec.~\ref{sec3.2} we saw that the accumulated orbital phase $\Psi(t)$ is (almost) nonmodulated, so the modulations of the waveform come mainly from the precession of the orbital plane. Under the FC convention, the modulations appear only in factor Q of Eq.\ \eqref{generich}, but they appear both in the phase $\Phi(t)$ and in the precession of the tensors $\mathbf{e}_{+,\times}(t)$. It would be nice to isolate the precessional effects in either element.
\end{enumerate}
Both issues would be solved if we could find a modification of the FC convention where $\Phi$ coincides with the accumulated orbital phase, $\Psi$. As it turns out, it is possible to do so: we need to redefine the vectors $\ve_{1,2}(t)$ so that they precess alongside $\hL_N$,
\beq
\label{eprecess}
\dot{\mathbf{e}}_i(t)=\mathbf{\Omega}_e(t) \times \mathbf{e}_i(t)\,,\quad
i=1,2\,,
\eeq
with
\beq
\mathbf{\Omega}_e(t) \equiv \mathbf{\Omega}_L(t)-[\mathbf{\Omega}_L(t)\cdot\hL_N(t)]\,\hL_N(t)\,,
\label{Omegae}
\eeq
where $\mathbf{\Omega}_L$ is obtained by collecting the terms
that (cross-product) multiply $\hL_N$ in Eq.~(\ref{Lhdot}).
In App.\ \ref{appy} we prove that this convention yields
$\dot{\Phi} = \omega = \dot{\Psi}$, as desired. Qualitatively, one can reason
as follows. The angular velocity of the binary lies along $\hL_{N}(t)$, and has
magnitude $\dot{\Psi}=\omega$. The reason why $\Phi$ and $\Psi$ differ is that the
orbital basis $\ve_{1,2}$, used to define $\Phi$, must rotate to keep up with the
precession of the orbital plane. However, the difference vanishes
if we constrain the angular velocity of $\ve_{1,2}$ to be orthogonal to $\hL_{N}$;
Eq.\ \eqref{Omegae} provides just the right constraint.
In the following, we shall refer to our new convention as the
\emph{precessing} convention.
\begin{table*}
\begin{tabular}{l|cc|c|c}
\hline
convention & \multicolumn{2}{c|}{factor P} & \multicolumn{2}{c}{factor Q} \\
& \multicolumn{2}{c|}{$\mathbf{T}_{+,\times} \quad\quad\quad F_{+,\times}$} & \multicolumn{1}{c}{$\Phi(t)$} & $\mathbf{e}_{+,\times}(t)$ \\
\hline \hline
ACST       & \multicolumn{2}{c|}{\begin{minipage}[t]{0.28\textwidth}\flushleft
                                function of basic, local,
                                and directional parameters; \\
                                time dependent
                                \end{minipage}}
                              & \begin{minipage}[t]{0.30\textwidth}\flushleft
                                function of basic, local,
                                and directional parameters
                                \end{minipage}
                              & \begin{minipage}[t]{0.23\textwidth}\flushleft
                                function of basic, local,
                                and directional parameters
                                \end{minipage} \\ \hline
FC         & \multicolumn{2}{c|}{\begin{minipage}[t]{0.28\textwidth}\flushleft
                                function of directional parameters;
                                time independent
                                \end{minipage}}
                              & \begin{minipage}[t]{0.30\textwidth}\flushleft
                                function of basic, local,
                                and directional parameters
                                \end{minipage}
                              & \begin{minipage}[t]{0.23\textwidth}\flushleft
                                function of basic, local,
                                and directional parameters
                                \end{minipage} \\ \hline
precessing & \multicolumn{2}{c|}{\begin{minipage}[t]{0.28\textwidth}\flushleft
                                function of directional parameters;
                                time independent
                                \end{minipage}}
                              & \begin{minipage}[t]{0.30\textwidth}\flushleft
                                function of basic and local
                                parameters only;
                                coincides with $\Psi(t)$
                                \end{minipage}
                              & \begin{minipage}[t]{0.23\textwidth}\flushleft
                                function of basic and local
                                parameters only
                                \end{minipage} \\
\hline
\end{tabular}
\caption{Parametric dependence of the building elements
of the detector response function $h_\mathrm{resp}$
[Eq.\ \eqref{generich}] under the ACST, FC, and precessing conventions.
\label{conventions}}
\end{table*}

In Tab.\ \ref{conventions} we summarize the parameter dependence of the terms that make up the detector response function [Eq.\ \eqref{generich}], under the three conventions. It is important to remark that in the precessing convention the polarization tensors $\mathbf{e}_{+,\,\times}(t)$, as geometric
objects, do \emph{not} depend on the source frame, but only on the basic and local parameters. In practice, however, we need to introduce an arbitrary choice of the source frame to relate the orientation of the binary to the direction and orientation of the detector (that is, to write explicitly the products $[\mathbf{e}_{+,\,\times}]_{ij} [\mathbf{T}_{+,\,\times}]_{ij}$). We can avoid this arbitrariness by setting the source frame according to the initial configuration of the binary at a fiducial orbital frequency;
for example, we can impose (without loss of generality)
\beq
\label{sourcedef}
\mathbf{e}_{x}^{S}\propto \vS_1(0)-[\vS_1(0)\cdot \hLN(0)]\hLN(0)
\,,\quad
\mathbf{e}_{y}^{S}=\hLN(0)\times\mathbf{e}_{x}^{S}
\,,\quad
\mathbf{e}_{z}^{S}=\hLN(0)\,,
\eeq
and
\beq
\label{eijdef}
\mathbf{e}_{1}(0)=\mathbf{e}_{x}^{S}\,,\quad
\mathbf{e}_{2}(0)=\mathbf{e}_{y}^{S}\,,\quad
\mathbf{e}_{3}(0)=\mathbf{e}_{z}^{S}\,.
\eeq
[If $\vS_1(0)$ and $\hLN(0)$ are parallel, $\mathbf{e}_x^S$ can be chosen to lie in any direction within the plane orthogonal to $\hLN(0)$.]  Then the initial conditions, as expressed by their components with respect to the source frame, are determined only by the local parameters,
\bea
\label{inithLN}
\hLN(0)&=&(0,0,1)\,,\\
\label{initS1}
\vS_1(0)&=&(\sin\theta_{S_1},0,\cos\theta_{S_1})\,,\\
\label{initS2}
\vS_2(0)&=&(\sin\theta_{S_2}\cos(\phi_{S_2}-\phi_{S_1}),
\sin\theta_{S_2}\sin(\phi_{S_2}-\phi_{S_1}),
\cos\theta_{S_2})\,,
\eea
along with an initial orbital phase $\Psi_0$ given by
\beq
\label{phiinit}
\mathbf{n}(0) = \mathbf{e}_1(0)\cos\Psi_0 + \mathbf{e}_2(0) \sin\Psi_0\,.
\eeq
With this choice, all the directional parameters are isolated in
factor P of Eq.\ \eqref{generich}, while the basic and local
parameters (which affect the dynamics of the binary) are isolated in
factor Q.  We will call upon this property of the precessing
convention in Sec.\ \ref{sec8.1}, where we propose a new family of
templates for NS--BH binaries built by writing a set of orthonormal
component templates that contain all the dynamical information
expressed by factor Q, and then using their linear combinations to
reproduce the projection operation expressed by factor P.

Going back to the main thrust of this section, we obtain the detector
response $h_\mathrm{resp}$ by setting the direction to the detector
$\hN$ (specified by the angles $\Theta$ and $\varphi$ with respect to
the source frame), and by introducing the radiation frame, oriented
along the axes
\bea
\mathbf{e}_x^R &=& -\ve_x^{S}\,\sin\varphi  + \ve_y^{S}\,\cos\varphi  \,,\\
\mathbf{e}_y^R &=& - \ve_x^{S}\,\cos\Theta\cos\varphi
- \ve_y^{S}\, \cos\Theta\,\sin\varphi  +  \ve_z^{S}\,\sin\Theta \,,\\
\mathbf{e}_z^R &=& + \ve_x^{S}\,\sin\Theta\,\cos\varphi
+ \ve_y^{S}\,\sin\Theta\,\sin\varphi  +  \ve_z^{S}\, \cos\Theta  = \hN\,;
\eea
we then get
\beq
\label{easyh}
h_{\rm resp}=-\frac{2\mu}{D}\frac{M}{r}\, \left(
[\ve_+]^{ij}\,\cos2\Psi+[\ve_\times]^{ij}\,\sin2\Psi \right) \left(
[\mathbf{T}_+]_{ij}\,F_+ + [\mathbf{T}_\times]_{ij}\,F_\times \right)
\,.
\eeq
Applying the stationary-phase approximation (SPA) at the leading order, we can write the Fourier transform of $h_{\rm resp}$ as
\beq
\label{hfeasy}
\tilde{h}_{\rm resp}(f)=
-\tilde{h}_C(f)\,\left( [\mathbf{e}_+(t_f)]^{jk}+i \, [\mathbf{e}_\times(t_f)]^{jk} \right)\,
\left( [\mathbf{T}_+]_{jk}\,F_+ + [\mathbf{T}_\times]_{jk} \, F_\times \right)
\quad\mathrm{for\;}f>0\,,
\eeq
where $\tilde{h}_C(f)$ is the SPA Fourier transform of the \emph{carrier signal},
\beq
h_C=\frac{2\mu}{D}\,\frac{M}{r}\,\cos2\Psi\,,
\eeq
and where $t_f$ is the time at which the carrier signal has
instantaneous frequency $f$.

\subsection{Definition of a new DTF for precessing binaries}
\label{newdtf}

By adopting the precessing convention, we isolate all the modulational
effects due to precession in the evolving polarization tensors
$[\ve_{+,\times}]^{ij}$ (these effects will show up both in the
amplitude \emph{and} in the phase of $h_\mathrm{resp}$). The
discussion of Sec.~\ref{sec3.2} shows that, to a very good
approximation, the carrier signal is not modulated, so we expect that
$\tilde{h}_C(f)$ should be approximated well by the nonspinning PN
templates studied in BCV1, or variations thereof.  As for the time
dependence of the tensors $[\ve_{+,\times}]^{ij}$, the discussion of
Secs.\ \ref{sec3.3} and \ref{sec3.4} suggests that we adopt the
Apostolatos' ansatz~\cite{ApostolatosFF}, and write expressions in the
generic forms
\beq
\label{ansatz}
[\mathrm{e}_{+,\times}]^{ij} [\mathbf{T}_{+,\times}]_{jk}
\propto \mathcal{C}_{+,\times}
\cos\left(\mathcal{B}\,f^{-2/3}+\delta_{+,\times}\right)
\quad \mbox{or} \quad
\propto \mathcal{C}_{+,\times}
\cos\left(\mathcal{B}\,f^{-1}+\delta_{+,\times}\right)\,.
\eeq
Indeed, our extended numerical investigations provide evidence
that expressions of the form \eqref{ansatz}
should work quite well for the binaries under consideration.

All these elements suggest that we introduce a family of detection
templates of the general (Fourier-domain) form
\begin{equation}
\label{efftempLC}
h(\psi_{\mathrm{NM}},\mathcal{A}_{k},t_{0},\alpha _{k};f)=\left
[\sum^{n}_{k=1}(\alpha_{k}+i\alpha _{k+n})\mathcal{A}_{k}(f)\right
]e^{2\pi ift_{0}}e^{i\psi_{\mathrm{NM}}(f)} \quad (\mathrm{for}\;f>0)
\end{equation}
[and $h(f)=h^{*}(-f)$ for $f<0$], where the $\mathcal{A}_{k}(f)$ are
real \emph{amplitude functions}, the $\alpha_{k}$ are their (real)
coefficients, and $t_{0}$ is the time of arrival of the GW signals.
The function $\psi_\mathrm{NM}$ represents the phase of the unmodulated carrier signal; we write it as a series in the powers of $f^{1/3}$,
\begin{equation}
\label{psiNS}
\psi_{\mathrm{NM}}(f) = f^{-5/3} \,
(\psi_{0} + \psi_{1/2} f^{1/3} + \psi_{1} f^{2/3} + \psi_{3/2} f + \ldots )\,.
\end{equation}
As discussed in BCV1, this phasing works well for relatively
high-mass, nonspinning BBHs, and for NS--BH binaries; in addition, as
anticipated in Sec.\ \ref{sec3.2}, the PN coefficients $\psi_i$ are able
to capture the nonmodulational effects of spin-orbit and spin-spin
couplings on the orbital phase. In this paper we examine three specific families of detection templates of this form, listed in Tab.\ \ref{tab:templates}. The
subscripts ``2'', ``4'', and ``6'' in our abbreviations for the
template families denote the number of $\alpha_k$ coefficients that
appear in Eq.\ \eqref{efftempLC}.

The families $(\psi_0\psi_{3/2})_2$ and $(\psi_0\psi_{3/2}\alpha)_4$
were already studied in Ref.~\cite{BCV1} for the case of nonspinning
binaries.  Both families contain the leading $f^{-7/6}$ Newtonian
dependence of the amplitude; however, $(\psi_0\psi_{3/2}\alpha)_4$
contains a correction to the Newtonian amplitude (introduced in BCV1,
where it was parametrized by $\alpha$) which can account for the
variation of the rate of inspiral in the late stages of orbital
evolution. The first family is given by
\begin{equation}
(\psi_0\psi_{3/2})_2 : \quad h(\ldots;f) =
(\alpha_1 + i \alpha_2) f^{-7/6} \, \theta(f_\mathrm{cut} - f) \,
e^{2 \pi i f t_0} \exp i [\psi_0 f^{-5/3} + \psi_{3/2} f^{-2/3}];
\end{equation}
here $\alpha_1 + i \alpha_2$ can also be written as $\mathcal{A} \exp
i \phi^\mathrm{GW}_0$, where $\phi^\mathrm{GW}_0$ is the initial GW
phase, and $\mathcal{A}$ is an overall normalization factor for the
template.  So the two $\alpha_k$ coefficients encode the initial
global phase of the waveform, plus a normalization factor.  The second
family is given by
\begin{equation}
(\psi_0\psi_{3/2}\alpha)_4 : \quad
h(\ldots;f) =
[(\alpha_1 + i \alpha_2) f^{-7/6} + (\alpha_3 + i \alpha_4) f^{-1/2}]
\, \theta(f_\mathrm{cut} - f) \, e^{2 \pi i f t_0} \exp i [\psi_0 f^{-5/3} +
\psi_{3/2} f^{-2/3}];
\end{equation}
another way to rewrite the coefficients $\alpha_{1\mbox{--}4}$ more
physically is $\mathcal{A} \exp [i \phi^\mathrm{GW}_0] f^{-7/6} (1 +
\alpha \exp [i \phi^\alpha] f^{2/3})$, where $\alpha$ is the additional
amplitude parameter and $\phi^\alpha$ is the relative phase of the
amplitude correction (as in BCV1, in this paper we always set $\phi^\alpha = 0$).
So the four coefficients $\alpha_k$ encode the global phase, the
strength of the correction to the Newtonian amplitude, and the
relative phase of this correction with respect to the Newtonian
amplitude, plus an overall normalization factor.

The third family, $(\psi_0\psi_{3/2}\mathcal{B})_6$, contains the
leading Newtonian amplitude, modified by two modulation terms [a
generalization of the Apostolatos' ansatz \eqref{ansatz}] that
account for the precession of the orbital angular momentum due to spin
effects. It is given by
\begin{eqnarray}
(\psi_0\psi_{3/2}\mathcal{B})_6 : \quad h(\ldots;f) &=& f^{-7/6}
[(\alpha_1 + i \alpha_2) + (\alpha_3 + i \alpha_4) \cos(\mathcal{B}
f^{-2/3}) + (\alpha_5 + i \alpha_6) \sin(\mathcal{B} f^{-2/3}) ] \nonumber \\
&\,& \quad \quad \quad \times \, \theta(f_\mathrm{cut} - f) \, e^{2
\pi i f t_0} \exp i [\psi_0 f^{-5/3} + \psi_{3/2} f^{-2/3}];
\end{eqnarray}
another way to rewrite the six coefficients $\alpha_{1\mbox{--}6}$ in
close analogy to Apostolatos' ansatz is
\begin{eqnarray}
&& \mathcal{A} \, e^{i \phi^\mathrm{GW}_0}
f^{-7/6}
\left[ 1 + \mathcal{C} \, e^{i \phi^\mathrm{mod}}
\cos (\beta f^{-2/3} + \delta_1 + i \delta_2) \right] \\
&\equiv&
\mathcal{A} \, e^{i \phi^\mathrm{GW}_0}
f^{-7/6}
\left[ 1 + \mathcal{C}_\mathrm{cos} \, e^{i \phi^\mathrm{cos}} \cos (\beta f^{-2/3})
         + \mathcal{C}_\mathrm{sin} \, e^{i \phi^\mathrm{sin}} \sin
(\beta f^{-2/3}) \right] \nonumber
\end{eqnarray}
(where all the coefficients are still real).  So the six coefficients
$\alpha_k$ encode the global phase, the strength of the amplitude
modulation, its relative phase with respect to the Newtonian
amplitude, and the internal (complex) phase of the modulation.  It is
clear that our family implements a generalization of Apostolatos'
ansatz, because we allow a \emph{complex} phase offset between the
Newtonian and the sinusoidal amplitude terms, and also between the
cosine and sine modulational terms.
We consider also a variant $(\psi_0\psi_{3/2}\mathcal{B}')_6$ of this
family where the $f^{-2/3}$ frequency dependence in the sinusoidal
amplitude functions is replaced by $f^{-1}$.
\begin{table*}
\begin{tabular}{l|c|c|l|l}
\hline Template family &
\multicolumn{1}{c|}{$\psi_\mathrm{NM}(f)$} &
\multicolumn{1}{c|}{$\mathcal{A}_1$(f)} &
\multicolumn{1}{c|}{$\mathcal{A}_2(f)$} &
\multicolumn{1}{c}{$\mathcal{A}_3(f)$} \\ \hline \hline
$(\psi_0\psi_{3/2})_2$ & $\psi_{0}
f^{-5/3} + \psi_{3/2} f^{-2/3}$ & $f^{-7/6}
\,\theta(f_\mathrm{cut}-f)$ & & \\ $(\psi_0\psi_{3/2}\alpha)_4$ & $\psi_{0} f^{-5/3} +
\psi_{3/2} f^{-2/3}$ & $f^{-7/6} \,\theta(f_\mathrm{cut}-f)$ &
$f^{-1/2} \,\theta(f_\mathrm{cut}-f)$ & \\
$(\psi_0\psi_{3/2}\mathcal{B})_6$ & $\psi_{0}
f^{-5/3} + \psi_{3/2} f^{-2/3}$ & $f^{-7/6}
\,\theta(f_\mathrm{cut}-f)$ & $f^{-7/6} \cos(\mathcal{B} f^{-2/3})
\,\theta(f_\mathrm{cut}-f)$ & $f^{-7/6} \sin(\mathcal{B} f^{-2/3})
\,\theta(f_\mathrm{cut}-f)$ \\
$(\psi_0\psi_{3/2}\mathcal{B}')_6$ & $\psi_{0}
f^{-5/3} + \psi_{3/2} f^{-2/3}$ & $f^{-7/6}
\,\theta(f_\mathrm{cut}-f)$ & $f^{-7/6} \cos(\mathcal{B} f^{-1})
\,\theta(f_\mathrm{cut}-f)$ & $f^{-7/6} \sin(\mathcal{B} f^{-1})
\,\theta(f_\mathrm{cut}-f)$ \\ \hline
\end{tabular}
\caption{Specification of the DTFs examined in this paper.
\label{tab:templates}}
\end{table*}

For all three families, the templates are terminated at a \emph{cut}
frequency $f_\mathrm{cut}$, above which the amplitude drops to zero;
this $f_\mathrm{cut}$ is in effect one of the (intrinsic) search
parameters. For all three families, the frequency dependence of the
phase includes the leading Newtonian term, $f^{-5/3}$, and a term
$f^{-2/3}$ that corresponds to the 1.5PN correction in the phase
evolution of nonspinning binaries (as obtained, in the SPA, by
integrating the energy-balance equation through an adiabatic sequence
of circular orbits, using PN expanded energy and flux). In BCV1 we
found that including either the 1PN or 1.5PN term is in general
sufficient to model the phase evolution of nonspinning binaries of
high mass.

\subsection{Definition of the standard SPA template families}
\label{refspa}

In this section we define two families of standard nonspinning-binary
templates, obtained by solving the Taylor-expanded energy-balance
equation for an adiabatic sequence of quasicircular orbits, and using
the stationary-phase approximation (SPA) to express the result as a
function of the GW frequency $f$ (see BCV1). In Sec.\ \ref{sec6} we
compare the matching performance of these templates to the
performance of our new DTFs, to show that the various tricks used to
build the new families do indeed improve their effectualness~\cite{DIS1}.
The standard SPA families are built from the analytic expressions of
Refs.\ \cite{2PN,2.5PNand3.5PN}. The frequency-domain phasing (under the
assumption of nonevolving orbital angular momentum and spins) is given
by \cite{apostolatos2}
\begin{eqnarray}
\label{psispa}
\psi_{\rm SPA}(f)=2\pi\,f\,t_c-\phi_c+\frac{3}{128}\,(\pi {\cal M}\,f)^{-5/3} && \!\!\!\!
\left[ 1 + \frac{20}{9}\, \left( \frac{743}{336}+\frac{11}{4}\eta \right) \,
(\pi M\,f)^{2/3}\,-4(4\pi - {\rm SO})\,(\pi M\,f) \right. \nonumber \\
&& \;\; \left. + 10\,\left ( \frac{3058673}{1016064}+\frac{5429}{1008}\eta +\frac{617}{144}\eta^2-
{\rm SS} \right) \,(\pi M\,f)^{4/3} \right] \,,
\end{eqnarray}
where ${\cal M} = M \eta^{3/5}$ is the \emph{chirp mass}, and where ${\rm SO}$ and ${\rm SS}$ are the spin-orbit and spin-spin terms, given explicitly by
\bea
\label{SO}
{\rm SO} &=& \frac{1}{M^2}\,\left [ \left (\frac{113}{12}+
\frac{25}{4}\,\frac{m_2}{m_1}\right )\,\vS_1+
\left (\frac{113}{12}+
\frac{25}{4}\,\frac{m_1}{m_2}\right )\,\vS_2\right ]
\cdot \hL_N\,,\\
\label{SS}
{\rm SS} &=& \frac{1}{48m_1\,m_2\,M^2}\,\left [
-247\,\vS_1 \cdot \vS_2 + 721 (\vS_1 \cdot \hL_N)\,
(\vS_2 \cdot \hL_N)\right ]\,.
\eea
We neglect all PN corrections to the amplitude, by adopting its
Newtonian functional form, $f^{-7/6}$; we also neglect all
precessional effects, by setting $\mathrm{SO} = \mathrm{SS} =
0$. Templates of this form are routinely used in searches for GW
signals from nonspinning binaries. In that case, the templates are
generally ended at the GW frequency corresponding to the Schwarzschild
ISCO $f_{\rm Schw} \simeq 0.022/M$. We denote such templates as
SPAs. We introduce also a variant of this family, SPAc, characterized
by the additional frequency-cut parameter $f_\mathrm{cut}$, used also
in our DTFs. Altogether, we get
\begin{equation}
\mathrm{SPAs} : h(\mathcal{M},\eta,t_0,\psi_0,\alpha_N;f) =
\alpha_N f^{-7/6} \theta(f_\mathrm{Schw} - f) e^{2 \pi i f t_0} \exp i[\psi_\mathrm{SPA} + \psi_0];
\end{equation}
\begin{equation}
\mathrm{SPAc} : h(\mathcal{M},\eta,f_\mathrm{cut},t_0,\psi_0,\alpha_N;f) =
\alpha_N f^{-7/6} \theta(f_\mathrm{cut} - f) e^{2 \pi i f t_0} \exp i[\psi_\mathrm{SPA} + \psi_0].
\end{equation}

\section{GW data analysis with the DTF}
\label{sec5}

In searching for GW signals using matched-filtering techniques, we
construct a discrete bank of templates that represent all the possible
signals that we expect to receive from a given class of sources.  We
then proceed to compare each stretch of detector output with each of
the templates, computing their \emph{overlap} (essentially, a weighted
correlation).  A high value of the overlap statistic for a given
stretch of detector output and for a particular template implies that
there is a high probability that during that time the detector
actually received a GW signal similar to the template.  This technique
is intrinsically probabilistic because, for any template, detector
noise alone can (rarely) yield high values of the statistic. In
general, the higher the value of the statistic, the harder it is to
obtain it from noise alone. So it is important to set the
\emph{detection threshold} (above which we confidently claim a
detection) by considering the resulting probability of the \emph{false alarms}
caused by noise.

To verify whether the DTFs developed in Sec.\ \ref{sec5} can be used
to search reliably and effectually for the GWs from spinning binaries,
we need to evaluate the \emph{fitting factor} FF of the DTFs in
matching the target signals for a variety of binary and detector
parameters. The FF is defined as the ratio between the overlap of the
target signal with the best possible template in the family and the
overlap of the target signal with itself \cite{note41}. So in Sec.\ \ref{sec5.1} we discuss the
maximization of the overlap over template
parameters for a given target signal. The other important element to
evaluate the reliability and effectualness~\cite{DIS1} of the DTFs are the
detection thresholds that the DTFs yield for a given false-alarm
probability. In Sec.\ \ref{sec5.3} we discuss these thresholds under
the simplifying hypothesis of Gaussian detector noise.
The material presented in this section builds on the treatment of
matched-filtering data analysis for GW sources given in Sec.\ II of
BCV1 (which is built on Refs.~\cite{DA,DIS1,DIS3}), and it uses the same notations.

\subsection{Maximization of the overlap over template parameters}
\label{sec5.1}

Among all the template parameters that appear in Eq.\
\eqref{efftempLC}, we are going to treat the $\psi_i$,
$f_\mathrm{cut}$ and $\mathcal{B}$ as \emph{intrinsic} parameters; and
the $\alpha_k$ and $t_0$ as \emph{extrinsic} parameters: that is, when
we look within one of our DTFs for the template that best matches a
given target signal, we will need to consider \emph{explicitly} many
different values of the $\psi_i$, of $f_\mathrm{cut}$, and of
$\mathcal{B}$; however, for any choice of these parameters, the best
$\alpha_k$ and $t_0$ are determined automatically by simple algebraic
expressions (see Sec.\ II B of BCV1). For the next few paragraphs, where we discuss the optimization of the coefficients $\alpha_k$, we shall
not indicate the dependence of the templates on the intrinsic parameters.

For a given signal $s$, we seek the maximum of the overlap,
\begin{equation}
\label{maximumdesired}
\max_{t_{0},\alpha_{k}} \, \langle s,h(t_{0},\alpha _{k})\rangle \,,
\end{equation}
under the normalization condition
\begin{equation}
\label{normalizedtemplate}
\langle h(t_{0},\alpha_{k}),h(t_{0},\alpha_{k})\rangle =1
\end{equation}
[this condition is necessary to set a scale for the statistic
distribution of the overlap between a given template and pure noise]. Here the
inner product $\langle g , h \rangle$ of two real signals with Fourier
transforms $\tilde{g}$, $\tilde{h}$ is defined by
\begin{equation}
\langle g, h \rangle =
2 \int_{-\infty}^{+\infty} \frac{\tilde{g}^*(f)
\tilde{h}(f)}{S_n(|f|)} df =
4 \, \mathrm{Re} \int_{0}^{+\infty}
\frac{\tilde{g}^*(f) \tilde{h}(f)}{S_n(f)} df
\end{equation}
(see BCV1). We proceed constructively: first, we build a new set of
amplitude functions $\hat{\mathcal{A}}_{k}(f)$ that are linear
combinations of the $\mathcal{A}_{k}(f)$, and that satisfy the
orthonormality condition $\langle
\hat{\mathcal{A}}_{i}(f),\hat{\mathcal{A}}_{j}(f)\rangle =\delta
_{ij}$ for $i,j=1,2,\ldots n$; we then define an orthonormal set of
single-$\hat{\mathcal{A}}_{k}$ templates,
\begin{equation}
\label{hitems}
\hat{h}_{k}(t_{0};f)\equiv \hat{\mathcal{A}}_{k}(f)e^{2\pi
ift_{0}}e^{i\psi _{\mathrm{NM}}}\,,\quad \quad
\hat{h}_{k+n}(t_{0};f)\equiv i\hat{\mathcal{A}}_{k}(f)e^{2\pi
ift_{0}}e^{i\psi _{\mathrm{NM}}} \quad (\mathrm{for}\;f>0)
\end{equation}
[and $\hat{h}_{k}(f)=\hat{h}_{k}^{*}(-f)$ for $f<0$], which satisfy
$\langle \hat{h}_{i}(t_{0}),\hat{h}_{j}(t_{0})\rangle =\delta _{ij}$
(with $i,j=1, 2, \ldots, 2n$) for any $t_0$. The maximized overlap
[Eq.\ \eqref{maximumdesired}] can now be rewritten as
\begin{equation}
\label{exmax}
\max_{t_{0},\alpha _{k}} \, \langle s,h(t_{0},\alpha _{k})\rangle =
\max_{t_{0}}\max _{\hat{\alpha }_{k}}\sum _{k=1}^{2n}\hat{\alpha
}_{k}\langle s,\hat{h}_{k}(t_{0})\rangle \,,
\end{equation}
while the condition \eqref{normalizedtemplate} is now simply
$\sum _{k=1}^{2n}\hat{\alpha }_{k}^{2}=1$.
The inner maximum of Eq.~(\ref{exmax}) (over the $\hat{\alpha}_{k}$)
is achieved when
\begin{equation}
\label{alphaopt}
\hat{\alpha}_{k}=\frac{\langle s,\hat{h}_{k}(t_{0})\rangle
}{\sqrt{\sum _{j=1}^{2n}\langle s,\hat{h}_{j}(t_{0})\rangle ^{2}}}\,,
\end{equation}
and the maximum overlap itself is
\begin{equation}
\label{singlesignal}
\max_{t_{0},\alpha _{k}} \, \langle s,h(t_{0},\alpha _{k})\rangle =
\max_{t_{0}}\max _{\hat{\alpha }_{k}}\sum _{k=1}^{2n}\hat{\alpha
}_{k}\langle s,\hat{h}_{k}(t_{0})\rangle = \sqrt{\max _{t_{0}}\sum
_{j=1}^{2n}\langle s,\hat{h}_{j}(t_{0})\rangle ^{2}}\,.
\end{equation}
This happens essentially because the sum in Eq.\ \eqref{exmax} can be seen
as a scalar product in a $2n$-dimensional Euclidean space, which is
maximized when the unit $2n$-vector $\hat{\alpha}_k$ lies along the
direction of the $2n$-vector $\langle s, \hat{h}_k(t) \rangle$.  The
quantities $\langle s,\hat{h}_{j}(t_{0})\rangle $ for
$j=1,2,3,\ldots,n$ are given by the two related Fourier integrals
\begin{eqnarray}
\langle s,\hat{h}_{j}\rangle & = & 2\, \mathrm{Re} \, \int ^{+\infty
}_{0}\frac{\hat{\mathcal{A}}_{j}(f)e^{i\psi
_{\mathrm{NM}}(f)}s^{*}(f)}{S_{h}(f)} \, e^{2\pi
ift_{0}}df\,,\label{innerprodexplicita} \\
\langle s,\hat{h}_{j+n}\rangle & = & -2\, \mathrm{Im} \, \int ^{+\infty
}_{0}\frac{\hat{\mathcal{A}}_{j}(f)e^{i\psi_{\mathrm{NM}}(f)}s^{*}(f)}{S_{h}(f)}
\, e^{2\pi ift_{0}}df\,.\label{innerprodexplicit}
\end{eqnarray}

We now go back to discussing the full set of template parameters. The
relevant measure of the \emph{effectualness}~\cite{DIS1} of a template family at
matching a physical signal $s$ is the \emph{fitting factor} FF,
\begin{equation}
\label{FFdef}
\mathrm{FF} = \max_{t_{0},\alpha_{k},f_\mathrm{cut},\psi_i} \,
\frac{\langle s,h(t_{0},\alpha _{k})\rangle}{\sqrt{\langle
s,s\rangle}} \,,
\end{equation}
(see, for instance, Sec.\ II of BCV1) which is maximized over the
$\alpha_{k}$, but also over the time of arrival $t_0$ (also an
extrinsic parameter), and over all the intrinsic parameters, $\psi_i$,
$f_\mathrm{cut}$, and $\mathcal{B}$. The fitting factor is a function
of the physical parameters of the physical signal $s$, and of course
of the template family used to match it. We define also the
\emph{signal amplitude} SA for a given signal,
\begin{equation}
\mathrm{SA} = \sqrt{\langle s,s \rangle}.
\end{equation}
SA gives the \emph{optimal} overlap obtained for a template that is exactly equal to the signal (except for its normalization), and it is inversely proportional to the luminosity distance to the source; where we do not indicate otherwise, we always assume the fiducial distance $d_0 = 100$ Mpc.

The maximization of the overlap over $t_0$ is easy to obtain, because the integrals \eqref{innerprodexplicita} and \eqref{innerprodexplicit} can be
evaluated at the same time for all the $t_0$ using Fast Fourier
Transform techniques~\cite{Schutz}. On the other hand, the
maximization over $f_\mathrm{cut}$ and over the other intrinsic
parameters is obtained by an explicit search over a multidimensional
parameter range, where we look for the maximum of the partially maximized (over extrinsic parameters) overlap, given by Eq.\ \eqref{maximumdesired}.
For all the actual searches discussed in this paper we employ with good results the simplicial algorithm \texttt{amoeba}~\cite{nrc}.

\subsection{False-alarm statistics of the DTFs}
\label{sec5.3}

In the practice of GW data analysis, template \emph{families} are used
to build discrete template \emph{banks} parametrized by a discrete set
of ntuples of the intrinsic parameters.  Then each of the templates is
correlated with the detector output, to see if the detection statistic
[in our case, the partially maximized correlation
\eqref{maximumdesired}] is greater than the detection threshold. It is
important to notice that the statistic is already maximized with
respect to the extrinsic parameters, while the intrinsic parameters
serve as labels for each of the templates.  Therefore, we are
effectively setting up a separate detection test for each of the
templates in the bank.

In this section we are going to evaluate the false-alarm probability
for one such test, defined as the probability that detector noise
alone will yield an overlap greater than the detection threshold. The
total false-alarm probability is then obtained by multiplying the
false-alarm probability for a single template by the number
$\mathcal{N}_\mathrm{shapes}$ of independent signal shapes (generally
of the same order of magnitude as the number of templates in the
bank), and by the number $\mathcal{N}_\mathrm{times}$ of possible
times of arrival $t_0$, distanced in such a way that the displaced
templates are essentially orthogonal \cite{note42}. At the end of this exercise, we are going to
set the detection threshold so that the total false-alarm probability
is acceptably low.

Under the assumption of Gaussian noise, the inner product $\langle n ,
\hat{h}_j \rangle$ of noise $n$ alone with a normalized template
component $\hat{h}_i$ is (by construction) a Gaussian random variable
with zero mean and unit variance (see, for instance, Sec.\ II of
BCV1). Because (for the same $t_0$ and for the same intrinsic
parameters) all the $\hat{h}_j$ are orthogonal, the inner products
$\langle n , \hat{h}_j \rangle$ (for $j = 1, \ldots, 2n$) are all
independent normal variables. It follows that the statistic $X =
\max_{t_{0},\alpha _{k}} \, \langle n,h(t_{0},\alpha _{k})\rangle$
[see Eq.\ \eqref{singlesignal}], given by the square root of the sum
of their squares, follows the $\chi$ distribution with $2n$
degrees of freedom, characterized by the probability density function
and cumulative distribution function
\begin{equation}
\mathrm{PDF}_{\chi(2n)}(X = x) = \frac{x^{2n - 1} e^{-x^2 / 2}}{2^{n-1} \Gamma(n)}, \quad
\mathrm{CDF}_{\chi(2n)}(X < x) = \frac{\Gamma(n,0,x^2/2)}{\Gamma(n)},
\end{equation}
where we have used the \emph{generalized incomplete gamma function}
$\Gamma(n,z_0,z_1) = \int_{z_0}^{z_1} t^{n-1} e^{-t} dt$. For $n=1$ we
obtain the \emph{Rayleigh distribution}, typical of the maximization
of the amplitude of signals with two quadratures.

In Tab.\ \ref{table:false} we show the thresholds needed to obtain a
total false alarm probability of $10^{-3}$, with
$\mathcal{N}_\mathrm{times} = 3 \, 10^{10}$ (typical of about three
years of observation with LIGO), and with the
$\mathcal{N}_\mathrm{shapes}$ given in the first column. We observe
that each time we increase $\mathcal{N}_\mathrm{shapes}$ by one order
of magnitude, the threshold increases by about 2\% (this happens
uniformly for all $n$'s). On the other hand, each step in $n$
increases the threshold by about 4\%. Thus, when we design DTFs we
should keep in mind that the best possible overlap increases with the
number of templates employed, and with the complexity of the templates
(clearly, the complexity of our DTFs increases with the number of
amplitude functions); but the detection threshold increases as well,
reducing the number of signals that pass the detection test. So in
principle we are justified in using more numerous and more complex
templates only if the gain in the overlap is larger than the increase
in the detection threshold.

The prospects shown in Tab.\ \ref{table:false} for the models with
$n=2$ and $n=3$ improve somewhat if we constrain the values that the
$\alpha_k$ can attain when they are (algebraically) maximized. We can
do this, for instance, if we judge that certain combinations of the
$\alpha_k$ correspond to unphysical waveforms, but then we must be
consistent and exclude any detections that cross the threshold within
the excluded parameter region.  In any case, we should remember that
our study of false-alarm statistics is based on the idealization of
Gaussian noise, which will not be realized in practice: real-world
data-analysis schemes relie on matched-filtering techniques
complemented by \emph{vetoing schemes} \cite{40meter}, which remove
detection candidates using nonlinear tests on the signal. Therefore,
any DTF should be evaluated in that context before it is excluded for
producing excessive detection thresholds within the Gaussian analysis.
\begin{table}
\begin{tabular}{c|c|c|c}
\hline
$\mathcal{N}_\mathrm{shapes}$ & \multicolumn{3}{c}{Threshold for
false-alarm probability $= 10^{-3}$} \\
& \multicolumn{1}{c|}{$(\psi_0\,\psi_{3/2})_2$}
& \multicolumn{1}{c|}{$(\psi_0\,\psi_{3/2}\,\alpha)_4$} &
\multicolumn{1}{c}{$(\psi_0\,\psi_{3/2}\,\mathcal{B})_6$} \\
& \multicolumn{1}{c|}{$n = 1$}
& \multicolumn{1}{c|}{$n = 2$} &
\multicolumn{1}{c}{$n = 3$} \\\hline \hline
$10^2$ & 8.44 & 8.87 & 9.22 \\
$10^3$ & 8.71 & 9.13 & 9.48 \\
$10^4$ & 8.97 & 9.39 & 9.73 \\
$10^5$ & 9.22 & 9.63 & 9.97 \\
$10^6$ & 9.47 & 9.87 & 10.21 \\
\hline
\end{tabular}
\caption{\label{table:false}Detection thresholds for a false-alarm
probability $= 10^{-3}$ for a $\chi$-distributed detection statistic
with $2n$ degrees of freedom, for $\mathcal{N}_\mathrm{times} = 3 \,
10^{10}$, and for the $\mathcal{N}_\mathrm{shapes}$ given in the first
column. The values given for $(\psi_0,\psi_{3/2}\,\alpha)_4$ do not take into account the $\phi^\alpha = 0$ constraint.}
\end{table}

\section{Evaluation of DTF performance}
\label{sec6}

We wish to investigate the effectualness~\cite{DIS1} of our DTFs in matching the
GW signals generated by precessing binaries of spinning compact
objects, at least as approximated by the target model described in
Sec.\ \ref{sec2}. To do so, we shall evaluate the fitting factor FF
[Eq.\ \eqref{FFdef}] of the DTFs over a population of binaries with a
variety of basic, local and directional parameters, and compare the
results with the FF obtained for the standard SPA families [Sec.\
\ref{refspa}].  In Sec.\ \ref{dirpar} we study the effect of the
directional parameters on FF (and SA), with the aim of reducing the
dimensionality of the test populations.  In Sec.\ \ref{montecarlo} we
describe the Monte Carlo scheme used to generate the populations, and
we identify two performance indices for the template families (namely,
the simple and SA-weighted averages of FF).  In Sec.\ \ref{sec8} we
give our results for these indices, focusing first on the BBHs
considered in this paper.  Finally, in Sec.\ \ref{sec8.1} we give our
results for NS--BH binaries, and we briefly describe a new, very
promising family of templates for these systems, suggested by the
insights accreted during the development of this paper.

\subsection{Effect of directional parameters on FF and SA}
\label{dirpar}

As we have seen in Secs.\ \ref{sec2.3} and \ref{sec2.4}, the detector
response $h_\mathrm{resp}$ is a function not only of the basic and
local parameters of the binary (which describe respectively the masses
and spin magnitudes, and the initial relative directions of the spins
and the orbital angular momentum, and therefore change the dynamical
evolution of the binary), but also of the directional parameters
(which describe the relative direction and orientation of binary and
detector, and alter the \emph{presentation} of the precessing orbital
plane of the binary with respect to the direction and orientation of
the detector). Thus, \emph{all} the parameters will affect \emph{both}
the amplitude $\mathrm{SA} = \langle h_\mathrm{resp} , h_\mathrm{resp}
\rangle^{1/2}$ of the signals received at the detector \emph{and} the
ability of our DTFs to match them, as codified in the fitting factor
FF; it is therefore clear that, in evaluating the effectualness of our
DTFs at matching the target signals, we will need to compute FF not
only for a range of binary masses and spins, but also for a suitable
sampling of the local and directional parameters.

In the case of nonspinning binaries (see BCV1), there are no local
parameters as we defined them in this paper; the directional
parameters \emph{do} change the GW signal, but only by multiplying its
amplitude by a constant factor, and by adding a constant offset to its
phase (as opposed to modulating amplitude and phase as in the case of
spinning binaries). In BCV1 (following a common practice in the GW
data-analysis literature), we included the variation of the amplitude
\emph{in the definition} of the target signals, by averaging the
amplitude factor over uniform solid-angle distributions of the
directional parameters [see Eq.\ (29) of Sec.\ II D]. As for the initial
phase of the signal, we defined the FF on the basis of \emph{minmax}
overlaps \cite{DIS1}, which
are maximized over the initial \emph{template} phase (and over all the
other extrinsic and intrinsic template parameters) but
\emph{minimized} over the initial signal phase; this minimization is
obtained algebraically, just as for extrinsic template parameters. In
fact, it turns out that minimizing or maximizing the overlap over the
initial signal phase changes the resulting FF by a very small
quantity.

In the case of the spinning binaries examined in this paper, this
picture changes radically, because minimizing the overlap over the
directional parameters yields very low FFs that are not representative
of the typical results that we would get in actual observations.
So we
take a different approach: we study the distribution of FF for a
population of binaries characterized by different basic, local, and
directional parameters. In particular, we select several
astrophysically relevant combinations of basic parameters, and we
sample randomly (but as uniformly as possible) the space spanned by
the local and directional parameters. In practice, we can exploit
certain symmetries of this space (that is, the fact that different
combinations of the local and directional parameters yield the same
signal) to reduce its effective dimensionality. Let us see how.

Under the FC convention, the complete specification of a target signal
requires (at least formally) 15 parameters: according to our
classification (Sec.\ \ref{sec2.4}), four of these are the basic
parameters ($M$, $\eta$, $S_1$, and $S_2$); three are the local binary
angles ($\theta_{S_1}$, $\theta_{S_2}$, and $\phi_{S_1}-\phi_{S_2}$);
three are the directional binary angles ($\theta_{\rm L_{\rm N}}$,
$\phi_{\rm L_{\rm N}}$, and $\phi_{S_1}+\phi_{S_2}$); and five are the
directional GW and detector angles ($\Theta$, $\varphi$, $\theta$,
$\phi$, and $\psi$). Of the latter, $\theta$, $\phi$, and $\psi$ come
into the waveform only through the antenna patterns $F_+$ and
$F_\times$ [see Eqs.\ \eqref{Fplus} and \eqref{Fcross}].  It is
redundant to specify \emph{both} the directional binary angles (which
determine the orientation of the binary as a whole in space)
\emph{and} the directional GW angles (which determine the direction
$\hN$ of GW propagation to the detector), because if we apply the same
rotation to $\hN$ and to the binary vectors $\hL_N$, $\hS_1$, and
$\hS_2$, we do not change the response of the detector
$h_\mathrm{resp}$. So we can use this freedom to set $\Theta = \pi/2$
and $\varphi=0$. Once this is done, we still have the freedom to
rotate the detector--binary system around the axis $\hN$. Such a
rotation (by an angle $\nu$) will transform the $F_+$ and $F_\times$
antenna patterns according to
\begin{eqnarray}
\label{fptrans}
F_+ &\rightarrow& F_+ \cos 2\nu - F_\times \sin 2\nu, \\
F_\times &\rightarrow& F_+ \sin 2\nu + F_\times \cos 2\nu.
\label{fttrans}
\end{eqnarray}
Looking at Eqs.\ \eqref{Fplus} and \eqref{Fcross}, we see that, for
any original $\theta$, $\phi$, and $\psi$, we can always find an angle
$\nu$ for which $F_+ = 0$. The corresponding new $F_\times$ becomes
\begin{equation}
\label{Ftdist}
F_\times = \pm \frac{1}{2} \sqrt{(1+\cos^2 \theta)^2 \cos^2 2 \phi + 4
\cos^2 \theta \sin^2 2 \phi} \, ;
\end{equation}
once again, the detector response does not change. For future use, let
us define as $p[F_\times]$ (with $\int_0^1 p[F_\times] dF_\times = 1$)
the probability density for $|F_\times|$ induced by uniform
solid-angle distributions for $\theta$ and $\phi$ [notice that $\psi$
does not appear in \eqref{Ftdist}].

Now, for a given DTF and for given basic parameters, consider the
distribution of FF and SA obtained for an 11-parameter
population of target signals specified by uniform solid-angle
distributions of $\theta_{\mathrm{L_\mathrm{N}},S_1,S_2}$,
$\phi_{\mathrm{L_\mathrm{N}},S_1,S_2}$, $\Theta$, $\varphi$, $\theta$,
$\phi$, and $\psi$. By the above arguments, we obtain the same
distribution of FF and SA from a 6-parameter population of target
signals specified by uniform solid-angle distributions of
$\theta_{\mathrm{L_\mathrm{N}},S_1,S_2}$,
$\phi_{\mathrm{L_\mathrm{N}},S_1,S_2}$, by $\Theta = \pi/2$,
$\varphi=0$, $F_+ = 0$, and by $F_\times$ distributed acoording to
$p[F_\times]$. Moreover, because $F_\times$ appears only as a
normalization factor in front of the expression \eqref{h} for the
signal (once $F_+=0$), we can simply set $F_\times = 1$: this
operation does not change FF [because $F_\times$ appears homogeneously
in the numerator and denominator of Eq.\ \eqref{FFdef}], while the
distribution of SA for the original 11-parameter population can be
recovered from its moments on the 6-parameter population:
\begin{equation}
\langle \mathrm{SA}^m \rangle_{11\mbox{-}\mathrm{par}} = \left\langle \int_0^1 (F_\times)^m \mathrm{SA}^m p[F_\times] dF_\times
\right\rangle_{6\mbox{-}\mathrm{par}} = \langle \mathrm{SA}^m
\rangle_{6\mbox{-}\mathrm{par}} \int_0^1 (F_\times)^m p[F_\times] dF_\times .
\end{equation}

\subsection{A Monte Carlo procedure to evaluate DTF performance}
\label{montecarlo}

We are going to evaluate the effectualness~\cite{DIS1} of our DTFs within a Monte
Carlo framework, by studying the distribution of FF (and
$\mathrm{FF}^3 \mathrm{SA}^3$, see below) over six sampled populations
of 1000 binaries each, specified as follows.  We study the binary
systems already examined in Sec.\ \ref{sec3}: BBHs with masses
$(20+10)M_\odot$, $(15 + 15)M_\odot$, $(20 + 5)M_\odot$, $(10 +
10)M_\odot$, and $(7 + 5)M_\odot$, and NS--BH binaries with masses
$(10+1.4)M_\odot$. All the BHs have maximal spin, while the NSs have
no spin. We integrate numerically the target-model equations starting
from initial configurations that correspond to instantaneous GW
frequencies of 30 Hz when $M > 20 M_\odot$, and 40 Hz otherwise. For
each set of masses, we use the Halton sequence with bases 2, 3, 5, 7,
11, and 13 to generate 1000 \emph{quasirandom} sets of the six angles
$\theta_{\mathrm{L_\mathrm{N}},S_1,S_2}$ and
$\phi_{\mathrm{L_\mathrm{N}},S_1,S_2}$; the directions of the
resulting orbital angular momentum and spins are uniformly distributed
over the solid angle. We denote each sestuple by the sequential index
$l$, for $l = 1, \ldots, \mathcal{N} = 1000$. We always set $\Theta =
\pi/2$, $\varphi = 0$, $F_+ = 0$, $F_\times = 1$, and we take $d_0 =
100$ Mpc.

For each set of masses, and for each DTF, we compute the Monte Carlo average of the FF,
\begin{equation}
\overline{\mathrm{FF}} = \langle \mathrm{FF} \rangle =
\frac{1}{\mathcal{N}} \sum_{l=1}^\mathcal{N} \mathrm{FF}[l],
\end{equation}
and its variance
\begin{equation}
\sigma^2_\mathrm{FF} = \langle \Delta \mathrm{FF}^2 \rangle =
\frac{1}{\mathcal{N}-1} \sum_{l=1}^\mathcal{N} \left( \mathrm{FF}[l] -
\overline{\mathrm{FF}} \right)^2,
\end{equation}
which can be used to estimate the sampling error of the Monte Carlo
average as $\Delta \mathrm{FF} \simeq
\sigma_\mathrm{FF}/\sqrt{\mathcal{N}}$.

There is another function of FF and SA that has a particular interest
for our purposes. Consider each configuration $l$ as a representative
of a subclass of physical signals that have the same binary, GW, and
detector parameters (except for the degenerate parameters discussed
above), but that are generated uniformly throughout the universe. The
rate of successful signal detections using a given DTF is then
\begin{equation}
\mathcal{R}_\mathrm{detect}[l,F_\times = 1] = \mathcal{R}_{d_0} \left(
\frac{\mathrm{FF}[l]\,\mathrm{SA}[l]}{\mathrm{threshold}[\mathrm{DTF}]
} \right)^3,
\end{equation}
where $\mathcal{R}_{d_0}$ is the rate of events out to the distance
$d_0$ from Earth. Here we assume that $\mathcal{R}_{d_0}$ is a
function of the basic parameters of the binary, but not of $l$. This
equation holds because FF$[l]$ SA$[l]$ is the signal-to-noise ratio
(that is, the overlap maximized over the DTF) for the signal $l$ at
the distance $d_0$; the ratio of FF$[l]$ SA$[l]$ to the DTF threshold
gives the fraction or multiple of the distance $d_0$ out to which
signals of the class $l$ will pass the detection test. Folding in
$p[F_\times]$ we get
\begin{equation}
\mathcal{R}_\mathrm{detect}[l] =
\mathcal{R}_\mathrm{detect}[l,F_\times = 1] \cdot \int_0^1
(F_\times)^3 p[F_\times] dF_\times = 0.293 \cdot
\mathcal{R}_\mathrm{detect}[l,F_\times = 1].
\end{equation}
Summing over the $l$, we get an estimate of the total detection rate,
$\mathcal{R}_\mathrm{detect} = (1/\mathcal{N}) \sum_{l=1}^\mathcal{N}
\mathcal{R}_\mathrm{detect}[l]$.  On the other hand, the
\emph{optimal} detection rate that we would obtain with a perfectly
faithful DTF is
\begin{equation}
\mathcal{R}_\mathrm{optimal} = \mathcal{R}_{d_0} \frac{1}{\mathcal{N}}
\sum_{l=1}^\mathcal{N}
\left( \frac{\mathrm{SA}[l]}{\mathrm{threshold}[\mathrm{DTF}] } \right)^3 \cdot
\int_0^1 (F_\times)^3 p[F_\times] dF_\times.
\end{equation}
We can therefore define the \emph{effective average fitting factor}
$\overline{\mathrm{FF}}_\mathrm{eff}$ (which is a function of the
basic parameters of the binary, but which is already integrated over
$l$) from the equation
\begin{equation}
\mathcal{R}_\mathrm{detect} = \overline{\mathrm{FF}}_\mathrm{eff}^3
\mathcal{R}_\mathrm{optimal}.
\end{equation}
We then get
\begin{equation}
\overline{\mathrm{FF}}_\mathrm{eff} = \left\{
\frac{\langle \mathrm{FF}^3 \mathrm{SA}^3 \rangle}{\langle \mathrm{SA}^3 \rangle}
\right\}^{1/3}.
\end{equation}
To compute the Monte Carlo results presented below we use the
\emph{jackknifed} \cite{jack} version of this statistic to remove
bias, and we estimate the error $\Delta
\overline{\mathrm{FF}}_\mathrm{eff}$ as the jackknifed sampling
variance. For each class of binaries and for a specific DTF, the
effective fitting factor $\overline{\mathrm{FF}}_\mathrm{eff}$
represents the reduction in the detection range due to the
imperfection of the DTF. The corresponding reduction in the detection
rate is $\overline{\mathrm{FF}}_\mathrm{eff}^3$.

\begin{figure*}[t]
\begin{center}
\includegraphics{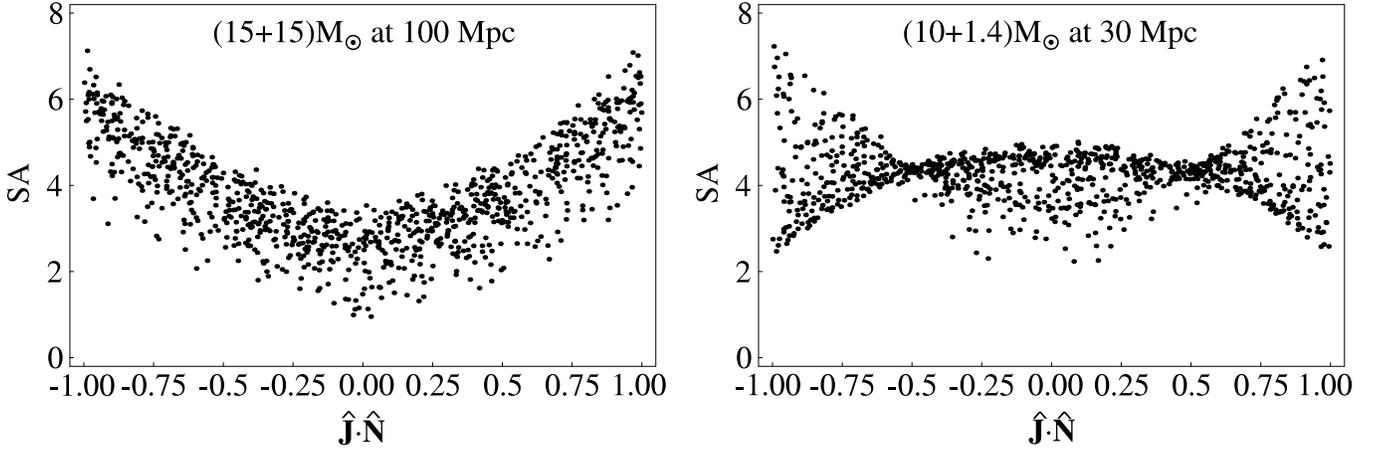}
\caption{\label{Ampl} GW signal amplitude SA as a function of the
initial $\hJ_N \cdot \hN$ (that is, the cosine of the angle between
the direction of GW propagation and the initial total angular momentum
at the Newtonian order), for our Monte Carlo populations of
$(15+15)M_\odot$ BBHs (in the left panel) and $(10 +1.4)M_\odot$
NS--BH binaries (in the right panel).  The signal amplitude is computed
for a LIGO-I noise curve [Eq.\ (28) of BCV1]; it is normalized at
fiducial distances of 100 and 30 Mpc, and averaged over the
probability distribution $p[F_\times]$.}
\end{center}
\end{figure*}
In Fig.~\ref{Ampl} we show two examples of the distribution of signal
amplitudes for the $(15+15) M_\odot$ BBHs and for the $(10+1.4)
M_\odot$ NS--BH binaries in our Monte Carlo population (as computed
with the 2PN target model). The plots show SA as a function of the
initial $\hJ_N \cdot \hN$, normalized at distances that yield SAs
comparable to typical detection thresholds, and averaged over the
probability distribution $p[F_\times]$.  For heavy, comparable-mass
BBHs (except perhaps for the last stages of the inspiral), the orbital
angular momentum $\vL_N$ is much larger than $\vS_{1,2}$, so the
initial total angular momentum $\vJ_N$ is almost perpendicular to the
orbital plane; furthermore, as seen in Sec.~\ref{sec3.3}, the
direction of $\vJ_N$ does not change much during evolution. Because in
the quadrupole approximation the emission of GWs is stronger along the
direction perpendicular to the orbital plane, values of $|\hJ_N \cdot
\hN|$ close to one give stronger signals, as seen in the left panel of
Fig.~\ref{Ampl}. For NS--BH binaries, where $\eta$ is
small, the BH spin $\vS_1$ is much larger than $\vL_N$, and $\vJ_N$
lies roughly along $\vS_1$. So the upward curve of the left panel
appears when $\vL_N$ is roughly parallel or antiparallel to $\vS_1$
and $\hJ_N$ (that is, when the conserved quantity $\kappa_\mathrm{eff}
\propto \hL_N \cdot \hS_1$ has a large absolute value), while a
downward curve appears when $\vL_N$ is orthogonal to $\vS_1$ and
$\hJ_N$ (that is, when $\kappa_\mathrm{eff}$ has a value close to zero
\cite{note43}). The mixture of these two tendencies creates
the shape seen in the right panel of Fig.~\ref{Ampl}.

\subsection{Performance indices for the standard SPA templates and for the modulated DTFs}
\label{sec8}

\begin{figure*}
\begin{center}
\includegraphics{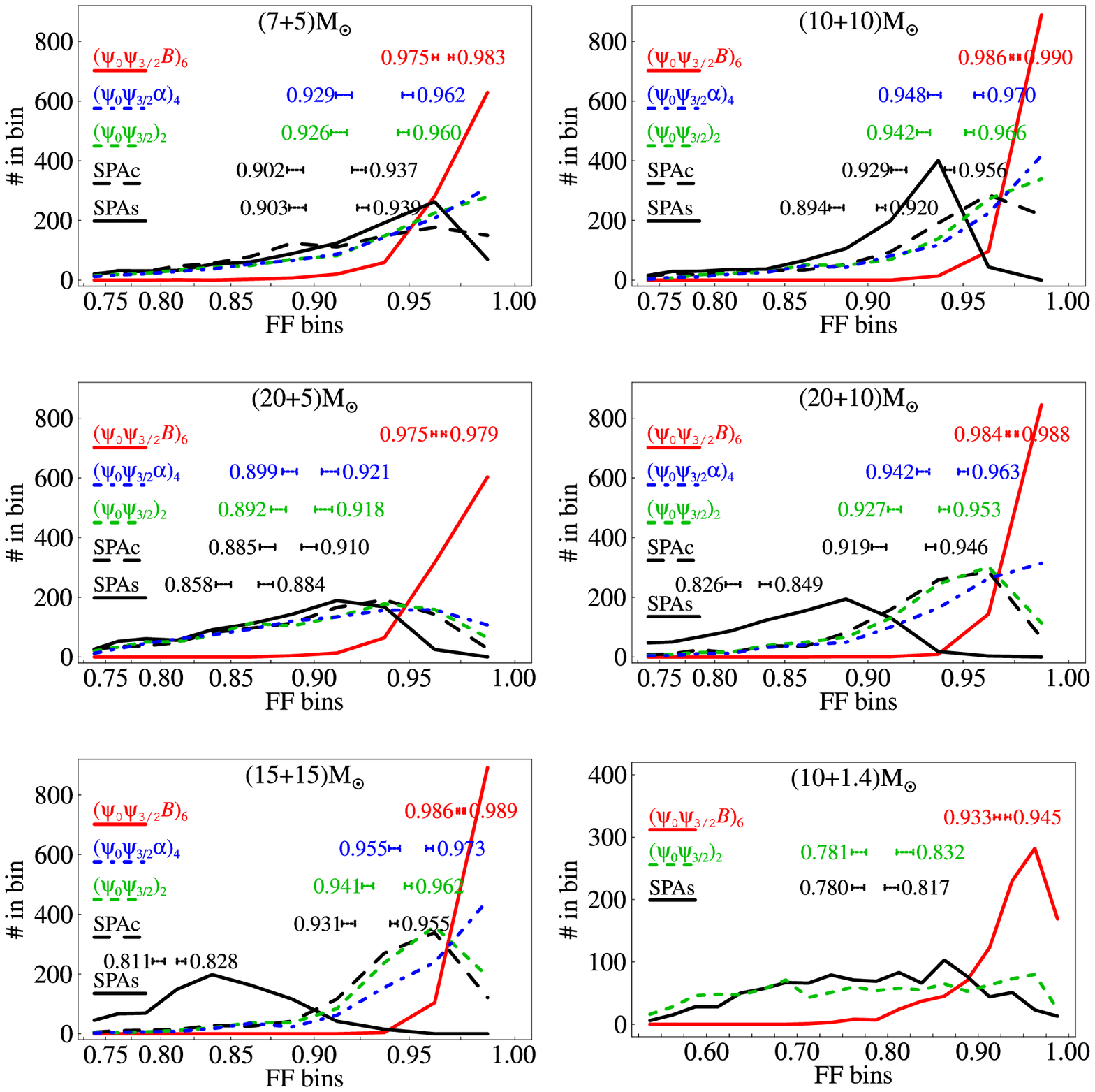}
\caption{Distribution of fitting factor FF against the 2PN target
model for the DTFs and for the standard SPA template families, for
our BBH and NS--BH Monte Carlo populations.  The vertices of the
segmented curves show the number of samples (out of 1000) for which
the FF falls within the equispaced bins $[0.725, 0.75)$, $[0.75,
0.775)$, \ldots\ (the bins are plotted logarithmically to emphasize the
region of FF close to one; notice that the NS--BH figure in the bottom
right corner shows a different bin range).  The vertical lines show
the averages $\overline{\mathrm{FF}}$ and
$\overline{\mathrm{FF}}_\mathrm{eff}$ with their $1\sigma$ error bars
($\overline{\mathrm{FF}}_\mathrm{eff}$ is always the larger number).
\label{allFF}}
\end{center}
\end{figure*}

Figure \ref{allFF} shows the distribution of FFs, evaluated for our
DTFs and for the SPA standard templates against the 2PN target model,
within the Monte Carlo populations of BBHs and NS--BH binaries
described in the previous section. The vertical lines show the Monte
Carlo estimates of $\overline{\mathrm{FF}}$ and
$\overline{\mathrm{FF}}_\mathrm{eff}$ (the latter is always larger), with their estimated errors; these numbers are given also in
Tabs.\ \ref{BHBH} and \ref{NSBH}.
We wish to discuss several features of the FFs.
\begin{table*}
\begin{tabular}{l|cc|cc|cc|cc|cc}
\hline
& \multicolumn{10}{c}{Fitting factors against 2PN target model} \\
 & \multicolumn{2}{c}{$(7+5)M_\odot$} & \multicolumn{2}{c}{$(10+10)M_\odot$} & \multicolumn{2}{c}{$(15+15)M_\odot$} & \multicolumn{2}{c}{$(20+5)M_\odot$} & \multicolumn{2}{c}{$(20+10)M_\odot$}\\ 
 & \multicolumn{1}{c}{$\overline{\mathrm{FF}}$} & \multicolumn{1}{c}{$\overline{\mathrm{FF}}_\mathrm{eff}$} & \multicolumn{1}{c}{$\overline{\mathrm{FF}}$} & \multicolumn{1}{c}{$\overline{\mathrm{FF}}_\mathrm{eff}$} & \multicolumn{1}{c}{$\overline{\mathrm{FF}}$} & \multicolumn{1}{c}{$\overline{\mathrm{FF}}_\mathrm{eff}$} & \multicolumn{1}{c}{$\overline{\mathrm{FF}}$} & \multicolumn{1}{c}{$\overline{\mathrm{FF}}_\mathrm{eff}$} & \multicolumn{1}{c}{$\overline{\mathrm{FF}}$} & \multicolumn{1}{c}{$\overline{\mathrm{FF}}_\mathrm{eff}$} \\ \hline \hline 
SPAs & $  0.9030(24)$ & $  0.9390(15)$  & $  0.8944(21)$ & $  0.9198(12)$  & $  0.8105(25)$ & $  0.8282(16)$  & $  0.8576(25)$ & $  0.8844(22)$  & $  0.8264(27)$ & $  0.8494(18)$  \\ 
SPAc & $  0.9018(23)$ & $  0.9367(18)$  & $  0.9294(20)$ & $  0.9558(12)$  & $  0.9313(18)$ & $  0.9548(10)$  & $  0.8854(23)$ & $  0.9096(21)$  & $  0.9186(20)$ & $  0.9461(12)$  \\ 
$(\psi_0 \psi_{3/2})_2$ & $  0.9262(22)$ & $  0.9595(13)$  & $  0.9423(17)$ & $  0.9657(10)$  & $  0.9414(15)$ & $  0.9620(08)$  & $  0.8921(22)$ & $  0.9178(23)$  & $  0.9270(17)$ & $  0.9529(12)$  \\ 
$(\psi_0 \psi_{3/2} \alpha)_4$ & $  0.9288(22)$ & $  0.9617(13)$  & $  0.9480(16)$ & $  0.9703(10)$  & $  0.9551(14)$ & $  0.9726(08)$  & $  0.8986(21)$ & $  0.9212(23)$  & $  0.9421(16)$ & $  0.9625(12)$  \\ 
$(\psi_0 \psi_{3/2} \mathcal{B})_6$ & $  0.9753(07)$ & $  0.9828(05)$  & $  0.9861(03)$ & $  0.9895(02)$  & $  0.9863(03)$ & $  0.9891(02)$  & $  0.9746(05)$ & $  0.9794(05)$  & $  0.9843(03)$ & $  0.9884(03)$ 
\\ \hline
& \multicolumn{10}{c}{Fitting factors against 3.5PN target model} \\
 & \multicolumn{2}{c}{$(7+5)M_\odot$} & \multicolumn{2}{c}{$(10+10)M_\odot$} & \multicolumn{2}{c}{$(15+15)M_\odot$} & \multicolumn{2}{c}{$(20+5)M_\odot$} & \multicolumn{2}{c}{$(20+10)M_\odot$}\\ 
 & \multicolumn{1}{c}{$\overline{\mathrm{FF}}$} & \multicolumn{1}{c}{$\overline{\mathrm{FF}}_\mathrm{eff}$} & \multicolumn{1}{c}{$\overline{\mathrm{FF}}$} & \multicolumn{1}{c}{$\overline{\mathrm{FF}}_\mathrm{eff}$} & \multicolumn{1}{c}{$\overline{\mathrm{FF}}$} & \multicolumn{1}{c}{$\overline{\mathrm{FF}}_\mathrm{eff}$} & \multicolumn{1}{c}{$\overline{\mathrm{FF}}$} & \multicolumn{1}{c}{$\overline{\mathrm{FF}}_\mathrm{eff}$} & \multicolumn{1}{c}{$\overline{\mathrm{FF}}$} & \multicolumn{1}{c}{$\overline{\mathrm{FF}}_\mathrm{eff}$} \\ \hline \hline 
$(\psi_0 \psi_{3/2} \mathcal{B})_6$ & $  0.9708(08)$ & $  0.9802(06)$  & $  0.9854(03)$ & $  0.9887(02)$  & $  0.9854(03)$ & $  0.9883(03)$  & $  0.9738(06)$ & $  0.9775(05)$  & $  0.9844(03)$ & $  0.9882(02)$ 
\\ \hline
\end{tabular}
\caption{Averages $\overline{\mathrm{FF}}$ and
$\overline{\mathrm{FF}}_\mathrm{eff}$ of the fitting factor FF against
the 2PN and 3.5PN target models, for the DTFs and for the standard
SPA template families, as computed on our BBH Monte Carlo
populations. The numbers in parentheses give the estimated Monte Carlo
errors on the last two digits of $\overline{\mathrm{FF}}$ and
$\overline{\mathrm{FF}}_\mathrm{eff}$.\label{BHBH}}
\end{table*}
\begin{table}
\begin{tabular}{l|cc}
\hline
& \multicolumn{2}{c}{2PN target model} \\
 & \multicolumn{2}{c}{$(10+1.4)M_\odot$}\\ 
 & \multicolumn{1}{c}{$\overline{\mathrm{FF}}$} & \multicolumn{1}{c}{$\overline{\mathrm{FF}}_\mathrm{eff}$} \\ \hline \hline 
SPAs & $  0.7800(34)$ & $  0.8169(37)$  \\ 
SPAc & $  0.7747(49)$ & $  0.8129(54)$  \\ 
$(\psi_0 \psi_{3/2})_2$ & $  0.7807(41)$ & $  0.8316(46)$  \\ 
$(\psi_0 \psi_{3/2} \mathcal{B})_6$ & $  0.9331(15)$ & $  0.9452(14)$ 
\\ \hline
& \multicolumn{2}{c}{3.5PN target model} \\
\hline \hline
$(\psi_0 \psi_{3/2} \mathcal{B})_6$ & $  0.9263(15)$ & $  0.9378(14)$
\\ \hline
\end{tabular}
\caption{Averages $\overline{\mathrm{FF}}$ and
$\overline{\mathrm{FF}}_\mathrm{eff}$ of the fitting factor FF against
the 2PN and 3.5PN target models, for the DTFs and for the standard
SPA template families, as computed on the $(10+1.4)M_\odot$ NS--BH
Monte Carlo populations. The numbers in parentheses give the estimated
Monte Carlo errors on the last two digits of $\overline{\mathrm{FF}}$
and $\overline{\mathrm{FF}}_\mathrm{eff}$.\label{NSBH}}
\end{table}
\begin{enumerate}
\item The SPA template families (solid and long-dashed black lines)
always give the worst performance. Except for the lighter systems,
$(7+5)M_\odot$ BBHs and $(10+1.4)M_\odot$ NS--BH binaries
\cite{note44}, the SPAs family (solid black line) is
consistently less effectual than SPAc, because the target-model ending
frequencies are usually different from the Schwarzchild-ISCO
frequencies used to terminate the SPAs templates (in the majority of
cases, they are higher).  The improvement [SPAs to SPAc] in
$\overline{\mathrm{FF}}$ is $\simeq$ 3\% for $M \simeq 20\mbox{--}25
M_\odot$, and $\gtrsim$ 10\% for $M = 30 M_\odot$.  As pointed out in
BCV1, it is important to add the frequency-cut parameter
$f_\mathrm{cut}$ whenever the ending frequency is not known very well,
but it is expected to fall within the band of good interferometer
sensitivity.
\item Although the $(\psi_0\,\psi_{3/2})_2$ DTF (short-dashed green
lines) is essentially a reparametrization of SPAc (both families have
the $f_\mathrm{cut}$ parameter), it is slightly more effectual. The
reason for this is that the \emph{physical} ranges of $M$ and $\eta$
used to optimize FF (and in particular the constraint $\eta < 0.25$)
limit the ability of the expression $\psi_\mathrm{SPA}(f)$ to
reproduce the phasing of the target. On the contrary, in the
$(\psi_0\,\psi_{3/2})_2$ DTF the coefficients of $f^{-5/3}$ and
$f^{-2/3}$ are not functions of $M$ and $\eta$, but free
phenomenological parameters that can achieve the best possible values
to match the target phasing.  This added freedom does not buy a
dramatic improvement for the spinning binaries studied in this paper,
because the SPAc templates are already rather close to the adiabatic
target model (except of course for precessional modulations). On the
contrary, in BCV1 we saw that using unconstrained phenomenological
parameters with extended ranges is very important to follow the
nonadiabatic dynamics of late inspiral, as predicted by some PN models
for nonspinning binaries.
\item The $(\psi_0\,\psi_{3/2}\,\alpha)_4$ DTF (dot-dashed blue lines)
introduces the amplitude-remodeling coefficient $\alpha$. In BCV1 we
found that $\alpha$ (together with the extension of parameter ranges)
helped follow the nonadiabatic dynamics of some target PN models [see
Tab.\ \ref{tab:spanotgood}]. In this paper, however, the only target
model is obtained in the adiabatic limit, so the frequency-domain
amplitude (except of course for the modulations due to precession) is
always very close to the Newtonian expression $f^{-7/6}$. As a result,
the improvement [$(\psi_0\,\psi_{3/2})_2$ to
$(\psi_0\,\psi_{3/2}\,\alpha)_4$] in $\overline{\mathrm{FF}}$ is only
$\simeq 0.3\mbox{--}1.6$\%, while (at least according to the simple
Gaussian analysis of Sec.\ \ref{sec5.3}) the detection threshold
increases by $\simeq$ 4\% (although this number does not 
take into account the $\phi^\alpha = 0$ constraint). It seems therefore that the $(\psi_0\,\psi_{3/2}\,\alpha)_4$ DTF is not a useful upgrade of $(\psi_0\,\psi_{3/2})_2$ for the purpose of detecting the signals
emitted by precessing binaries.
\item The $(\psi_0\,\psi_{3/2}\,\mathcal{B})_6$ DTF [lighter--red solid
lines] includes modulational corrections for both amplitude and
phase. The resulting improvement in $\overline{\mathrm{FF}}$ over the
SPA families is remarkable (for BBHs, 8--22\% over SPAs, and 6--10\%
over SPAc; for NS--BH binaries, 20\% over both).  However, the effect
of the modulational terms is seen best by comparing
$(\psi_0\,\psi_{3/2}\,\mathcal{B})_6$ to $(\psi_0\,\psi_{3/2})_2$: we
get an improvement of 5--9\% for BBHs, and 20\% for NS--BH
binaries. This numbers should be compared with the projected increase
$\simeq$ 8\% in the detection threshold (Sec.\ \ref{sec5.3}).
\item For the $(\psi_0\,\psi_{3/2}\,\mathcal{B}')_6$ DTF, where the
frequency dependence of the modulating terms is $f^{-2/3}$ rather than
$f^{-1}$, fitting factors are not significantly different from
$(\psi_0\,\psi_{3/2}\,\mathcal{B})_6$. Therefore we do not show these
numbers. Tables \ref{BHBH} and \ref{NSBH} also contain a few FFs
computed against the 3.5PN order target model (with
$\widehat{\theta}=0$). The FFs, shown for the
$(\psi_0\psi_{3/2}\mathcal{B})_6$ DTF, are essentially in line with
their 2PN counterparts.
\end{enumerate}
\begin{table*}
\begin{tabular}{r|cc|cc|cc|cc|cc|cc}
\hline & \multicolumn{12}{c}{FF against selected BCV1 PN models, for
the SPAc and $(\psi_0 \psi_{3/2} \alpha)$ template families} \\
 & \multicolumn{2}{c}{T(2,2)}& \multicolumn{2}{c}{T(3,3.5,$\hat{\theta}=2$)}&
 \multicolumn{2}{c}{P(2,2.5)}& \multicolumn{2}{c}{P(3,3.5,$\hat{\theta}=2$)}&
 \multicolumn{2}{c}{EP(2,2.5)}& \multicolumn{2}{c}{EP(3,3.5,$\hat{\theta}=2$)}\\ &
 \multicolumn{1}{c}{SPAc} & \multicolumn{1}{c}{$(\psi_0 \psi_{3/2}
 \alpha)_4$}& \multicolumn{1}{c}{SPAc} & \multicolumn{1}{c}{$(\psi_0
 \psi_{3/2} \alpha)_4$}& \multicolumn{1}{c}{SPAc} &
 \multicolumn{1}{c}{$(\psi_0 \psi_{3/2} \alpha)_4$}&
 \multicolumn{1}{c}{SPAc} & \multicolumn{1}{c}{$(\psi_0 \psi_{3/2}
 \alpha)_4$}& \multicolumn{1}{c}{SPAc} & \multicolumn{1}{c}{$(\psi_0
 \psi_{3/2} \alpha)_4$}& \multicolumn{1}{c}{SPAc} &
 \multicolumn{1}{c}{$(\psi_0 \psi_{3/2} \alpha)_4$}\\
\hline \hline 
(10+10)$M_\odot$ &   0.984 & 0.992 &   0.984 & 0.988 &   0.979 & 0.985 &   0.959 & 0.990 &   0.988 & 0.994 &   0.949 & 0.994 \\ 
(20+5)$M_\odot$ &   0.970 & 0.992 &   0.960 & 0.986 &   0.950 & 0.978 &   0.968 & 0.985 &   0.930 & 0.993 &   0.967 & 0.993 \\ 
(20+10)$M_\odot$ &   0.964 & 0.989 &   0.959 & 0.986 &   0.925 & 0.977 &   0.964 & 0.986 &   0.978 & 0.993 &   0.982 & 0.993 \\ 
(15+15)$M_\odot$ &   0.939 & 0.989 &   0.941 & 0.987 &   0.931 & 0.980 &   0.967 & 0.987 &   0.971 & 0.991 &   0.983 & 0.991 \\ 
\hline 

\end{tabular}
\caption{\label{tab:spanotgood}Fitting factors against selected PN
models of \emph{nonspinning} binaries (see BCV1), for the SPAc and
$(\psi_0\,\psi_{3/2}\,\alpha)_4$ template families. Notice that the
$(\psi_0\,\psi_{3/2}\,\alpha)_4$ DTF yields consistently higher FFs.
}
\end{table*}

Our results suggest two strategies to search for the signals from the
precessing BBHs examined in this paper. We can try to follow the
modulations induced by precession, using a DTF similar to
$(\psi_0\,\psi_{3/2}\,\mathcal{B})_6$; or we can just use
$(\psi_0\,\psi_{3/2})_2$, which is considerably better than SPAs
(mostly because of $f_\mathrm{cut}$), and slightly better than SPAc
(because of the extended parameter range).  The gain in FF when we
upgrade $(\psi_0\,\psi_{3/2})_2$ to
$(\psi_0\,\psi_{3/2}\,\mathcal{B})_6$ is offset by a similar increase
in the detection threshold, but the latter increase might be contained
by reducing the range of the allowed $\alpha_k$, or by other
data-analysis considerations that do come into the simple Gaussian
analysis of Sec.\ \ref{sec5.3}.

\begin{figure*}
\begin{center}
\includegraphics[width=6in]{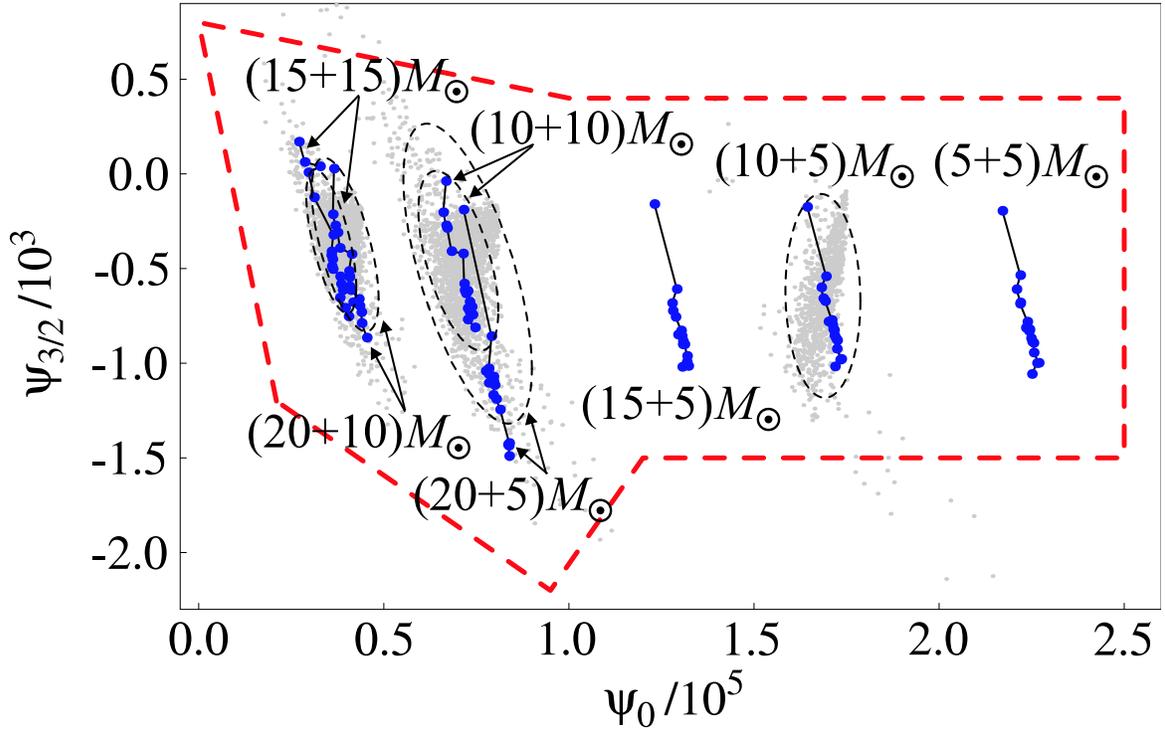}
\end{center}
\caption{\label{ad2maps} Projection of the 2PN target signals onto the
$(\psi_0\,\psi_{3/2})_2$ DTF.  For the $(10+10)M_\odot$,
$(15+15)M_\odot$, $(20+5)M_\odot$, $(7+5)M_\odot$, and
$(20+10)M_\odot$ BBHs in our Monte Carlo populations, the clusters of
gray dots show the projection of the 2PN target waveforms onto the
$(\psi_0,\psi_{3/2})$ parameter plane of the $(\psi_0\,\psi_{3/2})_2$
DTF (the projection of a given target signal is given by the values of
$\psi_0$ and $\psi_{3/2}$ that maximize the FF; here $f_\mathrm{cut}$
is not shown).
For each set of masses, we draw a dashed ellipse
centered on the parameter-space baricenter of the dots, and sized to
include 90\% of the dots (the proportions of the axes follow the
two-dimensional quadratic moments of the dots).
The larger blue dots, joined by the thin
lines (\emph{mass lines}), show the projections of the nonspinning PN
models studied in BCV1, for the same sets of masses plus
$(5+5)M_\odot$ and $(10+5)M_\odot$; each line joins signals with the
same binary masses, but obtained from different PN target models. As
we can see, for each set of masses, the projections of the
spinning-binary signals are clustered around the corresponding mass
line; moreover, all the projections fall within the region (delimited
by the thick dashed lines) suggested in BCV1 to match all the
nonspinning PN models.}
\end{figure*}

\begin{figure*}
\begin{center}
\includegraphics{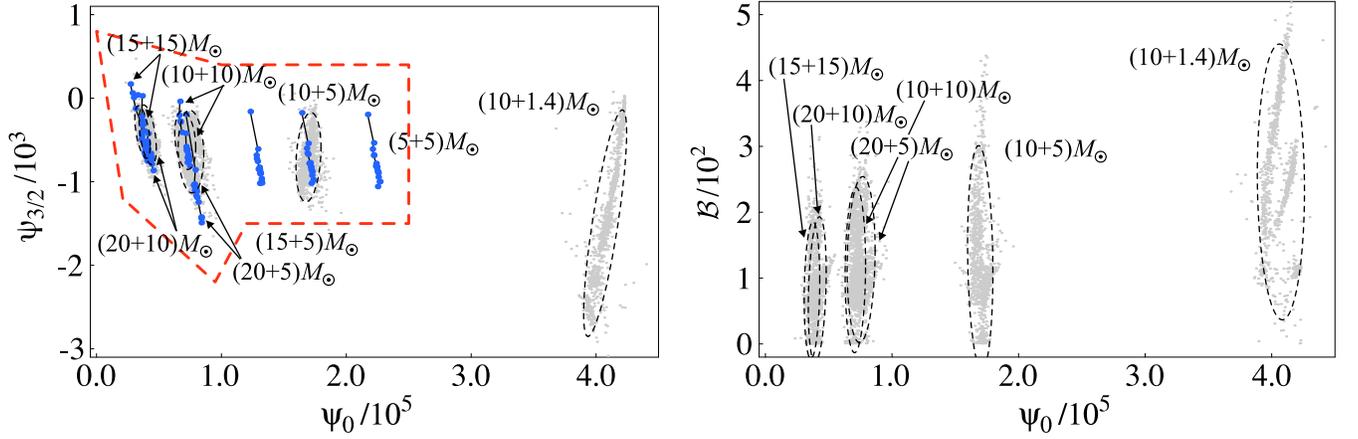}
\end{center}
\caption{\label{ad6maps} Projection of the 2PN target signals onto the
$(\psi_0\,\psi_{3/2}\,\mathcal{B})_6$ DTF.  For the $(10+10)M_\odot$,
$(15+15)M_\odot$, $(20+5)M_\odot$, $(7+5)M_\odot$, and
$(20+10)M_\odot$ BBHs, and for the $(10+1.4)M_\odot$ NS--BH binaries
in our Monte Carlo populations, the clusters of gray dots show the
projection of the 2PN target waveforms onto the $(\psi_0,\psi_{3/2})$
[on the left] and $(\psi_0,\mathcal{B})$ [on the right] parameter
plane of the $(\psi_0\,\psi_{3/2}\,\mathcal{B})_6$ DTF.  For each set
of masses, we draw a dashed ellipse centered on the parameter-space
baricenter of the dots, and sized to include 90\% of the dots (the
proportions of the axes follow the two-dimensional quadratic moments
of the dots).  The nonspinning-model mass lines and the boundary of
the suggested parameter ranges are shown as in Fig.\
\protect\ref{ad2maps}.}
\end{figure*}

Figure \ref{ad2maps} shows the projection of the 2PN target waveforms
onto the $(\psi_0,\psi_{3/2})$ section of the $(\psi_0 \psi_{3/2})_2$
parameter space; Fig.\ \ref{ad6maps} shows the projections of the
waveforms onto the $(\psi_0,\psi_{3/2})$ and $(\psi_0,\mathcal{B})$
sections of the $(\psi_0\,\psi_{3/2}\,\mathcal{B})_6$ parameter space.
It is interesting to notice that, with either strategy, the ranges of
$\psi_0$ and $\psi_{3/2}$ needed to match effectually the signals in
our populations are essentially the same found in BCV1 to match the
signals predicted by a variety of PN models for BBHs without spins.
In Figs.\ \ref{ad2maps} and \ref{ad6maps} these ranges are delimited
by the thick dashed lines; the thin \emph{mass lines} represent the
range of detection templates needed to match effectually the signals
predicted by different PN models for the same binary masses. As we can
see, the projections of the spinning-binary signals are smeared around
the nonspinning-binary mass lines with the same masses.

Thus, a signal search based on the $(\psi_0\,\psi_{3/2})_2$ DTF is a
good starting point for both nonspinning and spinning binaries.
It might also pay off, depending on the
results of a more realistic evaluation of false-alarm probabilities,
to upgrade this DTF to $(\psi_0\,\psi_{3/2}\,\alpha)_4$, with improved
performance for nonspinning but nonadiabatic BBHs, as shown in BCV1;
or even to $(\psi_0\,\psi_{3/2}\,\mathcal{B})_6$, with the best FFs
for spinning binaries and without any deterioration for nonspinning
ones.

\subsection{Modulated DTFs for NS--BH binaries}
\label{sec8.1}
\begin{figure}
\begin{center}
\includegraphics[width=\sizeonefig]{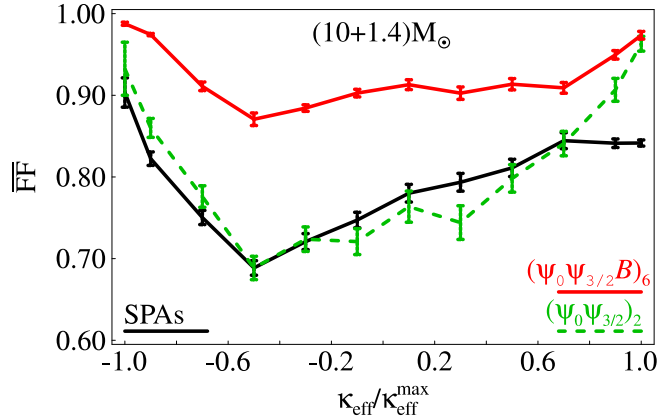}
\caption{Average fitting factor for the DTFs and for the SPAs template
families for $(10+1.4)M_\odot$ NS--BH binaries, plotted against the
initial $\kappa_\mathrm{eff} = \hL_N \cdot \mathbf{S}_\mathrm{eff}$.
The vertices of the segmented curves show the FF averaged on the sets
of samples that fall within the equispaced $\kappa_\mathrm{eff}$ bins
$[-1, -0.8)$, $[-0.8, -0.6)$, \ldots, $[0.8, 1]$. The error bars show
the sampling error on the bin averages. We plot also two additional
vertices, aligned with the abscissae $-1$ and $1$, which show the FF
averaged over the $\kappa_\mathrm{eff}$ bins $[-1, -0.98)$ and $[0.98,
1]$. \label{FFandK}}
\end{center}
\end{figure}

Let us now look in detail at the FFs achieved by the DTFs and
standard template families against the signals generated by
$(10+1.4)M_\odot$ NS--BH binaries where the BH is spinning rapidly
(see Tab.\ \ref{NSBH} and Figs.\ \ref{allFF} and \ref{FFandK}).  First
of all, we notice that there is little difference between the
performance of the SPAs and SPAc templates, because the ending
frequency lies outside of the band of good interferometer sensitivity.
Furthermore, the number of GW cycles within this band is very high, so
it is crucial that a DTF reproduce very accurately the evolution of
the GW phase; so using the $(\psi_0\,\psi_{3/2})_2$ DTF improves only
slightly on the performance of the SPA templates.  Introducing
precessional corrections brings about a dramatic change: for the
$(\psi_0\,\psi_{3/2}\,\mathcal{B})_6$ DTF, the increase in
$\overline{\rm FF}$ and $\overline{\rm FF}_\mathrm{eff}$ with respect
to SPA is respectively 20\% and 16\%, which is enough to justify the
introduction of six $\alpha_k$ coefficients, according the Gaussian
analysis of Sec.\ \ref{sec5.3}.

\begin{figure*}[t]
\begin{center}
\includegraphics[width=0.95\textwidth]{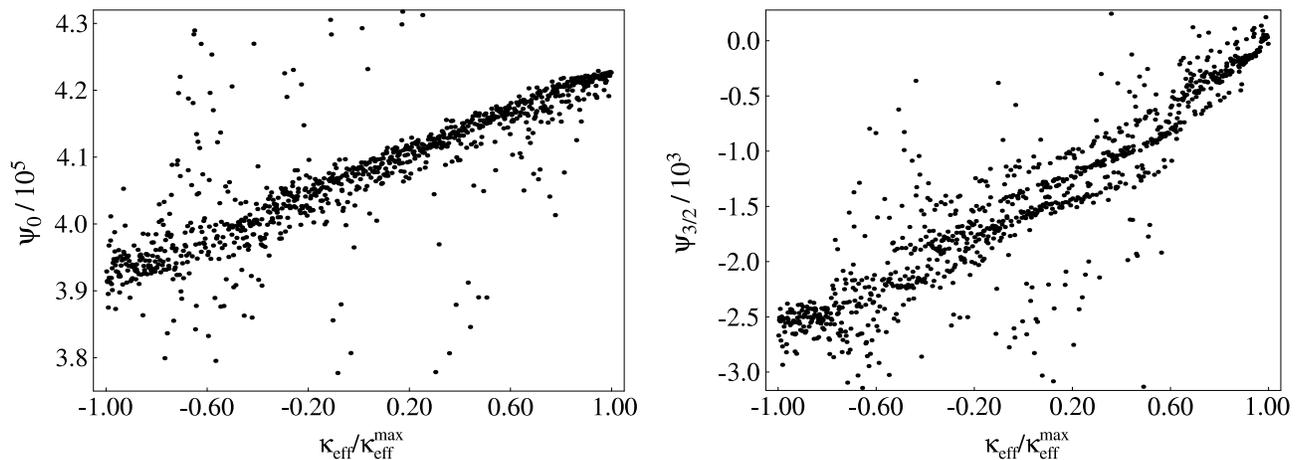}
\end{center}
\caption{\label{ADkappa} Projection of the $(10+1.4)M_\odot$ NS--BH
target signals (computed at 2PN order) onto the
$(\psi_0\,\psi_{3/2}\,\mathcal{B})_6$ DTF. The dots show the values of
the $\psi_0$ (left panel) and $\psi_{3/2}$ (right panel) target
parameters that yield maximum overlaps with the signals in the target
populations.}
\end{figure*}

\begin{figure}[t]
\begin{center}
\includegraphics[width=\sizeonefig]{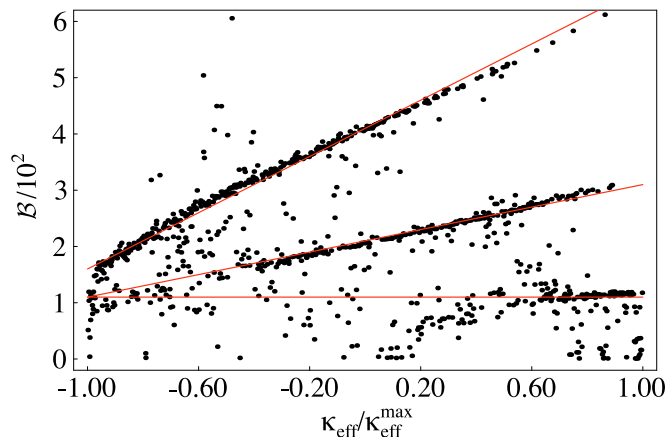}
\end{center}
\caption{\label{ADkappa2} Projection of the $(10+1.4)M_\odot$ NS--BH
target signals (computed at 2PN order) onto the
$(\psi_0\,\psi_{3/2}\,\mathcal{B})_6$ DTF. The dots show the values of
the $\mathcal{B}$ target parameter that yield maximum overlaps with
the signals in the target populations.}
\end{figure}

The dependence of the FF on the spin configuration is shown in
Fig.~\ref{FFandK}. For the NS--BH signals in our Monte Carlo
population, Figs.\ \ref{ADkappa} and \ref{ADkappa2} show the template
parameters $\psi_0$, $\psi_{3/2}$, and $\mathcal{B}$ that maximize the
overlap plotted against the initial $\kappa$ (conserved in NS--BH
binaries).  In the left panel, we see that the parameter $\psi_0$,
which is related to the Newtonian chirp mass, has only a weak
dependence on $\kappa$ (it varies by $\sim 8\%$); on the other hand,
the parameter $\psi_{3/2}$ has a strong dependence. A plausible
explanation is that the SO term in the SPA phasing is formally 1.5PN
[see Eqs.\ \eqref{psispa} and \eqref{SO}], and so is the term
$\psi_{3/2} f$ in $\psi_\mathrm{NM}(f)$, which takes on the job, as it
were, of reproducing the nonmodulational effects of the SO coupling.
In the right panel, we see that for most of the binary configurations
the values of $\mathcal{B}$ cluster around three lines
[$\mathcal{B}=100$, $\mathcal{B} = (1 + \kappa)\,110 + 110$, and
$\mathcal{B} = (1 + \kappa)\,240 + 160$].  Further analysis
are needed to provide an explanation for this interesting behavior.

Thus, the $(\psi_0\,\psi_{3/2}\,\mathcal{B})_6$ DTF is a good
candidate for the data-analysis problem of detecting GW signals from
NS--BH binaries with rapidly spinning BHs.  However, the analysis of
precessional dynamics and GW emission carried out in this paper
suggests an even more specialized DTF, which could be built with the
following guidelines.
\begin{enumerate}
\item The waveform can be computed directly from Eq.\ \eqref{easyh}
(obtained in the precessing convention): the necessary
ingredients are the time evolution of the orbital phase $\Psi$ and of
the binary polarization tensors $[\mathbf{e}_{+,\times}]_{ij}$, plus
the fixed detector polarization tensors
$[\mathbf{T}_{+,\times}]_{ij}$.
\item The evolution of $\Psi$
is obtained by solving Eq.\ \eqref{omegadot}, where $\vS_2$ can be set to zero, and $\vS_1$ enters
only in the \emph{conserved} term $\hL_N \cdot \vS_1$.  As a
consequence, Eq.\ \eqref{omegadot} is effectively uncoupled from the
evolution of $\hL_N$, Eq.\ \eqref{Lhdot}.
\item The evolution of the tensors $[\mathbf{e}_{+,\times}(t)]_{ij}$
is obtained from Eq.\ \eqref{eprecess}, after integrating Eqs.\
\eqref{S1dot} and \eqref{Lhdot} for the coupled evolution of $\hL_N$
and $\vS$, which depends only on $\hL_N \cdot \vS_1$, on $S_{1}$
(conserved), and on $\omega(t)$.
\item A source frame attached to the initial configuration of the
binary, similar to the frame constructed in Sec.~\ref{newconv}
[see Eqs.~(\ref{sourcedef})], can be used to carry out the explicit
construction. By way of the initial conditions (\ref{eijdef})--(\ref{initS2}), the tensors $\mathbf{e}_{+,\times}$ and the orbital phase $\Psi$ (up to an additive constant $\Psi_0$) are then well defined as functions of the basic and local binary parameters only. We have therefore completed the specification of the first part of Eq.\ \eqref{easyh}, which
expresses the components of the mass quadrupole moment.
\item The remaining part of Eq.\ \eqref{easyh}, which expresses
the projection on the polarization tensor of the detector,
\begin{equation}
P^{ij} \equiv [\mathbf{T}_{+}]^{ij} F_{+} + [\mathbf{T}_{\times}]^{ij}
F_{\times},
\end{equation}
is determined by the directional parameters $\Theta$,
$\varphi$, $\phi$, $\theta$, and $\psi$, which are now referred to the
source frame attached to the binary.  When we look for GWs using matched filtering, we can search rapidly over such a parametrization by treating the $P^{ij}$ as extrinsic parameters, along with the time of arrival and the initial orbital phase $\Psi_0$.  The only intrinsic parameters would then be $m_1$, $m_2$, $S_1$, and $\vS\cdot\hL_N$, all of which are conserved.
\end{enumerate}
This family of templates adds a further intrinsic parameter with respect to $(\psi_0\,\psi_{3/2}\,\mathcal{B})_6$, but it has the advantage of producing essentially exact waveforms (valid in the adiabatic regime, and up to the highest PN order included), and of expressing these waveforms directly in terms of the physical spin parameters $S_1$ and $\vS\cdot\hL_N$. We believe that the implementation and the false-alarm statistics of this family are worthy of further investigation \cite{BCV3}.

\section{Summary}
\label{sec9}

In BCV1, the nonmodulated DTFs $(\psi_0\psi_{3/2})_2$ and
$(\psi_0\psi_{3/2}\alpha)_4$ were shown to have FF $\gtrsim\, 0.95$
against several nonspinning-BBH target models, obtained under
different PN approximation schemes. In this paper, we have shown that
these two families are also rather effectual at matching the signals
from BH--BH and NS--BH precessing binaries with single-BH masses
between 5 and 20 $M_\odot$ and with maximal BH spins, at least if
these signals can be described by an adiabatic sequence of
quasicircular orbits up to 2PN order.

More specifically, for $(7+5)M_\odot$, $(10+10)M_\odot$,
$(20+10)M_\odot$, and $(15+15)M_\odot$ BBHs, we obtain
$\overline{\mathrm{FF}} \gtrsim 0.93$ and
$\overline{\mathrm{FF}}_\mathrm{eff} \gtrsim 0.95$. The improvement is
2--16\% over Schwarzschild-terminated SPAs templates, thanks largely
to the ending-frequency parameter $f_\mathrm{cut}$; and 1--2\% over
SPAc templates, thanks to the effective extension in the range of
parameters, released from their functional dependence on the masses of
the binary. Although the latter improvement seems negligible, we
should keep in mind that $(\psi_0 \, \psi_{3/2})_2$ DTFs are also
more suitable to match the nonspinning BH binaries studied in BCV1 with PN expanded and resummed models.
Results are worse for binaries that have smaller mass ratios $\eta$,
and therefore more GW cycles in the band of good interferometer
sensitivity. In this case the modulational effects due to precession
become important, and must be included in the detection
templates. Indeed, for $(20+5)M_\odot$ BBHs, the
$(\psi_0\psi_{3/2})_2$ and $(\psi_0\psi_{3/2}\alpha)_4$ DTFs have
$\overline{\mathrm{FF}} \simeq 0.89$ and
$\overline{\mathrm{FF}}_\mathrm{eff} \simeq 0.92$; for a
$(10+1.4)M_\odot$ NS--BH binary, we find $\overline{\mathrm{FF}}
\simeq 0.78$, and $\overline{\mathrm{FF}}_\mathrm{eff} \simeq 0.83$.

Motivated by these shortcomings, we have investigated in detail the
dynamics of precession in these binaries, and we have introduced a new
convention to write the GW signal (as computed in the quadrupole
approximation) as a function of binary and detector parameters,
isolating the oscillatory effects of precession in the evolution of
the polarization tensors $[\mathbf{e}_{+,\times}]_{ij}$. As a result,
the detector response to GWs can be written as the product of a
carrier signal, which very closely resembles the nonspinning signals
studied in BCV1, and a modulational correction, which can be handled
using an extension of Apostolatos' ansatz \eqref{ansatzone}. On the
basis of these observations, we build the modulated DTF
$(\psi_0,\psi_{3/2}\mathcal{B})_6$, which yields
$\overline{\mathrm{FF}}$ and $\overline{\mathrm{FF}}_\mathrm{eff}
\simeq$ 0.98--0.99 for the BBHs investigated, and
$\overline{\mathrm{FF}} \simeq 0.93$,
$\overline{\mathrm{FF}}_\mathrm{eff} \simeq 0.95$ for
$(10+1.4)M_\odot$ NS--BH binaries.  This DTF has the advantage that
all the modulational parameters (except for $\mathcal{B}$) can be
treated as extrinsic parameters, reducing considerably the
computational cost of signal searches.
According to the simple analysis of Sec.\ \ref{sec5.3},
the detection thresholds for this DTF should be set higher than 
those for simpler families; still, the gain
in FF is still somewhat larger than the increase in the threshold, and
more realistic analyses of false-alarm statistics might provide a way
to sidestep this difficulty.
The same arguments that lead to the $(\psi_0,\psi_{3/2}\mathcal{B})_6$
DTF suggest a new, very promising class of templates for NS--BH
binaries, which we discuss briefly in Sec.~\ref{sec8.1}, and which we
plan to investigate more thoroughly elsewhere~\cite{BCV3}.

We wish to make a few final remarks. First, in this paper we limited
our analysis to compact objects moving on quasicircular orbits;
from the results on the ending frequencies (see Fig.~\ref{finalf}) we see
that there exist spin initial conditions for which the ending frequencies
(end of inspiral) are in the LIGO--VIRGO band. So, in these cases
we should use spinning dynamics that goes beyond the adiabatic approximation.
This dynamics (without radiation-reaction effects) is already available
in the EOB framework~\cite{BD,EOB3PN} thanks to the work of Damour~\cite{TD}.
We plan to investigate the effects of nonadiabatic PN dynamics in the
near future.

Second, a few years ago Levin pointed out \cite{JL} that spin-spin
effects can introduce chaos into the trajectories; as a consequence,
the gravitational waveforms would come to depend sensitively on the
initial conditions. More studies followed \cite{SR,CL}. Considering
only conservative dynamics (no RR), Cornish and Levin~\cite{CL} found
some examples of rather eccentric ($e \sim 0.6$ or $0.9$) chaotic
orbits, and a few quasicircular chaotic orbits.
However, these authors observed that chaos would
be damped by RR effects, and that it would not affect the inspiral
waveforms, except (perhaps) at the very end (the plunge). Still, at
this time the dynamical structure of phase space has not been explored
systematically, and a more conclusive study tuned to the LIGO--VIRGO
detection problem remains desirable.  The analysis of this paper
assumes that, by the time the GW signal enters the band of good
detector sensitivity, RR effects have circularized the orbit, and have
brought the binary into the adiabatic regime, which is valid until the
MECO. We did not try to perturb the initial conditions slightly and to
investigate the resulting changes in the orbital evolution and in the
waveforms.

Third, we have evaluated the performance of our DTFs by averaging over
\emph{uniform} distributions of the initial spin angles. Of course it
would be preferable to assume more realistic, nonuniform distributions
derived from astrophysical considerations.  Some results for spin
distributions in BBHs (with only one spinning BH), and in NS--BH
binaries were obtained by Kalogera using population-synthesis
techniques \cite{VK}. In particular, Kalogera found that
$30\mbox{--}80\%$ of the NS--BH binaries that will coalesce within a
Hubble time can have a tilt angle (the angle between the spin and the
orbital angular momentum) larger than $30$ degrees.  These results
assume that the spinning BH in the binary forms first, and that its
spin is aligned with the orbital angular momentum; the tilt angle
originates from the supernova explosion that forms the NS.
For the case of the binaries
formed in globular clusters, there is no theoretical argument to
suggest any particular spin distribution.

Finally, recent analyses of spin-spin effects in the PN inspiral
equations \cite{JS} suggest that, for comparable-mass BBHs, by
the time the GW signal enters the band of good interferometer
sensitivity the two BH spins may have become roughly \emph{locked}
into a fixed relative configuration. If these results
are confirmed, they could provide preferred initial spin conditions,
and simplify the data-analysis problem for comparable mass binaries,
by reducing the variability of expected GW signals.

\acknowledgments
A special thank to David Chernoff for having started
the initial discussions and investigations that have led to this
project.  We thank Jolien Creighton, Thibault Damour, Philippe
Grandcl\'ement, Vicky Kalogera, Yi Pan, Sterl Phinney, Bangalore
Sathyaprakash, and Kip Thorne for interesting discussions.
We thank Thibault Damour for useful comments on this manuscript
and Cliff Will for clarifications on the material discussed in App.\
\ref{appendixA}. We acknowledge support from NSF grant PHY-0099568 and
NASA grant NAG5-10707. Separate acknowledgments for the three authors
follow. [A.\ B.] This research was also supported by Caltech's Richard
Chace Tolman Fund. [Y.\ C.] This research was supported by the David
and Barbara Groce Fund of the San Diego Foundation; the author thanks
the gravitational-wave group at the Australian National University for
their hospitality in October and November of 2002. [M.\ V.] Part of
this research was performed at the Jet Propulsion Laboratory,
California Institute of Technology, under contract with the National
Aeronautics and Space Administration; the author thanks the INFN group
at the University of Parma, Italy, for their hospitality during summer
2002.

\appendix
\section{Validity of the adiabatic sequence of spherical orbits}
\label{appendixA}

In the target model defined in Sec.\ \ref{sec2.1}, the inspiral of the
two compact bodies is described as an adiabatic sequence of spherical
orbits.  In this Appendix we wish to discuss the validity of this
assumption.  Introducing the orthonormal basis $(\hl,\hn,\hL_N)$,
where $\hn = \vx/r$, $\hL_N = \vL_N/L_N$, $\hl = \hL_N \times \hn$ and
$\vL_N = \mu \, \vx \times \vv$ (with $\mu$ the reduced mass), it is
straightforward to write the equations of motion as [see Eqs.~(4.1) of
Ref.~\cite{K}; we use the relations $\vv = \dot{r}\,\hn +
r\,\omega\,\hl, v^2 = \dot{r}^2 + r^2\,\omega^2$]:
\begin{eqnarray}
\label{2.1}
\hn \cdot \va &=& \ddot{r} - r\,\omega^2 \,, \\
\label{2.2}
\hl \cdot \va &=& r\,\dot{\omega} + 2 \dot{r}\,\omega\,, \\
\label{2.3}
\hL_N \cdot \va &=& - r\,\omega\,\frac{d \hL_N}{dt} \cdot \hl \,,
\end{eqnarray}
where $\va$ is the acceleration in harmonic gauge given by Eqs.~(2.2a), (2.2c) of Ref.~\cite{K}. If we impose $\dot{r} =0 = \ddot{r}$,
Eq.~(\ref{2.2}) then implies $\dot{\omega}=0$; and
from Eq.~(\ref{2.1}) we get
\beq
\label{2.4}
r^2\,\omega^2 = \frac{1}{r}\,\left (1 - \frac{2}{r^2}\,\vL_N \cdot \vS_{\omega}\right ) \,,
\quad \quad \vS_{\omega} \equiv  \left ( 1 + \frac{3}{2}\,\frac{m_2}{m_1} \right )\,\vS_1 +
\left ( 1 + \frac{3}{2}\,\frac{m_1}{m_2} \right )\,\vS_2\,,
\eeq
where for simplicity we have set $M=1$.
Although spherical orbits (orbits where both $r$ and $\omega$ remain constant) exist at any given instant, they are not preserved along dynamical evolution because the quantity $\vL_N \cdot \vS_{\omega}$ that appears in Eq.~(\ref{2.4}) is not conserved.
Indeed, averaging over an orbit \cite{note38} (and, for simplicity, neglecting spin-spin effects), we get
\beq
\label{2.5}
\left\langle \frac{d \vL_N}{dt} \right\rangle = \frac{2\mu}{r^3} \vS_{\rm eff} \times \vL_N \,,
\quad \quad \vS_{\rm eff} \equiv \left ( 1 + \frac{3}{4}\,\frac{m_2}{m_1} \right )\,\vS_1 +
\left ( 1 + \frac{3}{4}\,\frac{m_1}{m_2} \right )\,\vS_2\,,
\eeq
where $\langle A \rangle$ denotes the quantity $A$ when the spin-orbit (and spin-spin) terms have been averaged over an orbit. Using the precession equations for the spins we derive
\beq
\label{2.7}
\left\langle \frac{d (\vL_N \cdot \vS_{\omega})}{dt} \right\rangle = -3 \frac{(m_1^2 -m_2^2)}{m_1\,m_2}\,
\vL_N \cdot (\vS_1 \times \vS_2)\, \frac{1}{r^3}\,.
\eeq
Hence, because the circular-orbit condition is not preserved
during the evolution, either $\langle \dot{\omega} \rangle \neq
0$ or $\langle \dot{r} \rangle \neq 0$ (or both).

Let us now see how Eq.\ \eqref{omegadot} for
$\dot{\omega}$ changes if effects of this kind are included.
The usual argument \cite{K,KWW} used to obtain the adiabatic evolution of $\omega$ rests on the energy-balance equation,
\beq
\label{balance}
\dot{E}_{\rm RR}=\frac{d}{dt}E(\omega,\hL_N,\vS_1,\vS_2)=\frac{\partial
E}{\partial \omega}\dot{\omega}+
\left(
\frac{\partial
E}{\partial \hL_N}\cdot\dot{\hL}_N+\frac{\partial
E}{\partial \vS_1}\cdot\dot{\vS_1}+\frac{\partial
E}{\partial \vS_2}\cdot\dot{\vS_2}\right)\,,
\eeq
where
\bea
\label{E2PN}
E(\omega,\hL_N,\vS_1,\vS_2) 
= -\frac{\mu}{2}\,(M\omega)^{2/3}
\,\Bigg\{
&1& - \frac{(9+\eta)}{12}\,(M\omega)^{2/3} + \frac{8}{3M^2} \hL_N\cdot
\vS_{\rm eff}\,(M\omega) + \\
&& \bigg[\frac{1}{24}(-81 + 57 \eta - \eta^2)+\frac{1}{\eta M^4}
\left[(\vS_1\cdot\vS_2)-3(\hL_N\cdot\vS_1)(\hL_N\cdot\vS_2)\right]\bigg]
(M\omega)^{4/3} \Bigg\} \nonumber
\eea
is the orbital energy evaluated at Newtonian order, but
including spin-orbit and spin-spin effects, and
where $\dot{E}_{\rm RR}$
is the RR energy loss~\cite{KWW,K}. From Eqs.~(\ref{E2PN}), (\ref{Lhdot}),
(\ref{S1dot}) and (\ref{S2dot}), we notice that the sum of the last three
terms in parentheses in Eq.~(\ref{balance})
does not vanish: at leading order, its value is
\beq
\dot{E}_{\rm extra}
=\frac{1}{4}\frac{(m_1-m_2)}{M}\,\eta^2\,\chi_1\,\chi_2\,(M\omega)^{11/3}
\,\left[\left(\hS_1\times\hS_2\right)\cdot\hL_N\right]\,.
\label{extra}
\eeq
This expression is zero if masses are equal, or if spins are either aligned or antialigned.
Retaining the term (\ref{extra}) in the calculation yields
an additional contribution in the evolution of $\omega$,
with a leading order correction
\beq
\label{omegadotextra}
\frac{\dot{\omega}_{\rm
extra}}{\omega^2}=\frac{3}{4}\frac{(m_1-m_2)}{M}\,\eta\,\chi_1\,\chi_2\,(M\omega)^2
\,\left[\left(\hS_1\times\hS_2\right)\cdot\hL_N\right]\,.
\eeq
Thus, compared with the other terms in Eq.~(\ref{omegadot}),
$\dot{\omega}_{\rm extra}$ appears formally at 0.5 PN order (very
low!) in the expansion of $\dot{\omega}$.  Note that the spin-orbit
term in the energy (\ref{E2PN}), combined with the leading-order
precessions, does not produce such a term; this makes the adiabatic
approach fully consistent up to 1.5PN order.  In fact, $\dot{E}_{\rm
extra}$ originates from taking the derivative of $\dot{E}_{\rm SO}$
and using next-to-leading order terms in the precession equations,
and the derivative $\dot{E}_{\rm SS}$ while using
the leading-order terms in the precession equations.

However, the effect of this term in the regime that we consider is not
as large as suggested by its formal PN order. For example, under the worst possible assumption (that the geometric factor
$[(\hS_1 \times \hS_2 )\cdot\hL_N]$ has always
the maximum value of one, and that spins are maximal), we
get the correction
\beq
\label{naivecycle}
\frac{\Delta \Psi_{\rm extra}}{2\pi} =
\frac{1}{2\pi}\,\frac{25}{16384}\,\frac{\sqrt{1-4\eta}}{\eta}
\,\left[(M\omega_f)^{-4/3}-(M\omega_i)^{-4/3}\right]
\eeq
to the number of orbital cycles, where $\omega_i$ and $\omega_f$ are
the initial and final orbital frequencies under consideration.  This
is formally a 0.5PN correction, as can be seen by comparing it with
Eq.~(4.16) of Ref.~\cite{K}. Nevertheless, for (say) a $(20+5)M_\odot$
BBH, this correction will be at most $0.34$ orbital cycles from
$\omega_i = \pi\times 30$\,Hz to $\omega_i = \pi\times 400$\,Hz,
to be compared with a baseline of $52$ orbital cycles from
the Newtonian term and $8$ from the 1PN term.  For a $(10+1.4)M_\odot$
binary, the correction will be $1.6$ orbital cycles, to be compared
with $175$ orbital cycles from the Newtonian term and $30$ from the
1PN term.  The correction is small because, although the PN order is
formally low, the numerical coefficient of the geometric factor
$[(\hS_1\times\hS_2 )\cdot\hL_N ]$ is very small.

So far, we have assumed $[(\hS_1\times\hS_2 )\cdot\hL_N ] \sim 1$
along the evolution.  Let us now estimate the more important effect
that comes from the precession of $\hL_N$, $\vS_1$ and $\vS_2$, which
is especially important for binaries with small mass ratios, which
have longer RR time scales and more precessional cycles.  At the
leading order (with $M=1$)
\beq
\frac{d}{dt}\left[\left(\hS_1\times\hS_2\right)\cdot\hL_N\right] = \frac{3}{2}
(m_1-m_2)\,\omega^{5/3}\,\left[\hS_1\cdot\hS_2-(\hS_1\cdot\hL_N)\,(\hS_2\cdot\hL_N)\right]
+ {\cal O}(\omega^2)\,,
\eeq
and
\beq
\frac{d}{dt}
\left[\hS_1\cdot\hS_2-(\hS_1\cdot\hL_N)(\hS_2\cdot\hL_N)\right]=
-\frac{3}{2}(m_1 - m_2)\,\omega^{5/3}
\left[\left(\hS_1\times\hS_2\right)\cdot\hL_N\right] + {\cal O}(\omega^2) \,.
\eeq
Combining the above equations, we get (at leading order)
\beq
\frac{d^2}{dt^2}\left[\left(\hS_1\times\hS_2\right)\cdot\hL_N\right] \simeq
- \frac{9}{4}(m_1-m_2)^2\,\omega^{10/3}\,\left[\left(\hS_1\times\hS_2\right)\cdot\hL_N\right]\,.
\eeq
This means that the geometric factor $[(\hS_1\times\hS_2)\cdot\hL_N]$
oscillates around zero with a time scale $\sim \omega^{-5/3}$. Thus
the effect of $\dot{\omega}_{\rm extra}$ accumulates only within this
timescale, which is 1.5 PN orders shorter than the RR timescale.
Therefore, we expect that the real $\Delta \Psi_{\rm extra}$ will be
even smaller than the formal prediction given by
Eq.~(\ref{naivecycle}), and that it will contribute effectively at 2PN
order.  As a check, we evaluated the FF between the gravitational
waveforms obtained, for a $(10+1.4)M_\odot$ binary, by first including and
then dropping the extra term in $\dot{\omega}$. We found
that the FF is $\simeq 0.99$. On the basis of this last check and of
the analysis outlined above, we conclude that the adiabatic assumption
is quite adequate for the purposes of this paper.

\section{Proof that the precessing convention yields $\omega = \dot{\Phi}_S$}
\label{appy}

First of all, it is easy to confirm that, as long as
$\mathbf{e}_{1,2}(0)$ and $\hL_{N}(0)$ form an orthonormal basis at
some initial time, the evolution equation $\dot{\mathbf{e}}_{1,2} =
\mathbf{\Omega}_e \times \mathbf{e}_{1,2}$ will always keep the
triplet an orthonormal basis. It is then always possible to have a
$\Phi(t)$, such that
\beq
\hn(t)= \mathbf{e}_{1} \cos\Phi(t)+
\mathbf{e}_{2}\sin\Phi(t)\,,\quad
\hl(t)= -\mathbf{e}_{1} \sin\Phi(t)+ \mathbf{e}_{2}\cos\Phi(t)\,.
\eeq
Taking the time derivative of $\hn(t)$, we have
\beq
\dot{\hn} =  \dot{\Phi} \hl + \mathbf{\Omega}_e \times \hn, \\
\eeq
Now, the adiabatic condition for a sequence of circular orbits states
that $\dot{\hn} = \omega \hl$, so we have
\begin{equation}
\label{eq:desired}
\dot{\hn} = \omega \hl = \dot{\Phi} \hl + \mathbf{\Omega}_e \times \hn.
\end{equation}
By definition [Eq.\ \eqref{Omegae}], $\mathbf{\Omega}_e$ has no
components along $\mathbf{e}_3 \equiv \hL_N$. It also has no
components along $\hl$, because
\begin{eqnarray}
\mathbf{\Omega}_e \times \hL_N &=& \dot{\hL}_N = \dot{\hn} \times \hl +
\hn \times \dot{\hl} = \omega \hl \times \hl + \hn \times (-
\dot{\Phi} \hn + \mathbf{\Omega}_e \times \hl) \\
&=& \mathbf{\Omega}_e (\hn \cdot \hl) - \hl (\hn \cdot
\mathbf{\Omega}_e) \propto \hl,
\end{eqnarray}
where in the last step we used $\hn \cdot \hl = 0$ and the
vector--triple-product rule. It follows that $\mathbf{\Omega}_e$ lies
along $\hn$, and therefore $\mathbf{\Omega}_e \times \hn =
0$. Equation \eqref{eq:desired} then gives the desired result, $
\dot{\Phi}=\omega$, i.e.\ $\Phi(t)=\Psi(t)+\mathrm{const}$.

\renewcommand{\prd}{\emph{Phys. Rev. D}}


\begin{thebibliography}{999}
\bibitem{note37} Note that the MECO, as defined by Eq.~(\ref{MECO}) via a
total derivative, depends also on the evolution of $\vL$, $\vS_1$ and
$\vS_2$. However, later in this paper (Sec.~\ref{sec3.1}) we shall see
that this dependence is rather weak.
\bibitem{note39} When referring to results at 3PN and 3.5PN
orders we shall always set the arbitrary parameter
$\widehat{\theta}=0$ in Eq.~(\ref{omegadot}). In the nonspinning case,
the dependence of the GW signal on $\widehat{\theta}$ is rather mild,
at least if $\widehat{\theta}$ is limited to values of order
one~\cite{BFIJ}, as investigated in BCV1.
\bibitem{note40} Here
``quadrupole'' refers to the multipolar expansion used to compute GWs,
and not to the expansion used to introduce RR effects in the
inspiral.
\bibitem{note43} The downward-arcing envelope of Fig.\ \ref{Ampl}b actually
corresponds to $\kappa_\mathrm{eff} \simeq -0.3$, possibly because the
ending frequency (a function of $\kappa_\mathrm{eff}$) plays into
SA.
\bibitem{note44} Because the template family SPAs is a subset of SPAc, SPAs
should always be more effectual. In fact, for $(7+5)M_\odot$ BBHs and
$(10+1.4)M_\odot$ NS--BH binaries, we find slightly higher numbers for
SPAs. In these lighter systems, SPAc has no advantage over SPAs
because the ending frequency is above the range of good detector
sensitivity, so the FFs should be the same. However, evaluating the
effectualness of SPAc requires a three-parameter numerical
maximization of the overlap, which is inevitably less precise than the
two-parameter maximization needed for SPAs. So numerical error
explains the discrepancy.
\bibitem{note38} We are implicitly assuming that the precession
frequency $\omega_p$ is much smaller than both the orbital frequency
and the effective radiation-reaction frequency $1/T_\mathrm{RR}$,
where $T_\mathrm{RR}$ is the radiation-reaction folding time.
\end{thebibliography}
\end{document}